\documentclass[12pt,english]{article}
\usepackage[T1]{fontenc}
\usepackage[latin9]{inputenc}
\usepackage{amssymb}
\usepackage{graphicx}
\usepackage{babel}

\makeatletter

\providecommand{\tabularnewline}{\\}

 \@addtoreset{equation}{section}

\providecommand{\tabularnewline}{\\}

\newif\ifContLineOne
\newif\ifContLineTwo
\newif\ifContLineThree

\def\conC#1{\vbox{\ialign{##\crcr
  \ifContLineThree\hrulefill\else\vphantom{\hrulefill}\fi\crcr
  \noalign{\kern3.2pt\nointerlineskip}
  \ifContLineTwo\hrulefill\else\vphantom{\hrulefill}\fi\crcr
  \noalign{\kern3.2pt\nointerlineskip}
  \ifContLineOne\hrulefill\else\vphantom{\hrulefill}\fi\crcr
  \noalign{\nointerlineskip}
  $\hfil\textstyle{\vbox to 14pt{}#1}\hfil$\crcr}}}

\def\DrawLeg#1#2{
  \kern-.2pt              
  \dimen2 =#1             
  \advance\dimen2 by 2pt  
  \dimen3 = 10.6pt        
  \dimen4 =3.6pt          
  \advance\dimen3 by -\dimen2 
  \multiply\dimen4 by #2
  \advance\dimen3 by \dimen4
  \raise\dimen2 \hbox{\vrule height\dimen3 width .4pt} 
  \kern-.2pt}             

\def\begC#1#2{\setbox0 =\hbox{$\textstyle{#2}$}
  \dimen0=.5\wd0 \dimen1=\ht0
  \conC{\hskip\dimen0}
  \count255=#1
  \ifnum\count255 =1 \ContLineOnetrue\else
  \ifnum\count255 =2 \ContLineTwotrue\else
  \ifnum\count255 =3 \ContLineThreetrue\fi\fi\fi
  \DrawLeg{\dimen1}{\count255}
  \conC{\hskip\dimen0}
  \kern-\dimen0\kern-\dimen0 \box0}

\def\endC#1#2{\setbox0 =\hbox{$\textstyle{#2}$}
  \dimen0=.5\wd0 \dimen1=\ht0
  \conC{\hskip\dimen0}
  \count255=#1
  \ifnum\count255 =1 \ContLineOnefalse\else
  \ifnum\count255 =2 \ContLineTwofalse\else
  \ifnum\count255 =3 \ContLineThreefalse\fi\fi\fi
  \DrawLeg{\dimen1}{\count255}
  \conC{\hskip\dimen0}
  \kern-\dimen0\kern-\dimen0 \box0}

\makeatother

\begin{document}
\begin{titlepage}

\global\long\def\thefootnote{\fnsymbol{footnote}}

\begin{flushright}
\begin{tabular}{l}
UTHEP-683 \tabularnewline
\end{tabular}
\par\end{flushright}

\bigskip{}

\begin{center}
\textbf{\Large{}{}{}{}Worldsheet theory of light-cone gauge 
noncritical strings on higher genus
Riemann surfaces}{\Large{}{}{} } 
\par\end{center}

\bigskip{}


\begin{center}
{\large{}{}{}{}{}{}{}Nobuyuki Ishibashi}$^{a}$%
\footnote{e-mail: ishibash@het.ph.tsukuba.ac.jp%
} {\large{}{}{}{}{}{}{}and Koichi Murakami}$^{b}$%
\footnote{e-mail: koichi@kushiro-ct.ac.jp%
} 
\par\end{center}

\begin{center}
$^{a}$\textit{Graduate School of Pure and Applied Sciences, University
of Tsukuba, }\\
 \textit{ Tsukuba, Ibaraki 305-8571, Japan}\\

\par\end{center}

\begin{center}
$^{b}$\textit{National Institute of Technology, Kushiro College,}\\
 \textit{ Otanoshike-Nishi 2-32-1, Kushiro, Hokkaido 084-0916, Japan
} 
\par\end{center}

\bigskip{}

\bigskip{}

\bigskip{}

\begin{abstract}
It is possible to formulate light-cone gauge string field theory in
noncritical dimensions. Such a theory corresponds to conformal gauge
worldsheet theory with nonstandard longitudinal part. We study the
longitudinal part of the worldsheet theory on higher genus Riemann
surfaces. The results in this paper shall be used to study the dimensional
regularization of light-cone gauge string field theory. 
\end{abstract}
\setcounter{footnote}{0} \global\long\def\thefootnote{\arabic{footnote}}

\end{titlepage}

\section{Introduction}

Since light-cone gauge string field theory is a completely gauge fixed
theory, there is no problem in formulating it in noncritical dimensions.
It should be possible to find the worldsheet theory in the conformal
gauge describing such a string theory, in which the spacetime Lorentz
invariance shall be broken. In \cite{Baba:2009ns,Baba:2009fi}, we
have constructed the longitudinal part of the worldsheet theory which
we call the $X^{\pm}$ CFT. The $X^{\pm}$ CFT turns out to be a conformal
field theory with the right central charge so that the whole worldsheet
theory is BRST invariant. The light-cone gauge superstring field theory
in noncritical dimensions can be used \cite{Baba:2009kr,Baba:2009zm,Ishibashi:2010nq,Ishibashi:2011fy}
to regularize the so-called contact term divergences \cite{Greensite:1986gv,Greensite:1987hm,Greensite:1987sm,Green:1987qu,Wendt:1987zh},
in the case of tree level amplitudes. The supersymmetric $X^{\pm}$
CFT plays crucial roles in studying such a regularization.

In this paper, we would like to study the $X^{\pm}$ CFT on higher
genus Riemann surfaces. In a previous paper \cite{Ishibashi:2013nma},
we have dealt with the bosonic $X^{\pm}$ CFT on higher genus Riemann
surfaces, but we have not investigated its properties in detail. In
this paper, we will define and calculate the correlation functions
of bosonic and supersymmetric $X^{\pm}$ CFT on higher genus Riemann
surfaces and explore various properties of the theory. The results
in this paper will be used in a forthcoming publication, in which
we discuss the dimensional regularization of the multiloop amplitudes
of light-cone gauge superstring field theory.

The organization of this paper is as follows. In section \ref{sec:Bosonic--CFT},
the bosonic $X^{\pm}$ CFT is studied. We calculate the correlation
functions based on the results in \cite{Ishibashi:2013nma}. In section
\ref{sec:Supersymmetric--CFT}, we deal with the supersymmetric $X^{\pm}$
CFT. In \cite{Baba:2009fi}, we have given a way to calculate the
correlation functions of the supersymmetric $X^{\pm}$ CFT on a surface
of genus 0, but it is a bit unwieldy. In this paper, we develop an
alternative method to calculate them, apply it to higher genus case
and explore various properties of the supersymmetric $X^{\pm}$ CFT.
Section \ref{sec:Discussions} is devoted to discussions. In the appendices,
we give details of definitions and calculations which are not included
in the text.

\section{Bosonic $X^{\pm}$ CFT\label{sec:Bosonic--CFT}}

\begin{figure}
\begin{centering}
\includegraphics[scale=1.4]{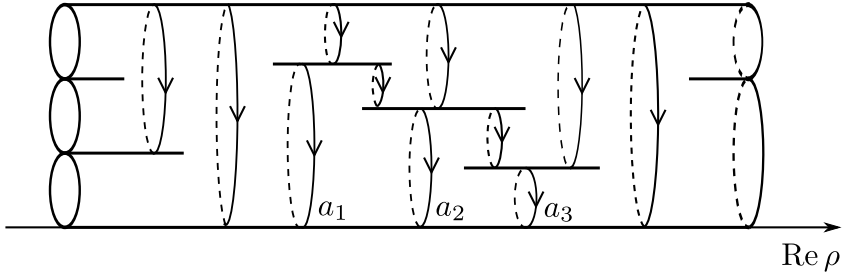} 
\par\end{centering}

\protect\protect\protect\protect\protect\protect\protect\caption{A string diagram with $3$ incoming, $2$ outgoing strings and $3$
loops. \label{fig:A-string-diagram}}
\end{figure}

It is straightforward to calculate the amplitudes of light-cone gauge
bosonic string field theory perturbatively by using the old-fashioned
perturbation theory and Wick rotation. Each term in the expansion
corresponds to a light-cone gauge Feynman diagram for strings. A typical
diagram is depicted in Figure \ref{fig:A-string-diagram}. A Wick
rotated $g$-loop $N$-string diagram is conformally equivalent to
an $N$ punctured genus $g$ Riemann surface $\Sigma$. The amplitudes
are given by an integral of correlation functions of vertex operators
on $\Sigma$ over the moduli parameters.

As has been shown in \cite{Baba:2009ns,Ishibashi:2013nma}, the amplitudes
in $d\ne26$ dimensions can be cast into the conformal gauge expression
using the worldsheet theory with the field contents 
\begin{equation}
X^{+},\quad X^{-},\quad X^{i},\quad b,\quad c,\quad\bar{b},\quad\bar{c}\,,\label{eq:conformalcontent}
\end{equation}
in which the reparametrization ghosts $b,c,\bar{b},\bar{c}$ and the
longitudinal variables $X^{\pm}$ are added to the original light-cone
variables $X^{i}$ $(i=1,\ldots,d-2)$. The worldsheet action for
the longitudinal variables is given by 
\begin{equation}
S\left[\hat{g}_{z\bar{z}},X^{\pm}\right]=-\frac{1}{4\pi}\int dz\wedge d\bar{z}i\left(\partial X^{+}\bar{\partial}X^{-}+\partial X^{-}\bar{\partial}X^{+}\right)+\frac{d-26}{24}\Gamma\left[\hat{g}_{z\bar{z}},X^{+}\right]\,.\label{eq:bosonicXpmaction}
\end{equation}
Here the metric on the worldsheet is taken to be $ds^{2}=2\hat{g}_{z\bar{z}}dzd\bar{z}$
and $\Gamma$ is the Liouville action 
\begin{equation}
\Gamma\left[\hat{g}_{z\bar{z}},X^{+}\right]=-\frac{1}{2\pi}\int dz\wedge d\bar{z}i\left(\partial\phi\bar{\partial}\phi+\hat{g}_{z\bar{z}}\hat{R}\phi\right)\,,\label{eq:Liouville}
\end{equation}
where the Liouville field $\phi$ is given by 
\begin{equation}
\phi\equiv\ln\left(-4\partial X^{+}\bar{\partial}X^{+}\right)-\ln\left(2\hat{g}_{z\bar{z}}\right)\,,
\end{equation}
and $\hat{R}$ is the scalar curvature derived from the metric $\hat{g}$.
The theory with the action (\ref{eq:bosonicXpmaction}) turns out
to be a conformal field theory which we call the $X^{\pm}$ CFT.

In order for the action to be well-defined, $ds^{2}=-4\partial X^{+}\bar{\partial}X^{+}dzd\bar{z}$
should be a well-defined metric on the worldsheet at least at generic
points. Hence we should always consider the theory in the presence
of the vertex operator insertions 
\begin{equation}
\prod_{r=1}^{N}e^{-ip_{r}^{+}X^{-}}\left(Z_{r},\bar{Z}_{r}\right)\,,\label{eq:eipX-}
\end{equation}
with $p_{r}^{+}\ne0$ $\left(r=1,\cdots,N\right)$ and $N\geq3$.
The amplitudes with such insertions correspond to light-cone diagrams
with external lines at $z=Z_{r}$ which have string length $\alpha_{r}=2p_{r}^{+}$.
With the insertion of these vertex operators, $X^{+}$ possesses a
classical background 
\begin{equation}
X_{\mathrm{cl}}^{+}\left(z,\bar{z}\right)=-\frac{i}{2}\left(\rho\left(z\right)+\bar{\rho}\left(\bar{z}\right)\right)\,,\label{eq:X+cl-bosonic}
\end{equation}
where $\rho\left(z\right)$ is given by 
\begin{equation}
\rho(z)=\sum_{r=1}^{N}\alpha_{r}\left[\ln E(z,Z_{r})-2\pi i\int_{P_{0}}^{z}\omega\frac{1}{\mathop{\mathrm{Im}}\Omega}\mathop{\mathrm{Im}}\int_{P_{0}}^{Z_{r}}\omega\right]\,.\label{eq:rhoz}
\end{equation}
Here $E(z,w)$ is the prime form, $\omega$ is the canonical basis
of the holomorphic abelian differentials and $\Omega$ is the period
matrix of the surface.%
\footnote{For the mathematical background relevant for string perturbation theory,
we refer the reader to \cite{D'Hoker:1988ta}. %
} The base point $P_{0}$ is arbitrary. For notational convenience,
we introduce 
\begin{equation}
g\left(z,w\right)\equiv\ln E(z,w)-2\pi i\int_{P_{0}}^{z}\omega\frac{1}{\mathop{\mathrm{Im}}\Omega}\mathop{\mathrm{Im}}\int_{P_{0}}^{w}\omega\,,\label{eq:gzw}
\end{equation}
so that (\ref{eq:rhoz}) can be expressed as 
\begin{equation}
\rho\left(z\right)=\sum_{r=1}^{N}\alpha_{r}g\left(z,Z_{r}\right)\,.
\end{equation}
Notice that $g\left(z,w\right)$ is a function of $z$ and not $\bar{z}$,
but that of both of $w$ and $\bar{w}$.

$\rho(z)$ coincides with the coordinate on the light-cone diagram
defined as follows. A light-cone diagram consists of cylinders which
correspond to propagators of closed strings. On each cylinder, one
can introduce a complex coordinate $\rho$ whose real part coincides
with the Wick rotated light-cone time $iX^{+}$ and imaginary part
parametrizes the closed string at each time. The $\rho$'s on the
cylinders are smoothly connected except at the interaction points
and we get a complex coordinate $\rho$ on $\Sigma$. $\rho$ is not
a good coordinate around the punctures and the interaction points
on the light-cone diagram. The interaction points $z=z_{I}$ $(I=1,\cdots,2g-2+N)$
are characterized by the equation 
\begin{equation}
\partial\rho(z_{I})=0\,.
\end{equation}
Since 
\begin{equation}
ds^{2}=-4\partial X_{\mathrm{cl}}^{+}\bar{\partial}X_{\mathrm{cl}}^{+}dzd\bar{z}=\left|\partial\rho\right|^{2}dzd\bar{z}\label{eq:drhometric}
\end{equation}
provides a well-defined metric on the worldsheet except for the points
$z=Z_{r}$, $z_{I}$, we can make $\Gamma$ well-defined.

\subsection{Correlation functions on higher genus Riemann surfaces}

As has been demonstrated in \cite{Baba:2009ns}, all the properties
of the worldsheet theory of the longitudinal variables can be deduced
from the correlation function of the form 
\begin{eqnarray}
\lefteqn{\left\langle \prod_{r=1}^{N}e^{-ip_{r}^{+}X^{-}}(Z_{r}.\bar{Z}_{r})\prod_{s=1}^{M}e^{-ip_{s}^{-}X^{+}}(w_{s}.\bar{w}_{s})\right\rangle _{\hat{g}_{z\bar{z}}}^{X^{\pm}}}\nonumber \\
 &  & \equiv\left(Z^{X}[\hat{g}_{z\bar{z}}]\right)^{-2}\int\left[dX^{+}dX^{-}\right]_{\hat{g}_{z\bar{z}}}e^{-S^{\pm}\left[\hat{g}_{z\bar{z}}\right]}\prod_{r=1}^{N}e^{-ip_{r}^{+}X^{-}}(Z_{r},\bar{Z}_{r})\prod_{s=1}^{M}e^{-ip_{s}^{-}X^{+}}(w_{s},\bar{w}_{s})\,.~~~~~~\label{eq:Xpmcorrdef}
\end{eqnarray}
The correlation functions are normalized by being divided by the factor
\begin{equation}
\left(Z^{X}[\hat{g}_{z\bar{z}}]\right)^{2}\equiv\left(\frac{8\pi^{2}\det'\left(-\hat{g}^{z\bar{z}}\partial_{z}\partial_{\bar{z}}\right)}{\int dz\wedge d\bar{z}\sqrt{\hat{g}}}\right)^{-1}\,,\label{eq:def-ZX}
\end{equation}
which coincides with the partition function of the worldsheet theory
when $d=26.$ As is explained in appendix~\ref{sec:Definition-of-the},
taking the integration contours of $X^{\pm}$ appropriately, we can
evaluate it and obtain 
\begin{eqnarray}
\lefteqn{\left\langle \prod_{r=1}^{N}e^{-ip_{r}^{+}X^{-}}(Z_{r},\bar{Z}_{r})\prod_{s=1}^{M}e^{-ip_{s}^{-}X^{+}}(w_{s},\bar{w}_{s})\right\rangle _{\hat{g}_{z\bar{z}}}^{X^{\pm}}}\nonumber \\
 &  & =(2\pi)^{2}\delta\left(\sum_{s}p_{s}^{-}\right)\delta\left(\sum_{r}p_{r}^{+}\right)\prod_{s}e^{-p_{s}^{-}\frac{\rho+\bar{\rho}}{2}}(w_{s},\bar{w}_{s})\, e^{-\frac{d-26}{24}\Gamma\left[\hat{g}_{z\bar{z}},\,-\frac{i}{2}\left(\rho+\bar{\rho}\right)\right]}~.\label{eq:Xpmcorr}
\end{eqnarray}
Therefore we need to calculate $\Gamma\left[\hat{g}_{z\bar{z}},\,-\frac{i}{2}\left(\rho+\bar{\rho}\right)\right]$
to get the correlation function.

\begin{figure}
\begin{centering}
\includegraphics[scale=0.6]{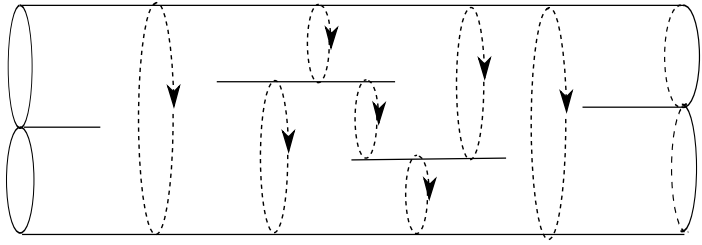} 
\par\end{centering}

\protect\protect\protect\protect\protect\protect\protect\caption{\label{fig:The-contours-.}The contours $C_{\mathcal{I}}$.}
\end{figure}

Since the metric (\ref{eq:drhometric}) is singular at $z=Z_{r},z_{I}$
as mentioned above, one gets a divergent result if one naively substitutes
$-\frac{i}{2}\left(\rho+\bar{\rho}\right)$ into $X^{+}$ in (\ref{eq:Liouville}).
One way to deal with the divergences may be to regularize them as
was done in \cite{Mandelstam:1985ww}. An alternative way is to integrate
the variation formula 
\begin{equation}
\delta\left(-\Gamma\right)=\sum_{\mathcal{I}}\delta\mathcal{T}_{\mathcal{I}}\oint_{C_{\mathcal{I}}}\frac{dz}{2\pi i}\frac{1}{\partial\rho}T^{\mathrm{Liouville}}(z)+\mathrm{c.c.}\,.\label{eq:deltaGamma}
\end{equation}
Here $\mathcal{I}$ labels the internal lines of the light-cone diagram
$\Sigma$ and $C_{\mathcal{I}}$ denotes the contour going around
it as depicted in Figure \ref{fig:The-contours-.}. $\mathcal{T}_{\mathcal{I}}$
is defined as 
\begin{equation}
\mathcal{T}_{\mathcal{I}}=T_{\mathcal{I}}+i\alpha_{\mathcal{I}}\theta_{\mathcal{I}}\,,\label{eq:bosonic-moduli-TI}
\end{equation}
where $T_{\mathcal{I}}$ denotes the length of the $\mathcal{I}$-th
internal line and $\alpha_{\mathcal{I}}$, $\theta_{\mathcal{I}}$
denote the string-length and the twist angle for the propagator. $\delta\mathcal{T}_{\mathcal{I}}$'s
and $\delta\bar{\mathcal{T}}_{\mathcal{I}}$'s should satisfy some
linear constraints so that the variation corresponds to that of the
shape of a light-cone diagram. $T^{\mathrm{Liouville}}(z)$ denotes
the energy-momentum tensor corresponding to the Liouville action (\ref{eq:Liouville})
given as 
\begin{equation}
T^{\mathrm{Liouville}}(z)=\left(\partial\phi(z)\right)^{2}-2\left(\partial-\partial\ln\hat{g}_{z\bar{z}}\right)\partial\phi(z)\,,\label{eq:TLiouvlle}
\end{equation}
where $\phi$ is now given as 
\begin{equation}
\phi=\ln\left|\partial\rho\right|^{2}-\ln\left(2\hat{g}_{z\bar{z}}\right)\,.\label{eq:chirho}
\end{equation}
It is possible to calculate the right hand side of (\ref{eq:deltaGamma})
and integrate it with respect to the variation to get $\Gamma$. By
doing so, we can fix the form of $\Gamma$ as a function of the parameter
$\mathcal{T}_{\mathcal{I}}$'s. Imposing the factorization conditions
in the limit where some of the $\mathcal{T}_{\mathcal{I}}$'s become
infinity, it is possible to fix $\Gamma$ completely. By this method,
we can calculate $\Gamma$ without encountering divergent constants.

Such a computation was performed in \cite{Ishibashi:2013nma} and
we can evaluate $\Gamma$ by using the results. The energy-momentum
tensor (\ref{eq:TLiouvlle}) with $\phi$ in (\ref{eq:chirho}) can
be rewritten as 
\begin{eqnarray}
T^{\mathrm{Liouville}} & = & \left(\partial\ln\left|\partial\rho\right|^{2}\right)^{2}-2\partial^{2}\ln\left|\partial\rho\right|^{2}-\left(\partial\ln\hat{g}_{z\bar{z}}\right)^{2}+2\partial^{2}\ln\hat{g}_{z\bar{z}}\nonumber \\
 & = & -2\left\{ \rho,z\right\} -\left(\left(\partial\ln\hat{g}_{z\bar{z}}\right)^{2}-2\partial^{2}\ln\hat{g}_{z\bar{z}}\right)\,,
\end{eqnarray}
where 
\begin{equation}
\left\{ \rho,z\right\} =\frac{\partial^{3}\rho}{\partial\rho}-\frac{3}{2}\left(\frac{\partial^{2}\rho}{\partial\rho}\right)^{2}
\end{equation}
is the Schwarzian derivative. In \cite{Ishibashi:2013nma}, we have
calculated $Z^{\mathrm{LC}}$ which satisfies 
\begin{equation}
\delta\ln Z^{\mathrm{LC}}=\sum_{\mathcal{I}}\delta\mathcal{T}_{\mathcal{I}}\oint_{C_{\mathcal{I}}}\frac{dz}{2\pi i}\frac{1}{\partial\rho(z)}\left(24\left\langle T^{X}(z)\right\rangle -2\left\{ \rho,z\right\} \right)+\mathrm{c.c.}\,,\label{eq:deltaZLC}
\end{equation}
where $\left\langle T^{X}(z)\right\rangle $ denotes the expectation
value of the energy-momentum tensor of a free boson $X$. On the other
hand, the partition function $Z^{X}[\hat{g}_{z\bar{z}}]$ satisfies
\begin{equation}
\delta\ln Z^{X}\left[\hat{g}_{z\bar{z}}\right]=\sum_{\mathcal{I}}\delta\mathcal{T}_{\mathcal{I}}\oint_{C_{\mathcal{I}}}\frac{dz}{2\pi i}\frac{1}{\partial\rho}\left(\left\langle T^{X}(z)\right\rangle +\frac{1}{24}\left(\left(\partial\ln\hat{g}_{z\bar{z}}\right)^{2}-2\partial^{2}\ln\hat{g}_{z\bar{z}}\right)\right)+\mathrm{c.c.}\,.\label{eq:deltaZX}
\end{equation}
Comparing (\ref{eq:deltaZLC}), (\ref{eq:deltaZX}) and (\ref{eq:deltaGamma}),
we get 
\begin{equation}
e^{-\Gamma}\propto Z^{\mathrm{LC}}\left(Z^{X}\right)^{-24}\,,
\end{equation}
up to a possibly divergent multiplicative factor.

Taking $\hat{g}_{z\bar{z}}$ to be the Arakelov metric $g_{z\bar{z}}^{\mathrm{A}}$,
$Z^{X}[\hat{g}_{z\bar{z}}]$ was calculated in \cite{AlvarezGaume:1987vm,Verlinde:1986kw,Dugan:1987qe,Sonoda:1987ra,Wentworth:1991,Wentworth:2008}
and its explicit form is 
\begin{equation}
Z^{X}[g_{z\bar{z}}^{\mathrm{A}}]^{24}=e^{c_{g}}e^{2\delta(\Sigma)}~,
\end{equation}
where $\delta(\Sigma)$ is the Faltings' invariant \cite{Faltings:1984}
defined by 
\begin{eqnarray}
e^{-\frac{1}{4}\delta(\Sigma)} & = & \left(\det\mathop{\mathrm{Im}}\Omega\right)^{\frac{3}{2}}\left|\theta[\zeta](0|\Omega)\right|^{2}\frac{\prod_{i=1}^{g}\left(2g_{\hat{z}_{i}\bar{\hat{z}}_{i}}^{\mathrm{A}}\right)}{\left|\det\omega_{j}(\hat{z}_{i})\right|^{2}}\nonumber \\
 &  & \ \times\exp\left[-\sum_{i<j}G^{\mathrm{A}}(\hat{z}_{i};\hat{z}_{j})+\sum_{i}G^{\mathrm{A}}(\hat{z}_{i};\hat{w})\right]\,,
\end{eqnarray}
and $c_{g}$ is a numerical constant which depends on $g$. Here $\hat{z}_{i}$
$(i=1,\ldots,g)$ and $\hat{w}$ are arbitrary points on $\Sigma$,
and 
\begin{equation}
\zeta\equiv\sum_{i=1}^{g}\int_{P_{0}}^{\hat{z}_{i}}\omega-\int_{P_{0}}^{\hat{w}}\omega-\Delta~.
\end{equation}
$\Delta$ denotes the vector of Riemann constants for $P_{0}$. The
definitions of the Arakelov metric $g_{z\bar{z}}^{\mathrm{A}}$ and
the Arakelov Green's function $G^{\mathrm{A}}(z;w)$ are given in
appendix~\ref{sec:Arakelov-metric-and}. Also taking $\hat{g}_{z\bar{z}}$
to be the Arakelov metric, we obtain \cite{Ishibashi:2013nma} 
\begin{equation}
Z^{\mathrm{LC}}=\frac{1}{(32\pi^{2})^{4h}}e^{2\delta(\Sigma)}e^{-W}\prod_{r}e^{-2\mathop{\mathrm{Re}}\bar{N}_{00}^{rr}}\prod_{I}\left|\partial^{2}\rho\left(z_{I}\right)\right|^{-3}\,,\label{eq:ZLC}
\end{equation}
where 
\begin{eqnarray}
-W & \equiv & -2\sum_{I<J}G^{\mathrm{A}}\left(z_{I};z_{J}\right)-2\sum_{r<s}G^{\mathrm{A}}\left(Z_{r};Z_{s}\right)+2\sum_{I,r}G^{\mathrm{A}}\left(z_{I};Z_{r}\right)\nonumber \\
 &  & {}-\sum_{r}\ln\left(2g_{Z_{r}\bar{Z}_{r}}^{\mathrm{A}}\right)+3\sum_{I}\ln\left(2g_{z_{I}\bar{z}_{I}}^{\mathrm{A}}\right)\,.\label{eq:-W}
\end{eqnarray}
$\bar{N}_{00}^{rr}$ denotes one of the Neumann coefficients and is
given by 
\begin{eqnarray}
\bar{N}_{00}^{rr} & \equiv & \lim_{z\to Z_{r}}\left[\frac{\rho(z_{I^{(r)}})-\rho(z)}{\alpha_{r}}+\ln(z-Z_{r})\right]\nonumber \\
 & = & \frac{\rho(z_{I^{(r)}})}{\alpha_{r}}-\sum_{s\neq r}\frac{\alpha_{s}}{\alpha_{r}}\ln E(Z_{r},Z_{s})+\frac{2\pi i}{\alpha_{r}}\int_{P_{0}}^{Z_{r}}\omega\frac{1}{\mathop{\mathrm{Im}}\Omega}\sum_{s=1}^{N}\alpha_{s}\mathop{\mathrm{Im}}\int_{P_{0}}^{Z_{s}}\omega~,\label{eq:Neumann00-holo-1}
\end{eqnarray}
and $z_{I^{(r)}}$ denotes the coordinate of the interaction point
at which the $r$-th external line interacts. Therefore we get 
\begin{eqnarray}
e^{-\Gamma} & \propto & Z^{\mathrm{LC}}e^{-2\delta(\Sigma)}\nonumber \\
 & = & \frac{1}{(32\pi^{2})^{4g}}e^{-W}\prod_{r}e^{-2\mathop{\mathrm{Re}}\bar{N}_{00}^{rr}}\prod_{I}\left|\partial^{2}\rho\left(z_{I}\right)\right|^{-3}\,,\label{eq:Gammabosonic}
\end{eqnarray}
and fix the right hand side of (\ref{eq:Xpmcorr}) to be 
\begin{eqnarray}
\lefteqn{\left\langle \prod_{r=1}^{N}e^{-ip_{r}^{+}X^{-}}(Z_{r},\bar{Z}_{r})\prod_{s=1}^{M}e^{-ip_{s}^{-}X^{+}}(w_{s},\bar{w}_{s})\right\rangle _{\hat{g}_{z\bar{z}}}^{X^{\pm}}}\nonumber \\
 & = & (2\pi)^{2}\delta\left(\sum_{s}p_{s}^{-}\right)\delta\left(\sum_{r}p_{r}^{+}\right)\prod_{s}e^{-p_{s}^{-}\frac{\rho+\bar{\rho}}{2}}(w_{s},\bar{w}_{s})\nonumber \\
 &  & \qquad\times\left(\frac{1}{(32\pi^{2})^{4g}}e^{-W}\prod_{r}e^{-2\mathop{\mathrm{Re}}\bar{N}_{00}^{rr}}\prod_{I}\left|\partial^{2}\rho\left(z_{I}\right)\right|^{-3}\right)^{\frac{d-26}{24}}~.\label{eq:Xpmcorr2}
\end{eqnarray}

Once we know the correlation function of the form (\ref{eq:Xpmcorr2}),
it is possible to calculate other correlation functions by differentiating
it with respect to $p_{r}^{+},\, p_{s}^{-}$. For example, 
\begin{eqnarray}
\lefteqn{\left\langle \partial X^{+}\left(w\right)\prod_{r=1}^{N}e^{-ip_{r}^{+}X^{-}}(Z_{r},\bar{Z}_{r})\prod_{s=1}^{M}e^{-ip_{s}^{-}X^{+}}(w_{s},\bar{w}_{s})\right\rangle _{\hat{g}_{z\bar{z}}}^{X^{\pm}}}\nonumber \\
 &  & =\left.i\partial_{w_{0}}\partial_{p_{0}^{-}}\left\langle \prod_{r=1}^{N}e^{-ip_{r}^{+}X^{-}}(Z_{r},\bar{Z}_{r})\prod_{s=0}^{M+1}e^{-ip_{s}^{-}X^{+}}(w_{s},\bar{w}_{s})\right\rangle _{\hat{g}_{z\bar{z}}}^{X^{\pm}}\right|_{p_{0}^{-}=0,w_{0}=w}\,,
\end{eqnarray}
where we take $p_{M+1}^{-}=-p_{0}^{-}$. From (\ref{eq:Xpmcorr2}),
we get 
\begin{eqnarray}
\lefteqn{\left\langle \partial X^{+}\left(w\right)\prod_{r=1}^{N}e^{-ip_{r}^{+}X^{-}}(Z_{r},\bar{Z}_{r})\prod_{s=1}^{M}e^{-ip_{s}^{-}X^{+}}(w_{s},\bar{w}_{s})\right\rangle _{\hat{g}_{z\bar{z}}}^{X^{\pm}}}\nonumber \\
 &  & =-\frac{i}{2}\partial\rho\left(w\right)\left\langle \prod_{r=1}^{N}e^{-ip_{r}^{+}X^{-}}(Z_{r},\bar{Z}_{r})\prod_{s=1}^{M}e^{-ip_{s}^{-}X^{+}}(w_{s},\bar{w}_{s})\right\rangle _{\hat{g}_{z\bar{z}}}^{X^{\pm}}\,.
\end{eqnarray}
It is easy to see that for any functional $F\left[X^{+}\right]$ that
can be expressed in terms of the derivatives of $X^{+}$ and the Fourier
modes $e^{-ipX^{+}}$ satisfying 
\begin{equation}
F[X^{+}+c]=F[X^{+}]\qquad\left(c=\mbox{const.}\right)\,,
\end{equation}
the following equation holds, 
\begin{eqnarray}
\lefteqn{\left\langle F\left[X^{+}\right]\prod_{r=1}^{N}e^{-ip_{r}^{+}X^{-}}(Z_{r},\bar{Z}_{r})\prod_{s=1}^{M}e^{-ip_{s}^{-}X^{+}}(w_{s},\bar{w}_{s})\right\rangle _{\hat{g}_{z\bar{z}}}^{X^{\pm}}}\nonumber \\
 &  & =F\left[-\frac{i}{2}\left(\rho+\bar{\rho}\right)\right]\left\langle \prod_{r=1}^{N}e^{-ip_{r}^{+}X^{-}}(Z_{r},\bar{Z}_{r})\prod_{s=1}^{M}e^{-ip_{s}^{-}X^{+}}(w_{s},\bar{w}_{s})\right\rangle _{\hat{g}_{z\bar{z}}}^{X^{\pm}}\,.\label{eq:FX+}
\end{eqnarray}
This implies that the expectation value of $X^{+}\left(z,\bar{z}\right)$
is equal to $-\frac{i}{2}\left(\rho\left(z\right)+\bar{\rho}\left(\bar{z}\right)\right)$.

The correlation functions involving $X^{-}$ can be evaluated in the
same way. For example, 
\begin{eqnarray}
\lefteqn{\left\langle \partial X^{-}\left(z\right)\prod_{r=1}^{N}e^{-ip_{r}^{+}X^{-}}(Z_{r},\bar{Z}_{r})\prod_{s=1}^{M}e^{-ip_{s}^{-}X^{+}}(w_{s},\bar{w}_{s})\right\rangle _{\hat{g}_{z\bar{z}}}^{X^{\pm}}}\nonumber \\
 &  & =\left.i\partial_{Z_{0}}\partial_{p_{0}^{+}}\left\langle \prod_{r=0}^{N+1}e^{-ip_{r}^{+}X^{-}}(Z_{r},\bar{Z}_{r})\prod_{s=1}^{M}e^{-ip_{s}^{-}X^{+}}(w_{s},\bar{w}_{s})\right\rangle _{\hat{g}_{z\bar{z}}}^{X^{\pm}}\right|_{p_{0}^{+}=0,Z_{0}=z}\,,
\end{eqnarray}
with $p_{N+1}^{+}=-p_{0}^{+}$ and we get from (\ref{eq:Xpmcorr2})
\begin{eqnarray}
\lefteqn{\left\langle \partial X^{-}\left(z\right)\prod_{r=1}^{N}e^{-ip_{r}^{+}X^{-}}(Z_{r},\bar{Z}_{r})\prod_{s=1}^{M}e^{-ip_{s}^{-}X^{+}}(w_{s},\bar{w}_{s})\right\rangle _{\hat{g}_{z\bar{z}}}^{X^{\pm}}}\nonumber \\
 & = & \left[\sum_{s=1}^{M}\left(-ip_{s}^{-}\right)\partial_{z}g\left(z,w_{s}\right)+\left.i\partial_{Z_{0}}\partial_{p_{0}^{+}}\left(-\frac{d-26}{24}\Gamma\left[\hat{g}_{z\bar{z}},\,-\frac{i}{2}\left(\rho^{\prime}+\bar{\rho}^{\prime}\right)\right]\right)\right|_{p_{0}^{+}=0,Z_{0}=z}\right]\nonumber \\
 &  & \quad\times\left\langle \prod_{r=1}^{N}e^{-ip_{r}^{+}X^{-}}(Z_{r},\bar{Z}_{r})\prod_{s=1}^{M}e^{-ip_{s}^{-}X^{+}}(w_{s},\bar{w}_{s})\right\rangle _{\hat{g}_{z\bar{z}}}^{X^{\pm}}\,.\label{eq:derX-z}
\end{eqnarray}
Here we have introduced 
\begin{eqnarray}
\rho'(z) & \equiv & \sum_{r=0}^{N+1}\alpha_{r}g\left(z,Z_{r}\right)\nonumber \\
 & = & \rho(z)+\alpha_{0}\left(g\left(z,Z_{0}\right)-g\left(z,Z_{N+1}\right)\right)\,.\label{eq:rhoprime}
\end{eqnarray}
(\ref{eq:derX-z}) can be rewritten as 
\begin{eqnarray}
\lefteqn{\left\langle \partial X^{-}\left(z\right)\prod_{r=1}^{N}e^{-ip_{r}^{+}X^{-}}(Z_{r},\bar{Z}_{r})\prod_{s=1}^{M}e^{-ip_{s}^{-}X^{+}}(w_{s},\bar{w}_{s})\right\rangle _{\hat{g}_{z\bar{z}}}^{X^{\pm}}}\nonumber \\
 &  & =\left[\sum_{s=1}^{M}\left(-ip_{s}^{-}\right)\begC1{\partial X^{-}}\conC{\left(z\right)}\endC1{X^{+}}\left(w_{s}\right)+\left\langle \partial X^{-}\left(z\right)\right\rangle _{\rho}\right]\nonumber \\
 &  & \hphantom{\quad=}\times\left\langle \prod_{r=1}^{N}e^{-ip_{r}^{+}X^{-}}(Z_{r},\bar{Z}_{r})\prod_{s=1}^{M}e^{-ip_{s}^{-}X^{+}}(w_{s},\bar{w}_{s})\right\rangle _{\hat{g}_{z\bar{z}}}^{X^{\pm}}\,,
\end{eqnarray}
where 
\begin{eqnarray}
\begC1{\partial X^{-}}\conC{\left(z\right)}\endC1{X^{+}}\left(w_{s}\right) & = & \partial_{z}g\left(z,w_{s}\right)\,,\nonumber \\
\left\langle \partial X^{-}\left(z\right)\right\rangle _{\rho} & = & \left.i\partial_{Z_{0}}\partial_{p_{0}^{+}}\left(-\frac{d-26}{24}\Gamma\left[\hat{g}_{z\bar{z}},\,-\frac{i}{2}\left(\rho^{\prime}+\bar{\rho}^{\prime}\right)\right]\right)\right|_{p_{0}^{+}=0,Z_{0}=z}\,.
\end{eqnarray}
$\left\langle \partial X^{-}\left(z\right)\right\rangle _{\rho}$
can be formally written as 
\begin{equation}
\left\langle \partial X^{-}\left(z\right)\right\rangle _{\rho}=\frac{\left\langle \partial X^{-}\left(z\right)\prod_{r=1}^{N}e^{-ip_{r}^{+}X^{-}}(Z_{r},\bar{Z}_{r})\right\rangle _{\hat{g}_{z\bar{z}}}^{X^{\pm}}}{\left\langle \prod_{r=1}^{N}e^{-ip_{r}^{+}X^{-}}(Z_{r},\bar{Z}_{r})\right\rangle _{\hat{g}_{z\bar{z}}}^{X^{\pm}}}\,.
\end{equation}
We can go ahead and calculate the correlation function involving two
$\partial X^{-}$'s as 
\begin{eqnarray}
\lefteqn{\left\langle \partial X^{-}\left(Z_{0}\right)\partial X^{-}\left(z\right)\prod_{r=1}^{N}e^{-ip_{r}^{+}X^{-}}(Z_{r},\bar{Z}_{r})\prod_{s=1}^{M}e^{-ip_{s}^{-}X^{+}}(w_{s},\bar{w}_{s})\right\rangle _{\hat{g}_{z\bar{z}}}^{X^{\pm}}}\nonumber \\
 & = & \left.i\partial_{Z_{0}}\partial_{p_{0}^{+}}\left\langle \partial X^{-}\left(z\right)\prod_{r=0}^{N+1}e^{-ip_{r}^{+}X^{-}}(Z_{r},\bar{Z}_{r})\prod_{s=1}^{M}e^{-ip_{s}^{-}X^{+}}(w_{s},\bar{w}_{s})\right\rangle _{\hat{g}_{z\bar{z}}}^{X^{\pm}}\right|_{p_{0}^{+}=0}\nonumber \\
 & = & \left[\left(\sum_{s=1}^{M}\left(-ip_{s}^{-}\right)\begC1{\partial X^{-}}\conC{\left(Z_{0}\right)}\endC1{X^{+}}\left(w_{s}\right)+\left\langle \partial X^{-}\left(Z_{0}\right)\right\rangle _{\rho}\right)\right.\nonumber \\
 &  & \qquad\times\left(\sum_{s=1}^{M}\left(-ip_{s}^{-}\right)\begC1{\partial X^{-}}\conC{\left(z\right)}\endC1{X^{+}}\left(w_{s}\right)+\left\langle \partial X^{-}\left(z\right)\right\rangle _{\rho}\right)\nonumber \\
 &  & \qquad\qquad\quad\left.\vphantom{\left(\sum_{s=1}^{M}\left(-ip_{s}^{-}\right)\begC1{\partial X^{-}}\conC{\left(Z_{0}\right)}\endC1{X^{+}}\left(w_{s}\right)+\left\langle \partial X^{-}\left(Z_{0}\right)\right\rangle _{\rho}\right)}+\left\langle \partial X^{-}\left(Z_{0}\right)\partial X^{-}\left(z\right)\right\rangle _{\rho}^{\mathrm{c}}\right]\nonumber \\
 &  & \times\left\langle \prod_{r=1}^{N}e^{-ip_{r}^{+}X^{-}}(Z_{r},\bar{Z}_{r})\prod_{s=1}^{M}e^{-ip_{s}^{-}X^{+}}(w_{s},\bar{w}_{s})\right\rangle _{\hat{g}_{z\bar{z}}}^{X^{\pm}}\,,
\end{eqnarray}
where 
\begin{equation}
\left\langle \partial X^{-}\left(Z_{0}\right)\partial X^{-}\left(z\right)\right\rangle _{\rho}^{\mathrm{c}}=\left.i\partial_{Z_{0}}\partial_{p_{0}^{+}}\left(\left\langle \partial X^{-}\left(z\right)\right\rangle _{\rho^{\prime}}\right)\right|_{p_{0}^{+}=0}\,,
\end{equation}
which can be formally written as 
\begin{eqnarray}
\lefteqn{\left\langle \partial X^{-}\left(Z_{0}\right)\partial X^{-}\left(z\right)\right\rangle _{\rho}^{\mathrm{c}}}\nonumber \\
 & = & \frac{\left\langle \partial X^{-}\left(Z_{0}\right)\partial X^{-}\left(z\right)\prod_{r=1}^{N}e^{-ip_{r}^{+}X^{-}}(Z_{r},\bar{Z}_{r})\right\rangle _{\hat{g}_{z\bar{z}}}^{X^{\pm}}}{\left\langle \prod_{r=1}^{N}e^{-ip_{r}^{+}X^{-}}(Z_{r},\bar{Z}_{r})\right\rangle _{\hat{g}_{z\bar{z}}}^{X^{\pm}}}\nonumber \\
 &  & {}-\frac{\left\langle \partial X^{-}\left(Z_{0}\right)\prod_{r=1}^{N}e^{-ip_{r}^{+}X^{-}}(Z_{r},\bar{Z}_{r})\right\rangle _{\hat{g}_{z\bar{z}}}^{X^{\pm}}}{\left\langle \prod_{r=1}^{N}e^{-ip_{r}^{+}X^{-}}(Z_{r},\bar{Z}_{r})\right\rangle _{\hat{g}_{z\bar{z}}}^{X^{\pm}}}\cdot\frac{\left\langle \partial X^{-}\left(z\right)\prod_{r=1}^{N}e^{-ip_{r}^{+}X^{-}}(Z_{r},\bar{Z}_{r})\right\rangle _{\hat{g}_{z\bar{z}}}^{X^{\pm}}}{\left\langle \prod_{r=1}^{N}e^{-ip_{r}^{+}X^{-}}(Z_{r},\bar{Z}_{r})\right\rangle _{\hat{g}_{z\bar{z}}}^{X^{\pm}}}\,.~~~~~~
\end{eqnarray}

In this way, it is possible in principle to evaluate all the correlation
functions of $X^{+}$, $X^{-}$ with the source terms $e^{-ip_{r}^{+}X^{-}}(Z_{r},\bar{Z}_{r})$,
$e^{-ip_{s}^{-}X^{+}}(w_{s},\bar{w}_{s})$ starting from (\ref{eq:Xpmcorr2}).
Therefore with the definition of the path integral (\ref{eq:Xpmcorrdef2})
given in appendix~\ref{sec:Definition-of-the} and the anomaly factor
$\Gamma$ given in (\ref{eq:Gammabosonic}), we can in principle evaluate
all the correlation functions of the worldsheet theory of $X^{\pm}$.

The worldsheet theory of $X^{\pm}$ thus defined turns out to be a
conformal field theory with the central charge $28-d$. It is possible
to show that the energy-momentum tensor defined as 
\begin{eqnarray}
T^{X^{\pm}}(z) & = & :\partial X^{-}(z)\partial X^{+}(z):-\frac{d-26}{12}\left\{ X^{+},z\right\} \nonumber \\
 & = & \lim_{z'\to z}\left(\partial X^{-}(z')\partial X^{+}(z)-\frac{1}{(z'-z)^{2}}\right)-\frac{d-26}{12}\left\{ X^{+},z\right\} \,,\label{eq:bosonicT-Xpm}
\end{eqnarray}
has the following properties: 
\begin{itemize}
\item $T^{X^{\pm}}(z)$ is regular at $z=z_{I}$, if there is no operator
insertion there. 
\item The OPE between $T^{X^{\pm}}(z)$ and $e^{-ip_{r}^{+}X^{-}-ip_{r}^{-}X^{+}}(Z_{r},\bar{Z}_{r})$
is given by 
\[
T^{X^{\pm}}(z)e^{-ip_{r}^{+}X^{-}-ip_{r}^{-}X^{+}}(Z_{r},\bar{Z}_{r})\sim\left[\frac{-p_{r}^{+}p_{r}^{-}}{\left(z-Z_{r}\right)^{2}}+\frac{1}{z-Z_{r}}\frac{\partial}{\partial Z_{r}}\right]e^{-ip_{r}^{+}X^{-}-ip_{r}^{-}X^{+}}(Z_{r}.\bar{Z}_{r})\,.
\]

\item The OPE between two $T^{X^{\pm}}$'s is given by 
\[
T^{X^{\pm}}\left(z\right)T^{X^{\pm}}\left(z^{\prime}\right)\sim\frac{\frac{28-d}{2}}{\left(z-z^{\prime}\right)^{4}}+\frac{1}{\left(z-z^{\prime}\right)^{2}}2T^{X^{\pm}}\left(z^{\prime}\right)+\frac{1}{z-z^{\prime}}\partial T^{X^{\pm}}\left(z^{\prime}\right)\,.
\]

\end{itemize}
These facts can be derived by calculating the expectation value of
the energy-momentum tensor and their OPE's as carried out in appendix~\ref{sec:Calculations-of-various}.

Therefore the worldsheet theory with the field contents (\ref{eq:conformalcontent})
becomes a CFT with vanishing central charge. This implies that together
with the energy-momentum tensor $T^{\mathrm{LC}}$ of the transverse
variables, we can construct a nilpotent BRST operator 
\begin{equation}
Q=\oint\frac{dz}{2\pi i}\left[c(T^{X^{\pm}}+T^{\mathrm{LC}})+bc\partial c\right]\,,
\end{equation}
and obtain a well-defined formulation in the conformal gauge.

\section{Supersymmetric $X^{\pm}$ CFT\label{sec:Supersymmetric--CFT}}

In the case of superstrings \cite{Baba:2009fi}, it is also possible
to consider the theory in noncritical dimensions and construct the
conformal gauge formulation corresponding to it. The action of the
supersymmetric $X^{\pm}$ CFT is taken to be 
\begin{equation}
S_{\mathrm{super}}^{\pm}\left[\hat{g}_{z\bar{z}},\mathcal{X}^{\pm}\right]=-\frac{1}{2\pi}\int d^{2}\mathbf{z}\left(\bar{D}\mathcal{X}^{+}D\mathcal{X}^{-}+\bar{D}\mathcal{X}^{-}D\mathcal{X}^{+}\right)+\frac{d-10}{8}\Gamma_{\mathrm{super}}\left[\hat{g}_{z\bar{z}},\mathcal{X}^{+}\right]\,.\label{eq:superSXpm}
\end{equation}
Here the supercoordinate $\mathbf{z}$ is given by 
\begin{equation}
\mathbf{z}=(z,\theta)\,,
\end{equation}
where $\theta$ is the Grassmann odd partner of $z$. The superfield
$\mathcal{X}^{\pm}\left(\mathbf{z},\bar{\mathbf{z}}\right)$ is defined
as 
\begin{equation}
\mathcal{X}^{\pm}\left(\mathbf{z},\bar{\mathbf{z}}\right)=X^{\pm}\left(z,\bar{z}\right)+i\theta\psi^{\pm}\left(z\right)+i\bar{\theta}\bar{\psi}^{\pm}\left(\bar{z}\right)+\theta\bar{\theta}F^{\pm}(z,\bar{z})\,,
\end{equation}
${\displaystyle d^{2}\mathbf{z}}$ is given by ${\displaystyle d^{2}\mathbf{z}\equiv d\left(\mathop{\mathrm{Re}}z\right)d\left(\mathop{\mathrm{Im}}z\right)d\theta d\bar{\theta}}$,
and 
\begin{equation}
D\equiv\frac{\partial}{\partial\theta}+\theta\frac{\partial}{\partial z}\,,\quad\bar{D}\equiv\frac{\partial}{\partial\bar{\theta}}+\bar{\theta}\frac{\partial}{\partial\bar{z}}\,.
\end{equation}
The interaction term $\Gamma_{\mathrm{super}}$ is introduced as 
\begin{eqnarray}
\Gamma_{\mathrm{super}}\left[\hat{g}_{z\bar{z}},\mathcal{X}^{+}\right] & = & -\frac{1}{2\pi}\int d^{2}\mathbf{z}\left(\bar{D}\Phi D\Phi+\theta\bar{\theta}\hat{g}_{z\bar{z}}\hat{R}\Phi\right)\,,\nonumber \\
\Phi\left(\mathbf{z},\bar{\mathbf{z}}\right) & = & \ln\left(\left(D\Theta^{+}\right)^{2}\left(\mathbf{z}\right)\left(\bar{D}\bar{\Theta}^{+}\right)^{2}\left(\bar{\mathbf{z}}\right)\right)-\ln\hat{g}_{z\bar{z}}\,,\nonumber \\
\Theta^{+}\left(\mathbf{z}\right) & = & \frac{D\mathcal{X}^{+}}{(\partial\mathcal{X}^{+})^{\frac{1}{2}}}\left(\mathbf{z}\right)\,,\label{eq:Phi}
\end{eqnarray}
which is the super Liouville action defined for variable $\Phi$ with
the background metric $ds^{2}=2\hat{g}_{z\bar{z}}dzd\bar{z}$. In
the same way as we have derived (\ref{eq:Xpmcorr}) in the bosonic
case, one can deduce 
\begin{eqnarray}
\lefteqn{\left\langle \prod_{r=1}^{N}e^{-ip_{r}^{+}\mathcal{X}^{-}}(\mathbf{Z}_{r},\bar{\mathbf{Z}}_{r})\prod_{s=1}^{M}e^{-ip_{s}^{-}\mathcal{X}^{+}}(\mathbf{w}_{s},\bar{\mathbf{w}}_{s})\right\rangle _{\hat{g}_{z\bar{z}}}^{\mathcal{X}^{\pm}}}\nonumber \\
 & \equiv & Z_{\mathrm{super}}^{\mathcal{X}}[\hat{g}_{z\bar{z}}]^{-2}\int\left[d\mathcal{X}^{+}d\mathcal{X}^{-}\right]_{\hat{g}_{z\bar{z}}}e^{-S_{\mathrm{super}}^{\pm}\left[\hat{g}_{z\bar{z}}\right]}\prod_{r=1}^{N}e^{-ip_{r}^{+}\mathcal{X}^{-}}(\mathbf{Z}_{r},\bar{\mathbf{Z}}_{r})\prod_{s=1}^{M}e^{-ip_{s}^{-}\mathcal{X}^{+}}(\mathbf{w}_{s},\bar{\mathbf{w}}_{s})\nonumber \\
 & = & (2\pi)^{2}\delta\left(\sum_{s}p_{s}^{-}\right)\delta\left(\sum_{r}p_{r}^{+}\right)\prod_{s}e^{-p_{s}^{-}\frac{\mathbf{\rho}+\bar{\mathbf{\rho}}}{2}}(\mathbf{w}_{s},\bar{\mathbf{w}}_{s})\, e^{-\frac{d-10}{8}\Gamma_{\mathrm{super}}\left[\hat{g}_{z\bar{z}},\,-\frac{i}{2}\left(\rho+\bar{\rho}\right)\right]}\,.\label{eq:superXpmcorr}
\end{eqnarray}
Here 
\begin{equation}
Z_{\mathrm{super}}^{\mathcal{X}}[\hat{g}_{z\bar{z}}]=\int\left[d\mathcal{X}\right]_{\hat{g}_{z\bar{z}}}\exp\left[-\frac{1}{2\pi}\int d^{2}\mathbf{z}\bar{D}\mathcal{X}D\mathcal{X}\right]\,,
\end{equation}
and 
\begin{equation}
\mathbf{Z}_{r}=(Z_{r},\Theta_{r})\,,\qquad\mathbf{w}_{s}=(w_{s},\eta_{s})\,.
\end{equation}
$\rho(\mathbf{z})$ which appears on the last line in (\ref{eq:superXpmcorr})
is the supersymmetric version of $\rho(z)$ in (\ref{eq:rhoz}). Its
explicit form will be given shortly. Introducing the fermionic partner
$\xi$ of $\rho$ given by 
\begin{equation}
\xi\equiv\frac{D\rho}{\left(\partial\rho\right)^{\frac{1}{2}}}\,,
\end{equation}
we can define the supercoordinate $\mbox{\mathversion{bold}\ensuremath{\rho}}$
on the light-cone diagram 
\begin{equation}
\mbox{\mathversion{bold}\ensuremath{\rho}}\equiv(\rho,\xi)\,.
\end{equation}

As in the bosonic case, in order to calculate the correlation functions
of the CFT, we need to obtain $\Gamma_{\mathrm{super}}\left[\hat{g}_{z\bar{z}},\,-\frac{i}{2}\left(\rho+\bar{\rho}\right)\right]$.

\subsection{Evaluation of $e^{-\Gamma_{\mathrm{super}}}$ on the sphere}

When the worldsheet is a Riemann surface of genus $0$, we can employ
the coordinate $z$ on the complex plane to describe the surface.
Taking the background metric to be $ds^{2}=dzd\bar{z}$, $\Gamma_{\mathrm{super}}\left[\frac{1}{2},\,-\frac{i}{2}\left(\rho+\bar{\rho}\right)\right]$
was derived in \cite{Baba:2009fi} by integrating the variation formula
\begin{equation}
\delta\left(-\Gamma_{\mathrm{super}}\left[\frac{1}{2},\,-\frac{i}{2}\left(\rho+\bar{\rho}\right)\right]\right)=\sum_{\mathcal{I}}\delta\mathcal{T}_{\mathcal{I}}\oint_{C_{\mathcal{I}}}\frac{d\mathbf{z}}{2\pi i}\frac{\left(-2S\left(\mathbf{z},\mbox{\mathversion{bold}\ensuremath{\rho}}\right)\right)}{\left(D\xi\right)^{2}}+\mathrm{c.c.}\,,
\end{equation}
which is the supersymmetric version of (\ref{eq:deltaGamma}). Here
$S\left(\mathbf{z},\mbox{\mathversion{bold}\ensuremath{\rho}}\right)$
is the super Schwarzian derivative given by 
\begin{equation}
S\left(\mathbf{z},\mbox{\mathversion{bold}\ensuremath{\rho}}\right)\equiv\frac{D^{4}\xi}{D\xi}-2\frac{D^{3}\xi D^{2}\xi}{(D\xi)^{2}}\,.
\end{equation}
In \cite{Berkovits:1985ji,Berkovits:1987gp}, an approach similar
to \cite{Mandelstam:1985ww} was employed to get $\Gamma_{\mathrm{super}}\left[\frac{1}{2},\,-\frac{i}{2}\left(\rho+\bar{\rho}\right)\right]$.
Since it is quite tedious to get $\Gamma_{\mathrm{super}}\left[\frac{1}{2},\,-\frac{i}{2}\left(\rho+\bar{\rho}\right)\right]$
by these methods, we will look for another approach.

In the sphere case, $\rho\left(\mathbf{z}\right)$ takes the form
\begin{equation}
\rho(\mathbf{z})=\sum_{r=1}^{N}\alpha_{r}\ln\left(\mathbf{z}-\mathbf{Z}_{r}\right)~,\qquad\mathbf{z}-\mathbf{Z}_{r}\equiv z-Z_{r}-\theta\Theta_{r}~.\label{eq:superpho-sphere}
\end{equation}
This can be written as 
\begin{equation}
\rho\left(\mathbf{z}\right)=\rho_{b}\left(z\right)+\theta f\left(z\right)\ ,
\end{equation}
where 
\begin{eqnarray}
\rho_{b}\left(z\right) & = & \sum_{r}\alpha_{r}\ln\left(z-Z_{r}\right)\ ,\nonumber \\
f\left(z\right) & = & -\sum_{r}\frac{\alpha_{r}\Theta_{r}}{z-Z_{r}}\ .
\end{eqnarray}
When all the Grassmann parameters $\Theta_{r}$ vanish, the worldsheet
theory can be described by an ordinary Riemann surface with coordinate
$\rho_{b}$ and obviously 
\begin{equation}
\Gamma_{\mathrm{super}}\left[\hat{g}_{z\bar{z}},\,-\frac{i}{2}\left(\rho_{b}+\bar{\rho}_{b}\right)\right]=\frac{1}{2}\Gamma\left[\hat{g}_{z\bar{z}},\,-\frac{i}{2}\left(\rho_{b}+\bar{\rho}_{b}\right)\right]\,,\label{eq:GammasuperhalfGamma}
\end{equation}
where $\Gamma\left[\hat{g}_{z\bar{z}},\,-\frac{i}{2}\left(\rho_{b}+\bar{\rho}_{b}\right)\right]$
is the anomaly factor in the bosonic case given in (\ref{eq:Gammabosonic}).
Since $\Gamma\left[\hat{g}_{z\bar{z}},\,-\frac{i}{2}\left(\rho_{b}+\bar{\rho}_{b}\right)\right]$
is known, the $\Gamma_{\mathrm{super}}\left[\hat{g}_{z\bar{z}},\,-\frac{i}{2}\left(\rho+\bar{\rho}\right)\right]$
can be obtained if we can derive how $\Gamma_{\mathrm{super}}\left[\hat{g}_{z\bar{z}},\,-\frac{i}{2}\left(\rho+\bar{\rho}\right)\right]$
changes under the deformation 
\begin{equation}
\left(Z_{r},0;\bar{Z}_{r},0\right)\to\left(Z_{r},\Theta_{r};\bar{Z}_{r},\bar{\Theta}_{r}\right)\qquad\left(r=1,\cdots,N\right)\ \label{eq:deformation}
\end{equation}
of the moduli parameters.

In the bosonic case, the moduli of the punctured Riemann surfaces
can be parametrized either by the coordinates $Z_{r},\bar{Z}_{r}$
$\left(r=1,\cdots N\right)$ of the punctures modulo the $SL(2,\mathbb{C})$
transformations or by 
\begin{equation}
\mathcal{T}_{\mathcal{I}}^{(b)}=\rho_{b}\left(z_{\mathcal{I}+1}^{(b)}\right)-\rho_{b}\left(z_{\mathcal{I}}^{(b)}\right)\qquad\left(\mathcal{I}=1,\cdots,N-3\right)\label{eq:TcalI}
\end{equation}
and their complex conjugates $\bar{\mathcal{T}}_{\mathcal{I}}^{(b)}$,
where $z_{\mathcal{I}}^{(b)}$ $\left(\mathcal{I}=1,\cdots N-2\right)$
are the coordinates of the interaction points which are labeled so
that $\mathcal{T}_{\mathcal{I}}^{(b)}$ coincide with $\mathcal{T}_{\mathcal{I}}$
in (\ref{eq:bosonic-moduli-TI}). $\mathcal{T}_{\mathcal{I}}^{(b)}$,$\bar{\mathcal{T}}_{\mathcal{I}}^{(b)}$
are the natural parameters describing the moduli space from the light-cone
gauge point of view and $\Gamma$ is derived by considering how it
changes under the variations of them. Therefore in order to calculate
$\Gamma_{\mathrm{super}}$, we need a supersymmetric version of (\ref{eq:TcalI}).
In \cite{Berkovits:1985ji,Berkovits:1987gp,Aoki:1990yn}, it is shown
that the moduli space of the punctured super Riemann surfaces can
be parametrized by $\left(\mathcal{T}_{\mathcal{I}},\bar{\mathcal{T}}_{\mathcal{I}},\xi_{I},\bar{\xi}_{I}\right)$
instead of $\left(Z_{r},\Theta_{r};\bar{Z}_{r},\bar{\Theta}_{r}\right)$.
$\mathcal{T}_{\mathcal{I}}\ \left(\mathcal{I}=1,\cdots,N-3\right)$
are given as 
\begin{equation}
\mathcal{T}_{\mathcal{I}}=\rho\left(\mathbf{z}_{\mathcal{I}+1}\right)-\rho\left(\mathbf{z}_{\mathcal{I}}\right)\,,
\end{equation}
where $\mathbf{z}_{\mathcal{I}}\ \left(\mathcal{I}=1,\cdots N-2\right)$
are the supercoordinates of the interaction points defined in 
appendix~\ref{sec:Interaction-points-and}, and 
$\bar{\mathcal{T}}_{\mathcal{I}}$
are their complex conjugates. The definitions of the odd supermoduli
$\xi_{I},\bar{\xi}_{I}$ are also given in appendix~\ref{sec:Interaction-points-and}.

\subsubsection*{Equivalence of the parametrizations}

Let us demonstrate how the parametrization $\left(Z_{r},\Theta_{r};\bar{Z}_{r},\bar{\Theta}_{r}\right)$
and $\left(\mathcal{T}_{\mathcal{I}},\bar{\mathcal{T}}_{\mathcal{I}},\xi_{I},\bar{\xi}_{I}\right)$
are related to each other. In order to do so, we consider a tree-level
light-cone gauge amplitude for type II superstring theory in $10$
dimensions with external lines corresponding to the states 
\begin{equation}
\alpha_{-n_{1}}^{i_{1}}\cdots\bar{\alpha}_{-\bar{n}_{1}}^{\bar{i}_{1}}\cdots\psi_{-s_{1}}^{j_{1}}\cdots\bar{\psi}_{-\bar{s}_{1}}^{\bar{j}_{1}}\cdots\left|0\right\rangle _{r}\quad\left(r=1,\cdots,N\right)\,.\label{eq:stater}
\end{equation}
The light-cone diagram corresponds to a sphere with $N$ punctures
which should be considered as a super Riemann surface. $\rho\left(\mathbf{z}\right)$
given in (\ref{eq:superpho-sphere}) is the map from the sphere to
the light-cone diagram. The amplitude is expressed as an integral
of the correlation function 
\begin{equation}
\left\langle V_{1}\left(Z_{1},\Theta_{1};\bar{Z}_{1},\bar{\Theta}_{1}\right)\cdots V_{N}\left(Z_{N},\Theta_{N};\bar{Z}_{N},\bar{\Theta}_{N}\right)\right\rangle _{\mathbb{C}\cup\infty}\label{eq:corrboldZ}
\end{equation}
on the sphere times the anomaly factor 
\begin{equation}
e^{-\Gamma_{\mathrm{super}}\left[\frac{1}{2},\,-\frac{i}{2}\left(\rho+\bar{\rho}\right)\right]}\,,
\end{equation}
over the moduli parameters. The moduli space of the punctured super
Riemann surface is parametrized by the coordinates $\left(Z_{r},\Theta_{r};\bar{Z}_{r},\bar{\Theta}_{r}\right)$
$(r=1,\ldots,N)$ modded out by the superprojective transformations.
Here $V_{r}$ denotes the vertex operator for the state (\ref{eq:stater})
and is given by 
\begin{equation}
V_{r}\left(\mathbf{Z}_{r},\bar{\mathbf{Z}}_{r}\right)\propto\mathcal{A}_{-n_{1}}^{i_{1}}\cdots\bar{\mathcal{A}}_{-\bar{n}_{1}}^{\bar{i}_{1}}\cdots\mathcal{B}_{-s_{1}}^{j_{1}}\cdots\bar{\mathcal{B}}_{-\bar{s}_{1}}^{\bar{j}_{1}}\cdots e^{i\vec{p}_{r}\cdot\vec{\mathcal{X}}}\left(\mathbf{Z}_{r},\bar{\mathbf{Z}}_{r}\right)e^{\frac{1}{2}\vec{p}_{r}^{\,2}\bar{N}_{00}^{rr}}\ ,
\end{equation}
where 
\begin{eqnarray}
\mathcal{A}_{-n}^{i} & = & \oint_{\mathbf{Z}_{r}}\frac{d\mathbf{z}}{2\pi i}iD\mathcal{X}^{i}\left(\mathbf{z}\right)e^{-\frac{n}{\alpha_{r}}\left(\rho\left(\mathbf{z}\right)-\rho\left(\mathbf{z}_{I^{\left(r\right)}}\right)\right)}\ ,\nonumber \\
\mathcal{B}_{-s}^{j} & = & \oint_{\mathbf{Z}_{r}}\frac{d\mathbf{z}}{2\pi i}\frac{D\rho}{\left(\partial\rho\right)^{\frac{1}{2}}}iD\mathcal{X}^{j}\left(\mathbf{z}\right)e^{-\frac{s}{\alpha_{r}}\left(\rho\left(\mathbf{z}\right)-\rho\left(\mathbf{z}_{I^{\left(r\right)}}\right)\right)}\ ,\nonumber \\
\bar{N}_{00}^{rr} & = & \frac{\rho\left(\tilde{\mathbf{z}}_{I^{\left(r\right)}}\right)}{\alpha_{r}}-\sum_{s\ne r}\frac{\alpha_{s}}{\alpha_{r}}\ln\left(\mathbf{Z}_{r}-\mathbf{Z}_{s}\right)\ ,
\end{eqnarray}
$\bar{\mathcal{A}},\bar{\mathcal{B}}$ are defined in a similar way,
and $\mathcal{X}^{i}(\mathrm{\mathbf{{z}}},\bar{{\mathbf{{z}}}})$
are the superfields for the transverse variables.

It is straightforward to show that $V_{r}\left(Z_{r},\Theta_{r};\bar{Z}_{r},\bar{\Theta}_{r}\right)$
is expressed as a superconformal transform of $V_{r}\left(Z_{r},0;\bar{Z}_{r},0\right)$:
\begin{equation}
U_{r}\bar{U}_{r}V_{r}\left(Z_{r},0;\bar{Z}_{r},0\right)=V_{r}\left(Z_{r},\Theta_{r};\bar{Z}_{r},\bar{\Theta}_{r}\right)\exp\left(-\mathop{\mathrm{Re}}\frac{\delta\rho\left(\tilde{\mathbf{z}}_{I^{\left(r\right)}}\right)}{\alpha_{r}}\right)\ ,\label{eq:UUbarV}
\end{equation}
with 
\begin{eqnarray}
U_{r} & \equiv & \exp\left[\oint_{\left(Z_{r},0\right)}\frac{d\mathbf{z}}{2\pi i}v\left(T^{\mathrm{{LC}}}\left(\mathbf{z}\right)-2S\left(\mathbf{z},\mbox{\mathversion{bold}\ensuremath{\rho}}_{b}\right)\right)\right]\nonumber \\
 &  & \quad\times\exp\left[\oint_{\left(Z_{r},0\right)}\frac{d\mathbf{z}}{2\pi i}\frac{\delta\rho\left(\tilde{\mathbf{z}}_{I^{\left(r\right)}}\right)}{\partial\rho_{b}\left(z\right)}\left(T^{\mathrm{{LC}}}\left(\mathbf{z}\right)-2S\left(\mathbf{z},\mbox{\mathversion{bold}\ensuremath{\rho}}_{b}\right)\right)\right]\ .\label{eq:Ur}
\end{eqnarray}
Here $T^{\mathrm{LC}}(\mathbf{z})=T_{B}^{\mathrm{LC}}(z)+\theta T_{F}^{\mathrm{LC}}(z)$
is the super energy-momentum tensor in the transverse directions,
\begin{eqnarray}
v\left(z,\theta\right) & = & -2\theta\frac{f\left(z\right)}{\partial\rho_{b}\left(z\right)}\ ,\nonumber \\
\delta\rho\left(\tilde{\mathbf{z}}_{I}\right) & = & \rho\left(\tilde{\mathbf{z}}_{I}\right)-\rho_{b}\left(z_{I}^{\left(b\right)}\right)\ ,\nonumber \\
\tilde{\mathbf{z}}_{I} & = & \left(z_{I}^{\left(b\right)},\tilde{\theta}_{I}\right)=\left(z_{I}^{\left(b\right)},-\frac{\partial f}{\partial^{2}\rho_{b}}\left(z_{I}^{\left(b\right)}\right)\right)\ ,\nonumber \\
\mbox{\mathversion{bold}\ensuremath{\rho}}_{b}\left(\mathbf{z}\right) & = & \left(\rho_{b},\theta\left(\partial\rho_{b}\right)^{\frac{1}{2}}\right)\ ,\nonumber \\
S\left(\mathbf{z},\mbox{\mathversion{bold}\ensuremath{\rho}}_{b}\right) & = & \theta\left(\frac{\partial^{3}\rho_{b}}{2\partial\rho_{b}}-\frac{3}{4}\left(\frac{\partial^{2}\rho_{b}}{\partial\rho_{b}}\right)^{2}\right)\ ,
\end{eqnarray}
and $z_{I}^{\left(b\right)}$ is the coordinate of an interaction
point which satisfies $\partial\rho_{b}\left(z_{I}^{\left(b\right)}\right)=0$.
Indeed, defining 
\begin{eqnarray}
\mathcal{A}_{-n}^{i\left(b\right)} & = & \oint_{\mathbf{Z}_{r}}\frac{d\mathbf{z}}{2\pi i}iD\mathcal{X}^{i}\left(\mathbf{z}\right)e^{-\frac{n}{\alpha_{r}}\left(\rho_{b}\left(z\right)-\rho_{b}\left(z_{I^{\left(r\right)}}^{\left(b\right)}\right)\right)}\ ,\nonumber \\
\mathcal{B}_{-s}^{j\left(b\right)} & = & \oint_{\mathbf{Z}_{r}}\frac{d\mathbf{z}}{2\pi i}\frac{D\rho_{b}}{\left(\partial\rho_{b}\right)^{\frac{1}{2}}}iD\mathcal{X}^{j}\left(\mathbf{z}\right)e^{-\frac{s}{\alpha_{r}}\left(\rho_{b}\left(z\right)-\rho_{b}\left(z_{I^{\left(r\right)}}^{\left(b\right)}\right)\right)}\ ,
\end{eqnarray}
it is straightforward to show 
\begin{equation}
U_{r}\mathcal{A}_{-n}^{i\left(b\right)}U_{r}^{-1}=\mathcal{A}_{-n}^{i}~,\qquad U_{r}\mathcal{B}_{-s}^{j\left(b\right)}U_{r}^{-1}=\mathcal{B}_{-s}^{j}\ ,
\end{equation}
and 
\begin{equation}
U_{r}e^{i\vec{p}_{r}\cdot\vec{\mathcal{X}}}\left(Z_{r},0;\bar{Z}_{r},0\right)e^{\frac{1}{2}\vec{p}_{r}^{\,2}\bar{N}_{00}^{rr\left(b\right)}}=e^{i\vec{p}_{r}\cdot\vec{\mathcal{X}}}\left(\mathbf{Z}_{r},\bar{\mathbf{Z}}_{r}\right)\exp\left(\frac{1}{2}\vec{p}_{r}^{\,2}\bar{N}_{00}^{rr}-\frac{1}{2}\frac{\delta\rho\left(\tilde{\mathbf{z}}_{I^{\left(r\right)}}\right)}{\alpha_{r}}\right)\ ,
\end{equation}
where 
\begin{equation}
\bar{N}_{00}^{rr\left(b\right)}=\frac{\rho_{b}\left(z_{I^{\left(r\right)}}^{\left(b\right)}\right)}{\alpha_{r}}-\sum_{s\ne r}\frac{\alpha_{s}}{\alpha_{r}}\ln\left(Z_{r}-Z_{s}\right)\ .
\end{equation}
It is easy to get (\ref{eq:UUbarV}) from these.

Substituting (\ref{eq:UUbarV}) into (\ref{eq:corrboldZ}), the correlation
function (\ref{eq:corrboldZ}) can be expressed as 
\begin{equation}
\left\langle \prod_{r}\left[U_{r}\bar{U}_{r}V_{r}\left(Z_{r},0;\bar{Z}_{r},0\right)\right]\right\rangle _{\mathbb{C}\cup\infty}\exp\left[\sum_{r}\mathop{\mathrm{Re}}\frac{\delta\rho\left(\tilde{\mathbf{z}}_{I^{\left(r\right)}}\right)}{\alpha_{r}}\right]\ .
\end{equation}
Deforming the contours of the the integrals in the exponent of $U_{r}$
in (\ref{eq:Ur}), we get 
\begin{equation}
\left\langle \prod_{r}\left[U_{r}\bar{U}_{r}V_{r}\left(Z_{r},0;\bar{Z}_{r},0\right)\right]\right\rangle _{\mathbb{C}\cup\infty}=\left\langle \prod_{I}\left(U_{I}\bar{U}_{I}\right)\prod_{\mathcal{I}}\left(U_{\mathcal{I}}\bar{U}_{\mathcal{I}}\right)\prod_{r}V_{r}\left(Z_{r},0;\bar{Z}_{r},0\right)\right\rangle _{\mathbb{C}\cup\infty}\ ,\label{eq:UrUrbarVr}
\end{equation}
where 
\begin{eqnarray}
U_{I} & = & \exp\left[-\oint_{\left(z_{I}^{\left(b\right)},0\right)}\frac{d\mathbf{z}}{2\pi i}\frac{\delta\rho\left(\tilde{\mathbf{z}}_{I}\right)}{\partial\rho_{b}\left(z\right)}\left(T^{\mathrm{LC}}\left(\mathbf{z}\right)-2S\left(\mathbf{z},\mbox{\mathversion{bold}\ensuremath{\rho}}_{b}\right)\right)\right]\nonumber \\
 &  & \quad\times\exp\left[-\oint_{\left(z_{I}^{\left(b\right)},0\right)}\frac{d\mathbf{z}}{2\pi i}v\left(T^{\mathrm{LC}}\left(\mathbf{z}\right)-2S\left(\mathbf{z},\mbox{\mathversion{bold}\ensuremath{\rho}}_{b}\right)\right)\right]\ ,\nonumber \\
U_{\mathcal{I}} & = & \exp\left[\delta\mathcal{T}_{\mathcal{I}}\oint_{C_{\mathcal{I}}}\frac{d\mathbf{z}}{2\pi i}\frac{1}{\partial\rho_{b}\left(z\right)}\left(T^{\mathrm{LC}}\left(\mathbf{z}\right)-2S\left(\mathbf{z},\mbox{\mathversion{bold}\ensuremath{\rho}}_{b}\right)\right)\right]\ .
\end{eqnarray}
Using the OPE's 
\begin{eqnarray}
T_{F}^{\mathrm{LC}}\left(z\right)T_{F}^{\mathrm{LC}}\left(w\right) & \sim & \frac{2}{\left(z-w\right)^{3}}+\frac{1}{z-w}\frac{1}{2}T_{B}^{\mathrm{LC}}\left(w\right)\,,\nonumber \\
T_{B}^{\mathrm{LC}}\left(z\right)T_{B}^{\mathrm{LC}}\left(w\right) & \sim & \frac{6}{\left(z-w\right)^{4}}+\frac{2}{\left(z-w\right)^{2}}T_{B}^{\mathrm{LC}}\left(w\right)+\frac{1}{z-w}\partial T_{B}^{\mathrm{LC}}\left(w\right)\,,\nonumber \\
T_{B}^{\mathrm{LC}}\left(z\right)T_{F}^{\mathrm{LC}}\left(w\right) & \sim & \frac{\frac{3}{2}}{\left(z-w\right)^{2}}T_{F}^{\mathrm{LC}}\left(w\right)+\frac{1}{z-w}\partial T_{F}^{\mathrm{LC}}\left(w\right)\,,\label{eq:TBTF}
\end{eqnarray}
it is straightforward to get 
\begin{eqnarray}
U_{I} & = & \exp\left(\xi_{I}\left(\partial^{2}\rho_{b}\right)^{-\frac{3}{4}}T_{F}^{\mathrm{LC}}\left(z_{I}^{\left(b\right)}\right)\right)\nonumber \\
 &  & \quad\times\left[1+\left(\frac{5}{12}\frac{\partial^{4}\rho_{b}}{\left(\partial^{2}\rho_{b}\right)^{3}}-\frac{3}{4}\frac{\left(\partial^{3}\rho_{b}\right)^{2}}{\left(\partial^{2}\rho_{b}\right)^{4}}\right)\partial ff\right.\nonumber \\
 &  & \hphantom{\qquad\times\quad}\left.{}-\frac{2}{3}\frac{\partial^{3}ff}{\left(\partial^{2}\rho_{b}\right)^{2}}+\frac{\partial^{3}\rho_{b}}{\left(\partial^{2}\rho_{b}\right)^{3}}\partial^{2}ff+\frac{1}{12}\frac{\partial^{3}f\partial^{2}f\partial ff}{\left(\partial^{2}\rho_{b}\right)^{4}}\right]\left(z_{I}^{\left(b\right)}\right)\ .\label{eq:UI}
\end{eqnarray}
Therefore the correlation function (\ref{eq:corrboldZ}) is expressed
as 
\begin{eqnarray}
\lefteqn{\left\langle V_{1}\left(Z_{1},\Theta_{1};\bar{Z}_{1},\bar{\Theta}_{1}\right)\cdots V_{N}\left(Z_{N},\Theta_{N};\bar{Z}_{N},\bar{\Theta}_{N}\right)\right\rangle _{\mathbb{C}\cup\infty}}\nonumber \\
 & = & \left\langle \prod_{I}\exp\left(\xi_{I}\left(\partial^{2}\rho_{b}\right)^{-\frac{3}{4}}T_{F}^{\mathrm{LC}}\left(z_{I}^{\left(b\right)}\right)+\mathrm{c.c.}\right)\prod_{\mathcal{I}}\left(U_{\mathcal{I}}\bar{U}_{\mathcal{I}}\right)\prod_{r}V_{r}\left(Z_{r},0;\bar{Z}_{r},0\right)\right\rangle _{\mathbb{C}\cup\infty}\nonumber \\
 &  & \quad\times\exp\left[\sum_{r}\mathop{\mathrm{Re}}\frac{\delta\rho\left(\tilde{\mathbf{z}}_{I^{\left(r\right)}}\right)}{\alpha_{r}}\right]\nonumber \\
 &  & \quad\times\prod_{I}\left|\left[1+\left(\frac{5}{12}\frac{\partial^{4}\rho_{b}}{\left(\partial^{2}\rho_{b}\right)^{3}}-\frac{3}{4}\frac{\left(\partial^{3}\rho_{b}\right)^{2}}{\left(\partial^{2}\rho_{b}\right)^{4}}\right)\partial ff\vphantom{\left.-\frac{2}{3}\frac{\partial^{3}ff}{\left(\partial^{2}\rho_{b}\right)^{2}}+\frac{\partial^{3}\rho_{b}}{\left(\partial^{2}\rho_{b}\right)^{3}}\partial^{2}ff+\frac{1}{12}\frac{\partial^{3}f\partial^{2}f\partial ff}{\left(\partial^{2}\rho_{b}\right)^{4}}\right)}\right.\right.\nonumber \\
 &  & \hphantom{\quad\times\qquad}\left.\left.-\frac{2}{3}\frac{\partial^{3}ff}{\left(\partial^{2}\rho_{b}\right)^{2}}+\frac{\partial^{3}\rho_{b}}{\left(\partial^{2}\rho_{b}\right)^{3}}\partial^{2}ff+\frac{1}{12}\frac{\partial^{3}f\partial^{2}f\partial ff}{\left(\partial^{2}\rho_{b}\right)^{4}}\right]\left(z_{I}^{\left(b\right)}\right)\right|^{2}\ .\label{eq:corrboldZ2}
\end{eqnarray}
From (\ref{eq:corrboldZ2}), we can see that the variation (\ref{eq:deformation})
of the moduli parameters can be realized by the insertions of operators
\begin{equation}
\exp\left(\xi_{I}\left(\partial^{2}\rho_{b}\right)^{-\frac{3}{4}}T_{F}^{\mathrm{LC}}\left(z_{I}^{\left(b\right)}\right)+\mathrm{c.c.}\right)\,,\label{eq:xiTF}
\end{equation}
and 
\begin{equation}
U_{\mathcal{I}}\bar{U}_{\mathcal{I}}\,.\label{eq:UcalI}
\end{equation}
(\ref{eq:xiTF}) is exactly the superconformal transformation which
induces the odd moduli $\xi_{I},\bar{\xi}_{I}$ \cite{Berkovits:1985ji,Berkovits:1987gp,Aoki:1990yn}.
The insertions of $T_{F}^{\mathrm{LC}}\bar{T}_{F}^{\mathrm{LC}}$
at the interaction points in light-cone gauge perturbation theory
arise by integrating superspace correlation function over the odd
moduli parameters $\xi_{I},\bar{\xi}_{I}$. (\ref{eq:UcalI}) corresponds
to the shift of $\mathcal{T}_{\mathcal{I}}$ and $\bar{\mathcal{T}}_{\mathcal{I}}$
because of the variation (\ref{eq:deformation}). Hence the parameters
$\mathcal{T}_{\mathcal{I}},\bar{\mathcal{T}}_{\mathcal{I}},\xi_{I},\bar{\xi}_{I}$
parametrize the punctured super Riemann surface and the variations
of these parameters are implemented by the insertions of the operators
(\ref{eq:xiTF}) and (\ref{eq:UcalI}).

\subsubsection*{Derivation of $\Gamma_{\mathrm{super}}$}

It is now possible to evaluate $\Gamma_{\mathrm{super}}$. In terms
of the parametrization $\left(\mathcal{T}_{\mathcal{I}},\bar{\mathcal{T}}_{\mathcal{I}},\xi_{I},\bar{\xi}_{I}\right)$,
the variation (\ref{eq:deformation}) corresponds to 
\begin{equation}
\left(\mathcal{T}_{\mathcal{I}}^{(b)},\bar{\mathcal{T}}_{\mathcal{I}}^{(b)},0,0\right)\to\left(\mathcal{T}_{\mathcal{I}},\bar{\mathcal{T}}_{\mathcal{I}},\xi_{I},\bar{\xi}_{I}\right)~.
\end{equation}
{}From the discussion above, we can see that such a variation can
be implemented by inserting operators (\ref{eq:xiTF}) and (\ref{eq:UcalI}).
Therefore, starting from the partition function (\ref{eq:GammasuperhalfGamma}),
we get 
\begin{eqnarray}
\lefteqn{\exp\left(-\Gamma_{\mathrm{super}}\left[\frac{1}{2},\,-\frac{i}{2}\left(\rho+\bar{\rho}\right)\right]\right)}\nonumber \\
 & = & \exp\left(-\frac{1}{2}\Gamma\left[\frac{1}{2},\,-\frac{i}{2}\left(\rho_{b}+\bar{\rho}_{b}\right)\right]\right)\nonumber \\
 &  & \quad\times\left\langle \prod_{I}\exp\left(\xi_{I}\left(\partial^{2}\rho_{b}\right)^{-\frac{3}{4}}T_{F}^{\mathrm{LC}}\left(z_{I}^{\left(b\right)}\right)+\mathrm{c.c.}\right)\prod_{\mathcal{I}}\left(U_{\mathcal{I}}\bar{U}_{\mathcal{I}}\right)\right\rangle _{\mathbb{C}\cup\infty}\,.
\end{eqnarray}
The second factor on the right hand side can easily be evaluated from
(\ref{eq:corrboldZ2}) with $V_{r}=1$ and we get the expression 
\begin{eqnarray}
\lefteqn{\exp\left(-\Gamma_{\mathrm{super}}\left[\frac{1}{2},\,-\frac{i}{2}\left(\rho+\bar{\rho}\right)\right]\right)}\nonumber \\
 & = & \exp\left(-\frac{1}{2}\Gamma\left[\frac{1}{2},\,-\frac{i}{2}\left(\rho_{b}+\bar{\rho}_{b}\right)\right]\right)\,\exp\left[-\sum_{r}\mathop{\mathrm{Re}}\frac{\delta\rho\left(\tilde{\mathbf{z}}_{I^{(r)}}\right)}{\alpha_{r}}\right]\nonumber \\
 &  & \quad\times\prod_{I}\left|\left[1+\left(\frac{5}{12}\frac{\partial^{4}\rho_{b}}{\left(\partial^{2}\rho_{b}\right)^{3}}-\frac{3}{4}\frac{\left(\partial^{3}\rho_{b}\right)^{2}}{\left(\partial^{2}\rho_{b}\right)^{4}}\right)\partial ff\vphantom{\left.-\frac{2}{3}\frac{\partial^{3}ff}{\left(\partial^{2}\rho_{b}\right)^{2}}+\frac{\partial^{3}\rho_{b}}{\left(\partial^{2}\rho_{b}\right)^{3}}\partial^{2}ff+\frac{1}{12}\frac{\partial^{3}f\partial^{2}f\partial ff}{\left(\partial^{2}\rho_{b}\right)^{4}}\right)^{-1}}\right.\right.\nonumber \\
 &  & \hphantom{\quad\times\qquad}\ \left.\left.-\frac{2}{3}\frac{\partial^{3}ff}{\left(\partial^{2}\rho_{b}\right)^{2}}+\frac{\partial^{3}\rho_{b}}{\left(\partial^{2}\rho_{b}\right)^{3}}\partial^{2}ff+\frac{1}{12}\frac{\partial^{3}f\partial^{2}f\partial ff}{\left(\partial^{2}\rho_{b}\right)^{4}}\right]^{-1}\left(z_{I}^{\left(b\right)}\right)\right|^{2}\,.\label{eq:Gammasuper1}
\end{eqnarray}
Since 
\begin{equation}
\delta\rho\left(\tilde{\mathbf{z}}_{I}\right)=-\frac{\partial ff}{\partial^{2}\rho_{b}}\left(z_{I}^{\left(b\right)}\right)\,,
\end{equation}
we can further rewrite (\ref{eq:Gammasuper1}) as 
\begin{eqnarray}
\lefteqn{\exp\left(-\Gamma_{\mathrm{super}}\left[\frac{1}{2},\,-\frac{i}{2}\left(\rho+\bar{\rho}\right)\right]\right)}\nonumber \\
 &  & =\exp\left(-\frac{1}{2}\Gamma\left[\frac{1}{2},\,-\frac{i}{2}\left(\rho_{b}+\bar{\rho}_{b}\right)\right]-\sum_{r}\Delta\Gamma_{r}-\sum_{I}\Delta\Gamma_{I}\right)\,,\label{eq:Gammasuper2}
\end{eqnarray}
where 
\begin{eqnarray}
-\Delta\Gamma_{r} & = & \frac{1}{2\alpha_{r}}\frac{\partial ff}{\partial^{2}\rho_{b}}\left(z_{I^{\left(r\right)}}^{\left(b\right)}\right)+\mathrm{c.c.}\,,\nonumber \\
-\Delta\Gamma_{I} & = & \left\{ -\left(\frac{5}{12}\frac{\partial^{4}\rho_{b}}{\left(\partial^{2}\rho_{b}\right)^{3}}-\frac{3}{4}\frac{\left(\partial^{3}\rho_{b}\right)^{2}}{\left(\partial^{2}\rho_{b}\right)^{4}}\right)\partial ff+\frac{2}{3}\frac{\partial^{3}ff}{\left(\partial^{2}\rho_{b}\right)^{2}}\right.\nonumber \\
 &  & \quad\ \left.{}-\frac{\partial^{3}\rho_{b}}{\left(\partial^{2}\rho_{b}\right)^{3}}\partial^{2}ff-\frac{1}{12}\frac{\partial^{3}f\partial^{2}f\partial ff}{\left(\partial^{2}\rho_{b}\right)^{4}}\right\} \left(z_{I}^{\left(b\right)}\right)+\mathrm{c.c.}\,.
\end{eqnarray}

In this case, (\ref{eq:Gammasuper1}) can be further simplified using
the superspace variables. Indeed, substituting 
\begin{equation}
\exp\left(-\frac{1}{2}\Gamma\left[\frac{1}{2},\,-\frac{i}{2}\left(\rho_{b}+\bar{\rho}_{b}\right)\right]\right)=\left|\sum_{r}\alpha_{r}Z_{r}\right|^{2}\prod_{I}\left|\partial^{2}\rho_{b}\left(z_{I}^{\left(b\right)}\right)\right|^{-\frac{1}{2}}\prod_{r}e^{-\mathop{\mathrm{Re}}\bar{N}_{00}^{rr\left(b\right)}}\,,
\end{equation}
one can show 
\begin{eqnarray}
\lefteqn{\exp\left(-\Gamma_{\mathrm{super}}\left[\frac{1}{2},\,-\frac{i}{2}\left(\rho+\bar{\rho}\right)\right]\right)}\nonumber \\
 & = & \left|\sum_{r}\alpha_{r}Z_{r}\right|^{2}\exp\left[-\sum_{r}\mathop{\mathrm{Re}}\left(\bar{N}_{00}^{rr\left(b\right)}+\frac{\delta\rho\left(\tilde{\mathbf{z}}_{I^{\left(r\right)}}\right)}{\alpha_{r}}\right)\right]\prod_{I}\left|\partial^{2}\rho_{b}\left(z_{I}^{\left(b\right)}\right)\right|^{-\frac{1}{2}}\nonumber \\
 &  & \quad\times\prod_{I}\left|\left[1+\left(\frac{5}{12}\frac{\partial^{4}\rho_{b}}{\left(\partial^{2}\rho_{b}\right)^{3}}-\frac{3}{4}\frac{\left(\partial^{3}\rho_{b}\right)^{2}}{\left(\partial^{2}\rho_{b}\right)^{4}}\right)\partial ff\vphantom{\left.-\frac{2}{3}\frac{\partial^{3}ff}{\left(\partial^{2}\rho_{b}\right)^{2}}+\frac{\partial^{3}\rho_{b}}{\left(\partial^{2}\rho_{b}\right)^{3}}\partial^{2}ff+\frac{1}{12}\frac{\partial^{3}f\partial^{2}f\partial ff}{\left(\partial^{2}\rho_{b}\right)^{4}}\right)^{-1}}\right.\right.\nonumber \\
 &  & \hphantom{\quad\times\qquad}\left.\left.-\frac{2}{3}\frac{\partial^{3}ff}{\left(\partial^{2}\rho_{b}\right)^{2}}+\frac{\partial^{3}\rho_{b}}{\left(\partial^{2}\rho_{b}\right)^{3}}\partial^{2}ff+\frac{1}{12}\frac{\partial^{3}f\partial^{2}f\partial ff}{\left(\partial^{2}\rho_{b}\right)^{4}}\right]^{-1}\left(z_{I}^{\left(b\right)}\right)\right|^{2}\nonumber \\
 & = & \left|\sum_{r}\alpha_{r}Z_{r}\right|^{2}\exp\left[-\sum_{r}\mathop{\mathrm{Re}}\left(\bar{N}_{00}^{rr\left(b\right)}+\frac{\delta\rho\left(\tilde{\mathbf{z}}_{I^{\left(r\right)}}\right)}{\alpha_{r}}\right)\right]\nonumber \\
 &  & \hphantom{\quad}\times\prod_{I}\left|\left(\partial^{2}\rho-\frac{5}{3}\frac{\partial^{3}D\rho D\rho}{\partial^{2}\rho}+3\frac{\partial^{3}\rho\partial^{2}D\rho D\rho}{\left(\partial^{2}\rho\right)^{2}}\right)\left(\tilde{\mathbf{z}}_{I}\right)\right|^{-\frac{1}{2}}\nonumber \\
 &  & \hphantom{\quad}\times\exp\left(\frac{1}{2}\sum_{I}\mathop{\mathrm{Re}}\left[\frac{\partial^{3}ff}{\left(\partial^{2}\rho_{b}\right)^{2}}+\frac{\partial^{2}f\partial f}{\left(\partial^{2}\rho_{b}\right)^{2}}-\frac{\partial^{3}\rho_{b}}{\left(\partial^{2}\rho_{b}\right)^{3}}\partial^{2}ff\right]\left(z_{I}^{\left(b\right)}\right)\right)\ .\label{eq:Gammasuper1-2}
\end{eqnarray}
The exponent in the last factor can be given in a more explicit form
as 
\begin{eqnarray}
\lefteqn{\sum_{I}\left[\frac{\partial^{3}ff}{\left(\partial^{2}\rho_{b}\right)^{2}}+\frac{\partial^{2}f\partial f}{\left(\partial^{2}\rho_{b}\right)^{2}}-\frac{\partial^{3}\rho_{b}}{\left(\partial^{2}\rho_{b}\right)^{3}}\partial^{2}ff\right]\left(z_{I}^{\left(b\right)}\right)}\nonumber \\
 & = & \sum_{I}\oint_{z_{I}^{\left(b\right)}}\frac{dz}{2\pi i}\frac{\partial^{2}ff}{\left(\partial\rho_{b}\right)^{2}}\left(z\right)\nonumber \\
 & = & {}-\sum_{r}\oint_{Z_{r}}\frac{dz}{2\pi i}\frac{\partial^{2}ff}{\left(\partial\rho_{b}\right)^{2}}\left(z\right)-\oint_{\infty}\frac{dz}{2\pi i}\frac{\partial^{2}ff}{\left(\partial\rho_{b}\right)^{2}}\left(z\right)\nonumber \\
 & = & {}-\sum_{r}\frac{2\Theta_{r}}{\alpha_{r}}\sum_{s\ne r}\frac{\alpha_{s}\Theta_{s}}{Z_{r}-Z_{s}}-\frac{4\sum_{r}\alpha_{r}\Theta_{r}\sum_{s}\alpha_{s}\Theta_{s}Z_{s}}{\left(\sum_{t}\alpha_{t}Z_{t}\right)^{2}}\ .\label{eq:partialff}
\end{eqnarray}
Notice that the integrand in the second line is not a one-form. Therefore
in deforming the contour in the third line, we should take account
of the contribution from $z=\infty$, around which we need to introduce
$w=\frac{1}{z}$ as a good coordinate. Substituting (\ref{eq:partialff})
into (\ref{eq:Gammasuper1-2}), we eventually obtain 
\begin{eqnarray}
\lefteqn{\exp\left(-\Gamma_{\mathrm{super}}\left[\frac{1}{2},\,-\frac{i}{2}\left(\rho+\bar{\rho}\right)\right]\right)}\nonumber \\
 & = & \left|\sum_{r}\alpha_{r}Z_{r}-\frac{\sum_{r}\alpha_{r}\Theta_{r}\sum_{s}\alpha_{s}\Theta_{s}Z_{s}}{\sum_{t}\alpha_{t}Z_{t}}\right|^{2}\exp\left[-\sum_{r}\mathop{\mathrm{Re}}\bar{N}_{00}^{rr}\right]\nonumber \\
 &  & \quad\times\prod_{I}\left|\left(\partial^{2}\rho-\frac{5}{3}\frac{\partial^{3}D\rho D\rho}{\partial^{2}\rho}+3\frac{\partial^{3}\rho\partial^{2}D\rho D\rho}{\left(\partial^{2}\rho\right)^{2}}\right)\left(\tilde{\mathbf{z}}_{I}\right)\right|^{-\frac{1}{2}}\,.\label{eq:Gamma-super-result}
\end{eqnarray}
Here we have used 
\begin{equation}
\bar{N}_{00}^{rr}=\bar{N}_{00}^{rr(b)}+\frac{\delta\rho\left(\tilde{\mathbf{z}}_{I^{\left(r\right)}}\right)}{\alpha_{r}}+\frac{\Theta_{r}}{\alpha_{r}}\sum_{s\ne r}\frac{\alpha_{s}\Theta_{s}}{Z_{r}-Z_{s}}~.
\end{equation}
The result (\ref{eq:Gamma-super-result}) coincides with the one from
the calculations in \cite{Baba:2009fi,Berkovits:1985ji,Berkovits:1987gp}.

\subsection{$e^{-\Gamma_{\mathrm{super}}}$ for higher genus Riemann surfaces}

Let us now consider $e^{-\Gamma_{\mathrm{super}}}$ on a higher genus
Riemann surface $\Sigma$. All that we need to evaluate (\ref{eq:superXpmcorr})
is the partition function with Grassmann odd parameters $\Theta_{r}\ne0$.%
\footnote{In the higher genus case, $\Theta_{r}$'s, i.e. the Grassmann odd
components of the punctures are not enough to parametrize the Grassmann
odd directions of the supermoduli space. Therefore what we evaluate
here is the partition function on a submanifold in the supermoduli
space. %
} $\mathbf{\rho}\left(\mathbf{z}\right)$ is now given by 
\begin{eqnarray}
\rho\left(\mathbf{z}\right) & = & \rho_{b}(z)+\theta f(z)
\nonumber \\
 & = & \sum_{r=1}^{N}\alpha_{r}g(\mathbf{z},\mathbf{Z}_{r})\,,
\end{eqnarray}
where 
\begin{eqnarray}
f(z) & = & -\sum_{r}\alpha_{r}\Theta_{r}S_{\delta}(z,Z_{r})~,\nonumber \\
g(\mathbf{z},\mathbf{z}^{\prime}) & \equiv & g\left(z,z^{\prime}\right)-\theta\theta^{\prime}S_{\delta}\left(z,z^{\prime}\right)\,.
\end{eqnarray}
$g(z,z^{\prime})$ is defined in (\ref{eq:gzw}) and $S_{\delta}\left(z,z^{\prime}\right)$
is the Green's function of the worldsheet fermions of the spin structure
$\delta$. Here we deal with the case in which all the external lines
are in the NS-NS sector and $\delta$ is an even spin structure. $S_{\delta}\left(z,w\right)$
is therefore equal to the so-called Szego kernel 
\begin{equation}
S_{\delta}\left(z,w\right)=\frac{1}{E\left(z,w\right)}\frac{\vartheta\left[\delta\right]\left(z-w,\Omega\right)}{\vartheta\left[\delta\right]\left(0,\Omega\right)}\,.
\end{equation}
In the same way as in the sphere case, we can derive 
\begin{eqnarray}
\lefteqn{\left\langle V_{1}\left(Z_{1},\Theta_{1};\bar{Z}_{1},\bar{\Theta}_{1}\right)\cdots V_{N}\left(Z_{N},\Theta_{N};\bar{Z}_{N},\bar{\Theta}_{N}\right)\right\rangle _{\Sigma}}\nonumber \\
 & = & \left\langle \prod_{I}\exp\left(\xi_{I}\left(\partial^{2}\rho_{b}\right)^{-\frac{3}{4}}T_{F}^{\mathrm{LC}}\left(z_{I}^{\left(b\right)}\right)+\mathrm{c.c.}\right)\prod_{\mathcal{I}}\left(U_{\mathcal{I}}\bar{U}_{\mathcal{I}}\right)\prod_{r}V_{r}\left(Z_{r},0;\bar{Z}_{r},0\right)\right\rangle _{\Sigma}\nonumber \\
 &  & \quad\times\exp\left[\sum_{r}\mathop{\mathrm{Re}}\frac{\delta\rho\left(\tilde{\mathbf{z}}_{I^{\left(r\right)}}\right)}{\alpha_{r}}\right]\nonumber \\
 &  & \quad\times\prod_{I}\left|\left[1+\left(\frac{5}{12}\frac{\partial^{4}\rho_{b}}{\left(\partial^{2}\rho_{b}\right)^{3}}-\frac{3}{4}\frac{\left(\partial^{3}\rho_{b}\right)^{2}}{\left(\partial^{2}\rho_{b}\right)^{4}}\right)\partial ff\vphantom{\left.-\frac{2}{3}\frac{\partial^{3}ff}{\left(\partial^{2}\rho_{b}\right)^{2}}+\frac{\partial^{3}\rho_{b}}{\left(\partial^{2}\rho_{b}\right)^{3}}\partial^{2}ff+\frac{1}{12}\frac{\partial^{3}f\partial^{2}f\partial ff}{\left(\partial^{2}\rho_{b}\right)^{4}}\right)}\right.\right.\nonumber \\
 &  & \hphantom{\quad\times\qquad}\left.\left.-\frac{2}{3}\frac{\partial^{3}ff}{\left(\partial^{2}\rho_{b}\right)^{2}}+\frac{\partial^{3}\rho_{b}}{\left(\partial^{2}\rho_{b}\right)^{3}}\partial^{2}ff+\frac{1}{12}\frac{\partial^{3}f\partial^{2}f\partial ff}{\left(\partial^{2}\rho_{b}\right)^{4}}\right]\left(z_{I}^{\left(b\right)}\right)\right|^{2}\,.
\end{eqnarray}
We also have 
\begin{eqnarray}
\lefteqn{\exp\left(-\Gamma_{\mathrm{super}}\left[\hat{g}_{z\bar{z}},\,-\frac{i}{2}\left(\rho+\bar{\rho}\right)\right]\right)}\nonumber \\
 & = & \exp\left(-\frac{1}{2}\Gamma\left[\hat{g}_{z\bar{z}},\,-\frac{i}{2}\left(\rho_{b}+\bar{\rho}_{b}\right)\right]\right)\nonumber \\
 &  & \quad\times\left\langle \prod_{I}\exp\left(\xi_{I}\left(\partial^{2}\rho_{b}\right)^{-\frac{3}{4}}T_{F}^{\mathrm{LC}}\left(z_{I}^{\left(b\right)}\right)+\mathrm{c.c.}\right)\prod_{\mathcal{I}}\left(U_{\mathcal{I}}\bar{U}_{\mathcal{I}}\right)\right\rangle _{\Sigma}\,,
\end{eqnarray}
and eventually get the same form as (\ref{eq:Gammasuper2}) 
\begin{eqnarray}
\lefteqn{\exp\left(-\Gamma_{\mathrm{super}}\left[\hat{g}_{z\bar{z}},\,-\frac{i}{2}\left(\rho+\bar{\rho}\right)\right]\right)}\nonumber \\
 &  & =\exp\left(-\frac{1}{2}\Gamma\left[\hat{g}_{z\bar{z}},\,-\frac{i}{2}\left(\rho_{b}+\bar{\rho}_{b}\right)\right]-\sum_{r}\Delta\Gamma_{r}-\sum_{I}\Delta\Gamma_{I}\right)\,,\label{eq:Gammasupermulti2}
\end{eqnarray}
where 
\begin{eqnarray}
-\Delta\Gamma_{r} & = & \frac{1}{2\alpha_{r}}\frac{\partial ff}{\partial^{2}\rho_{b}}\left(z_{I^{\left(r\right)}}^{\left(b\right)}\right)+\mathrm{c.c.}\,,\nonumber \\
-\Delta\Gamma_{I} & = & \left\{ -\left(\frac{5}{12}\frac{\partial^{4}\rho_{b}}{\left(\partial^{2}\rho_{b}\right)^{3}}-\frac{3}{4}\frac{\left(\partial^{3}\rho_{b}\right)^{2}}{\left(\partial^{2}\rho_{b}\right)^{4}}\right)\partial ff+\frac{2}{3}\frac{\partial^{3}ff}{\left(\partial^{2}\rho_{b}\right)^{2}}\right.\nonumber \\
 &  & \quad\ \left.{}-\frac{\partial^{3}\rho_{b}}{\left(\partial^{2}\rho_{b}\right)^{3}}\partial^{2}ff-\frac{1}{12}\frac{\partial^{3}f\partial^{2}f\partial ff}{\left(\partial^{2}\rho_{b}\right)^{4}}\right\} \left(z_{I}^{\left(b\right)}\right)+\mathrm{c.c.}\,.
\end{eqnarray}
We do not know a good identity like (\ref{eq:partialff}) to rewrite
this expression further using superspace variables.

The normalization of the correlation function (\ref{eq:superXpmcorr})
is defined by the third line of (\ref{eq:superXpmcorr}) with $e^{-\Gamma_{\mathrm{super}}}$
given in (\ref{eq:Gammasupermulti2}).

\subsection{Correlation functions of supersymmetric $X^{\pm}$ CFT\label{sub:Correlation-functions-of}}

We can proceed in the same say as in the bosonic case and calculate
the correlation functions of supersymmetric $X^{\pm}$ CFT, starting
from (\ref{eq:superXpmcorr}). We get equations such as 
\begin{eqnarray}
\lefteqn{\left\langle F\left[\mathcal{X}^{+}\right]\prod_{r=1}^{N}e^{-ip_{r}^{+}\mathcal{X}^{-}}(\mathbf{Z}_{r},\bar{\mathbf{Z}}_{r})\prod_{s=1}^{M}e^{-ip_{s}^{-}\mathcal{X}^{+}}(\mathbf{w}_{s},\bar{\mathbf{w}}_{s})\right\rangle _{\hat{g}_{z\bar{z}}}^{\mathcal{X}^{\pm}}}\nonumber \\
 &  & =F\left[-\frac{i}{2}\left(\rho+\bar{\rho}\right)\right]\left\langle \prod_{r=1}^{N}e^{-ip_{r}^{+}\mathcal{X}^{-}}(\mathbf{Z}_{r},\bar{\mathbf{Z}}_{r})\prod_{s=1}^{M}e^{-ip_{s}^{-}\mathcal{X}^{+}}(\mathbf{w}_{s},\bar{\mathbf{w}}_{s})\right\rangle _{\hat{g}_{z\bar{z}}}^{\mathcal{X}^{\pm}}\,,\\
\lefteqn{\left\langle D\mathcal{X}^{-}\left(\mathbf{z}\right)\prod_{r=1}^{N}e^{-ip_{r}^{+}\mathcal{X}^{-}}(\mathbf{Z}_{r},\bar{\mathbf{Z}}_{r})\prod_{s=1}^{M}e^{-ip_{s}^{-}\mathcal{X}^{+}}(\mathbf{w}_{s},\bar{\mathbf{w}}_{s})\right\rangle _{\hat{g}_{z\bar{z}}}^{\mathcal{X}^{\pm}}}\nonumber \\
 &  & =\left[\sum_{s=1}^{M}\left(-ip_{s}^{-}\right)\begC1{D\mathcal{X}^{-}}\conC{\left(\mathbf{z}\right)}\endC1{\mathcal{X}^{+}}\left(\mathbf{w}_{s}\right)+\left\langle D\mathcal{X}^{-}\left(\mathbf{z}\right)\right\rangle _{\rho}\right]\nonumber \\
 &  & \hphantom{=\quad}\times\left\langle \prod_{r=1}^{N}e^{-ip_{r}^{+}\mathcal{X}^{-}}(\mathbf{Z}_{r},\bar{\mathbf{Z}}_{r})\prod_{s=1}^{M}e^{-ip_{s}^{-}\mathcal{X}^{+}}(\mathbf{w}_{s},\bar{\mathbf{w}}_{s})\right\rangle _{\hat{g}_{z\bar{z}}}^{\mathcal{X}^{\pm}}\,,\\
\lefteqn{\left\langle D\mathcal{X}^{-}\left(\mathbf{Z}_{0}\right)D\mathcal{X}^{-}\left(\mathbf{z}\right)\prod_{r=1}^{N}e^{-ip_{r}^{+}\mathcal{X}^{-}}(\mathbf{Z}_{r},\bar{\mathbf{Z}}_{r})\prod_{s=1}^{M}e^{-ip_{s}^{-}\mathcal{X}^{+}}(\mathbf{w}_{s},\bar{\mathbf{w}}_{s})\right\rangle _{\hat{g}_{z\bar{z}}}^{\mathcal{X}^{\pm}}}\nonumber \\
 &  & =\left[\left(\sum_{s=1}^{M}\left(-ip_{s}^{-}\right)\begC1{D\mathcal{X}^{-}}\conC{\left(\mathbf{Z_{0}}\right)}\endC1{\mathcal{X}^{+}}\left(\mathbf{w}_{s}\right)+\left\langle D\mathcal{X}^{-}\left(\mathbf{Z}_{0}\right)\right\rangle _{\rho}\right)\right.\nonumber \\
 &  & \hphantom{=\qquad\quad}\times\left(\sum_{s=1}^{M}\left(-ip_{s}^{-}\right)\begC1{D\mathcal{X}^{-}}\conC{\left(\mathbf{z}\right)}\endC1{\mathcal{X}^{+}}\left(\mathbf{w}_{s}\right)+\left\langle D\mathcal{X}^{-}\left(\mathbf{z}\right)\right\rangle _{\rho}\right)\nonumber \\
 &  & \hphantom{=\qquad}\left.\vphantom{\left(\sum_{s=1}^{M}\left(-ip_{s}^{-}\right)\begC1{\partial X^{-}}\conC{\left(Z_{0}\right)}\endC1{X^{+}}\left(w_{s}\right)+\left\langle \partial X^{-}\left(Z_{0}\right)\right\rangle _{\rho}\right)}+\left\langle D\mathcal{X}^{-}\left(\mathbf{Z}_{0}\right)D\mathcal{X}^{-}\left(\mathbf{z}\right)\right\rangle _{\rho}^{\mathrm{c}}\right]\nonumber \\
 &  & \hphantom{=}\times\left\langle \prod_{r=1}^{N}e^{-ip_{r}^{+}\mathcal{X}^{-}}(\mathbf{Z}_{r},\bar{\mathbf{Z}}_{r})\prod_{s=1}^{M}e^{-ip_{s}^{-}\mathcal{X}^{+}}(\mathbf{w}_{s},\bar{\mathbf{w}}_{s})\right\rangle _{\hat{g}_{z\bar{z}}}^{X^{\pm}}\,,
\end{eqnarray}
where 
\begin{eqnarray}
\begC1{D\mathcal{X}^{-}}\conC{\left(\mathbf{z}\right)}\endC1{\mathcal{X}^{+}}\left(\mathbf{w}_{s}\right) & = & D_{\mathbf{z}}\left[g\left(w_{s},z\right)-\Theta_{s}\theta S_{\delta}\left(w_{s},z\right)\right]\,,\nonumber \\
\left\langle D\mathcal{X}^{-}\left(z\right)\right\rangle _{\rho} & = & \left.iD_{\mathbf{Z}_{0}}\partial_{p_{0}^{+}}\left(-\frac{d-10}{8}\Gamma_{\mathrm{super}}\left[\hat{g}_{z\bar{z}},\,-\frac{i}{2}\left(\rho+\bar{\rho}\right)\right]\right)\right|_{p_{0}^{+}=0,\mathbf{Z}_{0}=\mathbf{z}}\,,\nonumber \\
\left\langle D\mathcal{X}^{-}\left(\mathbf{Z}_{0}\right)D\mathcal{X}^{-}\left(\mathbf{z}\right)\right\rangle _{\rho}^{\mathrm{c}} & = & \left.iD_{\mathbf{Z}_{0}}\partial_{p_{0}^{+}}\left(\left\langle D\mathcal{X}^{-}\left(\mathbf{z}\right)\right\rangle _{\rho^{\prime}}\right)\right|_{p_{0}^{+}=0}\,,
\nonumber \\
\rho^{\prime}\left(\mathbf{z}\right) & = & \rho\left(\mathbf{z}\right)+\alpha_{0}\Bigl(g\left(z,Z_{0}\right)-\theta\Theta_{0}S_{\delta}\left(z,Z_{0}\right)\nonumber \\
 &  & \hphantom{\rho\left(\mathbf{z}\right)+\alpha_{0}\left(\right.}{}-g\left(z,Z_{N+1}\right)+\theta\Theta_{N+1}S_{\delta}\left(z,Z_{N+1}\right)\Bigr)\,.\label{eq:DX-rho}
\end{eqnarray}

In this way, it is possible to evaluate all the correlation functions
of $\mathcal{X}^{+},\mathcal{X}^{-}$ with the source terms $e^{-ip_{r}^{+}\mathcal{X}^{-}}(\mathbf{Z}_{r},\bar{\mathbf{Z}}_{r})$,
$e^{-ip_{s}^{-}\mathcal{X}^{+}}(\mathbf{w}_{s},\bar{\mathbf{w}}_{s})$
starting from (\ref{eq:superXpmcorr}). The worldsheet theory of $\mathcal{X}^{\pm}$
thus defined turns out to be a superconformal field theory with the
central charge $\hat{c}=12-d$. The energy-momentum tensor is defined
as 
\begin{eqnarray}
T^{\mathcal{X}^{\pm}}\left(\mathbf{z}\right) & = & \frac{1}{2}:\partial\mathcal{X}^{+}D\mathcal{X}^{-}\left(\mathbf{z}\right):+\frac{1}{2}:D\mathcal{X}^{+}\partial\mathcal{X}^{-}\left(\mathbf{z}\right):-\frac{d-10}{4}S\left(\mathbf{z},\mbox{\mathversion{bold}\ensuremath{\mathcal{X}}}^{+}\right)\nonumber \\
 & = & \frac{1}{2}\lim_{\mathbf{w}\to\mathbf{z}}\left(\partial\mathcal{X}^{+}\left(\mathbf{w}\right)D\mathcal{X}^{-}\left(\mathbf{z}\right)-\partial_{\mathbf{w}}D_{\mathbf{z}}\ln\left(\mathbf{w}-\mathbf{z}\right)\right)\nonumber \\
 &  & \quad+\frac{1}{2}\lim_{\mathbf{w}\to\mathbf{z}}\left(D\mathcal{X}^{+}\left(\mathbf{w}\right)\partial\mathcal{X}^{-}\left(\mathbf{z}\right)-D_{\mathbf{w}}\partial_{\mathbf{z}}\ln\left(\mathbf{w}-\mathbf{z}\right)\right)\nonumber \\
 &  & \quad-\frac{d-10}{4}S\left(\mathbf{z},\mbox{\mathversion{bold}\ensuremath{\mathcal{X}}}^{+}\right)\,,\label{eq:superT-Xpm}
\end{eqnarray}
where 
\begin{equation}
S\left(\mathbf{z},\mbox{\mathversion{bold}\ensuremath{\mathcal{X}}}^{+}\right)=\frac{\partial^{2}\Theta^{+}}{D\Theta^{+}}-\frac{2\partial D\Theta^{+}\partial\Theta^{+}}{\left(D\Theta^{+}\right)^{2}}\,,
\end{equation}
and $\Theta^{+}(\mathbf{z})$ is given in (\ref{eq:Phi}). It is possible
to show that this has the following properties: 
\begin{itemize}
\item $T^{\mathcal{X}^{\pm}}\left(\mathbf{z}\right)$ is regular at $\mathbf{z}=\tilde{\mathbf{z}}_{I}$,
if there is no operator insertion there. 
\item The OPE between $T^{\mathcal{X}^{\pm}}\left(\mathbf{z}\right)$ and
$e^{-ip_{r}^{+}\mathcal{X}^{-}-ip_{r}^{-}\mathcal{X}^{+}}(\mathbf{Z}_{r},\bar{\mathbf{Z}}_{r})$
is given by 
\begin{eqnarray*}
\lefteqn{T^{\mathcal{X}^{\pm}}\left(\mathbf{z}\right)e^{-ip_{r}^{+}\mathcal{X}^{-}-ip_{r}^{-}\mathcal{X}^{+}}(\mathbf{Z}_{r},\bar{\mathbf{Z}}_{r})}\\
 &  & \sim\frac{\theta-\Theta_{r}}{\left(\mathbf{z}-\mathbf{Z}_{r}\right)^{2}}\left(-p_{r}^{+}p_{r}^{-}\right)e^{-ip_{r}^{+}\mathcal{X}^{-}-ip_{r}^{-}\mathcal{X}^{+}}(\mathbf{Z}_{r},\bar{\mathbf{Z}}_{r})\\
 &  & \hphantom{\sim}{}+\frac{1}{\mathbf{z}-\mathbf{Z}_{r}}\frac{1}{2}De^{-ip_{r}^{+}\mathcal{X}^{-}-ip_{r}^{-}\mathcal{X}^{+}}(\mathbf{Z}_{r},\bar{\mathbf{Z}}_{r})\\
 &  & \hphantom{\sim}{}+\frac{\theta-\Theta_{r}}{\mathbf{z}-\mathbf{Z}_{r}}\partial e^{-ip_{r}^{+}\mathcal{X}^{-}-ip_{r}^{-}\mathcal{X}^{+}}(\mathbf{Z}_{r},\bar{\mathbf{Z}}_{r})\,.
\end{eqnarray*}

\item The OPE between two $T^{\mathcal{X}^{\pm}}$'s is given by 
\begin{eqnarray*}
\lefteqn{T^{\mathcal{X}^{\pm}}\left(\mathbf{z}\right)T^{\mathcal{X}^{\pm}}\left(\mathbf{z}^{\prime}\right)}\\
 &  & \sim\frac{12-d}{4\left(\mathbf{z}-\mathbf{z}^{\prime}\right)^{3}}+\frac{\theta-\theta^{\prime}}{\left(\mathbf{z}-\mathbf{z}^{\prime}\right)^{2}}\frac{3}{2}T^{\mathcal{X}^{\pm}}\left(\mathbf{z}^{\prime}\right)+\frac{1}{\mathbf{z}-\mathbf{z}^{\prime}}\frac{1}{2}DT^{\mathcal{X}^{\pm}}\left(\mathbf{z}^{\prime}\right)+\frac{\theta-\theta^{\prime}}{\mathbf{z}-\mathbf{z}^{\prime}}\partial T^{\mathcal{X}^{\pm}}\left(\mathbf{z}^{\prime}\right),
\end{eqnarray*}
which corresponds to the super Virasoro algebra with the central charge
$\hat{c}=12-d$. 
\end{itemize}
These facts can be derived by calculating the expectation value of
the energy-momentum tensor and their OPE's. Some of the details of
these calculations are given in appendix~\ref{sec:Calculations-of-the}.
It follows that combined with the transverse variables $\mathcal{X}^{i}\left(\mathbf{z},\mathbf{\bar{z}}\right)$
$(i=1,\ldots,d-2)$, the total central charge of the system becomes
$\hat{c}=10$. This implies that with the ghost superfields $B\left(\mathbf{z}\right)$
and $C\left(\mathbf{z}\right)$ defined as 
\begin{equation}
B(\mathbf{z})=\beta(z)+\theta b(z)\;,\qquad C(\mathbf{z})=c(z)+\theta\gamma(z)\;,
\end{equation}
it is possible to construct a nilpotent BRST charge 
\begin{equation}
Q_{\mathrm{B}}=\oint\frac{d\mathbf{z}}{2\pi i}\left[-C\left(T^{\mathcal{X}^{\pm}}+T^{\mathrm{LC}}\right)+\left(C\partial C-\frac{1}{4}\left(DC\right)^{2}\right)B\right].
\end{equation}

\section{Discussions\label{sec:Discussions}}

In this paper, we have studied the longitudinal part of the worldsheet
theory corresponding to the light-cone gauge string field theory in
noncritical dimensions, on higher genus Riemann surfaces. We have
defined and calculated the correlation functions of both bosonic and
supersymmetric cases and shown that they have the right properties
to be used to describe the theory in noncritical dimensions.

In order to analyze the supersymmetric case, we have proposed a way
to calculate $\Gamma_{\mathrm{super}}$ which is much simpler than
those proposed in \cite{Berkovits:1985ji,Berkovits:1987gp} or \cite{Baba:2009fi}.
The correlation functions in the higher genus case are quite complicated
because we have not been able to find a way to express it in terms
of the superfield $\rho(\mathbf{z})$ so far. It seems easy to generalize
our method to the $N=2$ case and calculate the anomaly factor on
higher genus Riemann surfaces. Such a calculation will be useful for
studying the amplitudes involving Ramond sector fields \cite{Berkovits1990}
or amplitudes in Green-Schwarz formalism \cite{Berkovits1992a,Berkovits1993,Berkovits2002,Green1983b,Green1983}.

With the $X^{\pm}$ CFT studied in this paper, it is possible to describe
the conformal gauge formulation of the light-cone gauge theory in
noncritical dimensions. We can construct the nilpotent BRST charge
and rewrite the light-cone gauge amplitudes in terms of the conformal
gauge worldsheet theory. The amplitudes in noncritical dimensions
can be used to regularize various divergences in a gauge invariant
way. In \cite{Baba:2009kr,Baba:2009zm,Ishibashi:2010nq,Ishibashi:2011fy},
we have dealt with the contact term divergences of the tree amplitudes.
The results in this paper shall be used in analyzing the higher-loop
superstring amplitudes in separate publications.

\section*{Acknowledgments}

One of the authors (K.M.) would like to acknowledge the hospitality
of Particle Theory Group at University of Tsukuba, where part of this
work was done. This work was supported in part by Grant-in-Aid for
Scientific Research (C) (25400242) and (15K05063) from MEXT.

\appendix

\section{Definition of the path integral of $X^{\pm}$ variables\label{sec:Definition-of-the}}

In this appendix, we explain how to define the path integral on the
right hand side of (\ref{eq:Xpmcorrdef}) following \cite{Ishibashi:2013nma}.
Since the action of $X^{\pm}$ variables is not bounded below, we
need to take the integration contours of $X^{\pm}$ carefully to define
the path integral. Let us first recapitulate how we do so in the critical
case. The action for $X^{\pm}$ is given as 
\begin{equation}
S_{d=26}^{\pm}=-\frac{1}{4\pi}\int dz\wedge d\bar{z}\, i\left(\partial X^{+}\bar{\partial}X^{-}+\partial X^{-}\bar{\partial}X^{+}\right)\,.
\end{equation}
The path integrals to be defined are of the form 
\begin{equation}
\int\left[dX^{+}dX^{-}\right]_{\hat{g}_{z\bar{z}}}e^{-S_{d=26}^{\pm}}\prod_{r=1}^{N}e^{-ip_{r}^{+}X^{-}}(Z_{r},\bar{Z}_{r})\prod_{s=1}^{M}e^{-ip_{s}^{-}X^{+}}(z_{s},\bar{z}_{s})\,,\label{eq:pathpm}
\end{equation}
which is supposed to be equal to 
\begin{equation}
(2\pi)^{2}\delta\left(\sum_{r}p_{r}^{+}\right)\delta\left(\sum_{s}p_{s}^{-}\right)\prod_{s=1}^{M}e^{-ip_{s}^{-}X_{\mathrm{cl}}^{+}}(z_{s},\bar{z}_{s})Z^{X}[\hat{g}_{z\bar{z}}]^{2}~,\label{eq:ctiricalxpm}
\end{equation}
where $X_{\mathrm{cl}}^{\pm}(z,\bar{z})$ are solutions to the equations
of motion with the source terms 
\begin{eqnarray}
\partial\bar{\partial}X_{\mathrm{cl}}^{+}(z,\bar{z}) & = & -i\sum_{r}p_{r}^{+}(-2\pi i)\delta^{2}(z-Z_{r})\,,\nonumber \\
\partial\bar{\partial}X_{\mathrm{cl}}^{-}(z,\bar{z}) & = & -i\sum_{s}p_{s}^{-}(-2\pi i)\delta^{2}(z-z_{s})\,.
\end{eqnarray}

In order to derive (\ref{eq:ctiricalxpm}), we decompose the variable
$X^{\pm}$ as 
\begin{equation}
X^{\pm}(z,\bar{z})=X_{\mathrm{cl}}^{\pm}(z,\bar{z})+x^{\pm}+\delta X^{\pm}(z,\bar{z})~,
\end{equation}
where $x^{\pm}+\delta X^{\pm}(z,\bar{z})$ are the fluctuations around
the solutions with 
\begin{equation}
\int dz\wedge d\bar{z}\sqrt{\hat{g}}\delta X^{\pm}=0~.
\end{equation}
Integrals over $X^{\pm}$ are expressed as those over $x^{\pm}$ and
$\delta X^{\pm}$. In (\ref{eq:pathpm}), we take the integration
contours of $x^{\pm}$ and $\delta X^{+}-\delta X^{-}$ to be along
the real axis and that of $\delta X^{+}+\delta X^{-}$ to be along
the imaginary axis. Then (\ref{eq:pathpm}) becomes well-defined and
is evaluated to be (\ref{eq:ctiricalxpm}).

Based on this definition of the path integral (\ref{eq:pathpm}),
it is possible to evaluate (\ref{eq:Xpmcorrdef}) as follows. The
right hand side of (\ref{eq:Xpmcorrdef}) can be expanded as 
\begin{eqnarray}
\lefteqn{\left(Z^{X}[\hat{g}_{z\bar{z}}]\right)^{-2}\int\left[dX^{+}dX^{-}\right]_{\hat{g}_{z\bar{z}}}e^{-S^{\pm}\left[\hat{g}_{z\bar{z}}\right]}\prod_{r=1}^{N}e^{-ip_{r}^{+}X^{-}}(Z_{r},\bar{Z}_{r})\prod_{s=1}^{M}e^{-ip_{s}^{-}X^{+}}(w_{s},\bar{w}_{s})}\nonumber \\
 &  & =\left(Z^{X}[\hat{g}_{z\bar{z}}]\right)^{-2}\int\left[dX^{+}dX^{-}\right]_{\hat{g}_{z\bar{z}}}e^{-S_{d=26}^{\pm}\left[\hat{g}_{z\bar{z}}\right]}\prod_{r=1}^{N}e^{-ip_{r}^{+}X^{-}}(Z_{r},\bar{Z}_{r})\prod_{s=1}^{M}e^{-ip_{s}^{-}X^{+}}(w_{s},\bar{w}_{s})\nonumber \\
 &  & \hphantom{\quad=\left(Z^{X}[\hat{g}_{z\bar{z}}]\right)^{-2}\int\left[dX^{+}dX^{-}\right]_{\hat{g}_{z\bar{z}}}}\times\sum_{n=0}^{\infty}\frac{1}{n!}\left(-\frac{d-26}{24}\Gamma\left[\hat{g}_{z\bar{z}},X^{+}\right]\right)^{n}\,,\label{eq:Xpmcorrdef2}
\end{eqnarray}
and expressed in terms of the correlation functions of the theory
in the critical dimensions. Using the prescription for the contours
of the integration, it is straightforward to prove 
\begin{eqnarray}
0 & = & \int\left[dX^{+}dX^{-}\right]_{\hat{g}_{z\bar{z}}}e^{-S_{d=26}^{\pm}}\nonumber \\
 &  & \quad\times\prod_{r=1}^{N}e^{-ip_{r}^{+}X^{-}}(Z_{r}.\bar{Z}_{r})\prod_{s=1}^{M}e^{-ip_{s}^{-}X^{+}}(z_{s}.\bar{z}_{s})\prod_{p=1}^{n}\partial\delta X^{+}(z_{p})\prod_{q=1}^{m}\bar{\partial}\delta X^{+}(\bar{z}_{q})~,
\end{eqnarray}
if $n\mbox{ or }m\geq1$. Using this and (\ref{eq:ctiricalxpm}),
it is possible to show that the right hand side of (\ref{eq:Xpmcorrdef2})
is equal to 
\begin{equation}
(2\pi)^{2}\,\delta\left(\sum_{s}p_{s}^{-}\right)\delta\left(\sum_{r}p_{r}^{+}\right)\prod_{s}e^{-p_{s}^{-}\frac{\rho+\bar{\rho}}{2}}(z_{s},\bar{z}_{s})\, e^{-\frac{d-26}{24}\Gamma\left[\hat{g}_{z\bar{z}},\,-\frac{i}{2}\left(\rho+\bar{\rho}\right)\right]}~.
\end{equation}

\section{Arakelov metric and Arakelov Green's function\label{sec:Arakelov-metric-and}}

In this appendix, we give the definitions of the Arakelov metric and
the Arakelov Green's function, following \cite{Dugan:1987qe,D'Hoker:1989ae}.

Let us define $\mu_{z\bar{z}}$ as 
\begin{equation}
\mu_{z\bar{z}}\equiv\frac{1}{2g}\omega(z)\frac{1}{\mathop{\mathrm{Im}}\Omega}\bar{\omega}(\bar{z})~.\label{eq:Bergman}
\end{equation}
We note that 
\begin{equation}
\int_{\Sigma}dz\wedge d\bar{z}\, i\mu_{z\bar{z}}=1~,\label{eq:norm-Bergman}
\end{equation}
which follows from 
\begin{equation}
\int_{\Sigma}\omega_{\mu}\wedge\bar{\omega}_{\nu}=-2i\mathop{\mathrm{Im}}\Omega_{\mu\nu}~.
\end{equation}
The Arakelov metric on $\Sigma$, 
\begin{equation}
ds_{\mathrm{A}}^{\;2}=2g_{z\bar{z}}^{\mathrm{A}}dzd\bar{z}~,\label{eq:Arakelov}
\end{equation}
is defined so that its scalar curvature $R^{\mathrm{A}}\equiv-2g^{\mathrm{A}z\bar{z}}\partial\bar{\partial}\ln g_{z\bar{z}}^{\mathrm{A}}$
satisfies 
\begin{equation}
g_{z\bar{z}}^{\mathrm{A}}R^{\mathrm{A}}=-8\pi(g-1)\mu_{z\bar{z}}~.\label{eq:ArakelovR}
\end{equation}
This condition determines $g_{z\bar{z}}^{\mathrm{A}}$ only up to
an overall constant, which we will choose later.

The Arakelov Green's function $G^{\mathrm{A}}(z,\bar{z};w,\bar{w})$
with respect to the Arakelov metric is defined to satisfy 
\begin{eqnarray}
 &  & -\partial_{z}\partial_{\bar{z}}G^{\mathrm{A}}(z,\bar{z};w,\bar{w})=2\pi\delta^{2}(z-w)-2\pi\mu_{z\bar{z}}~,\nonumber \\
 &  & \int_{\Sigma}dz\wedge d\bar{z}\, i\mu_{z\bar{z}}G^{\mathrm{A}}(z,\bar{z};w,\bar{w})=0~.\label{eq:G-B}
\end{eqnarray}
One can obtain a more explicit form of $G^{\mathrm{A}}(z,\bar{z};w,\bar{w})$
by solving (\ref{eq:G-B}) for $G^{\mathrm{A}}(z,\bar{z};w,\bar{w})$.
Let $F(z,\bar{z};w,\bar{w})$ be the $\left(-\frac{1}{2},-\frac{1}{2}\right)\times\left(-\frac{1}{2},-\frac{1}{2}\right)$
form on $\Sigma\times\Sigma$ defined as 
\begin{equation}
F(z,\bar{z};w,\bar{w})=\exp\left[-2\pi\mathop{\mathrm{Im}}\int_{w}^{z}\omega\frac{1}{\mathop{\mathrm{Im}}\Omega}\mathop{\mathrm{Im}}\int_{w}^{z}\omega\right]\left|E(z,w)\right|^{2}~.\label{eq:propagator2}
\end{equation}
It is easy to show 
\begin{equation}
\partial_{z}\partial_{\bar{z}}\ln F(z,\bar{z};w,\bar{w})=-2\pi i\delta^{2}(z-w)-2\pi g\mu_{z\bar{z}}~.\label{eq:diffeq-propagator}
\end{equation}
Putting eqs.(\ref{eq:diffeq-propagator}) and (\ref{eq:ArakelovR})
together, we find that $G^{\mathrm{A}}(z,\bar{z};w,\bar{w})$ is given
by 
\begin{equation}
G^{\mathrm{A}}(z,\bar{z};w,\bar{w})=-\ln F(z,\bar{z};w,\bar{w})-\frac{1}{2}\ln\left(2g_{z\bar{z}}^{\mathrm{A}}\right)-\frac{1}{2}\ln\left(2g_{w\bar{w}}^{\mathrm{A}}\right)~,\label{eq:G-F-g-g}
\end{equation}
up to an additive constant independent of $z,\bar{z}$ and $w,\bar{w}$.
This possible additive constant can be absorbed into the ambiguity
in the overall constant of $g_{z\bar{z}}^{\mathrm{A}}$ mentioned
above. It is required that (\ref{eq:G-F-g-g}) holds exactly as it
is \cite{Dugan:1987qe,Sonoda:1987ra,D'Hoker:1989ae}. This implies
that 
\begin{equation}
2g_{z\bar{z}}^{\mathrm{A}}=\lim_{w\to z}\exp\left[-G^{\mathrm{A}}(z,\bar{z};w,\bar{w})-\ln|z-w|^{2}\right]~,\label{eq:gA-expGA}
\end{equation}
and the overall constant of $g_{z\bar{z}}^{\mathrm{A}}$ is, in principle,
determined by the second relation in (\ref{eq:G-B}).

\section{Calculations of correlation functions of bosonic $X^{\pm}$ CFT \label{sec:Calculations-of-various}}

In this appendix, we calculate the correlation functions involving
the energy-momentum tensor of the bosonic $X^{\pm}$ CFT and show
various properties of the theory.

\subsection{Evaluation of $\left\langle \partial X^{-}(z)\right\rangle _{\rho}$}

We evaluate the relevant quantities starting from the expectation
value $\left\langle \partial X^{-}\left(z\right)\right\rangle _{\rho}$
which is given as 
\begin{equation}
\left\langle \partial X^{-}(z)\right\rangle _{\rho}=\left.2i\partial_{Z_{0}}\partial_{\alpha_{0}}\left(-\frac{d-26}{24}\Gamma\left[g_{z\bar{z}}^{\mathrm{A}},\,-\frac{i}{2}\left(\rho^{\prime}+\bar{\rho}^{\prime}\right)\right]\right)\right|_{\alpha_{0}=0,Z_{0}=z}\,.
\end{equation}
In the following, various quantities defined by using $\rho^{\prime}$
given in (\ref{eq:rhoprime}) instead of $\rho$ in (\ref{eq:rhoz})
will be denoted by attaching a prime. We also define 
\begin{eqnarray}
\partial_{Z_{0}}\tilde{g}\left(Z_{0},Z_{N+1}\right) & \equiv & \frac{\partial}{\partial Z_{0}}\ln E(Z_{0},Z_{N+1})-2\pi i\omega\left(Z_{0}\right)\frac{1}{\mathop{\mathrm{Im}}\Omega}\mathop{\mathrm{Im}}\int_{Z_{0}}^{Z_{N+1}}\omega\,,\nonumber \\
\partial_{Z_{N+1}}\tilde{g}\left(Z_{N+1},Z_{0}\right) & \equiv & \frac{\partial}{\partial Z_{N+1}}\ln E(Z_{0},Z_{N+1})-2\pi i\omega\left(Z_{N+1}\right)\frac{1}{\mathop{\mathrm{Im}}\Omega}\mathop{\mathrm{Im}}\int_{Z_{N+1}}^{Z_{0}}\omega\,,
\end{eqnarray}
which make the following calculations look simpler.

In order to get $\left\langle \partial X^{-}\left(z\right)\right\rangle $,
we need the following expansions in terms of $\alpha_{0}$: 
\begin{eqnarray}
z'_{I}-z_{I} & = & -\frac{\alpha_{0}}{\partial^{2}\rho(z_{I})}\frac{\partial}{\partial z_{I}}\left(g\left(z_{I},Z_{0}\right)-g\left(z_{I},Z_{N+1}\right)\right)+\mathcal{O}(\alpha_{0}^{2})\,,\nonumber \\
z'_{I^{(0)}}-Z_{0} & = & -\frac{\alpha_{0}}{\partial\rho(Z_{0})}\nonumber \\
 &  & {}-\alpha_{0}^{2}\left(\frac{\partial^{2}\rho(Z_{0})}{\left(\partial\rho(Z_{0})\right)^{3}}+\frac{1}{\left(\partial\rho\left(Z_{0}\right)\right)^{2}}\partial_{Z_{0}}\tilde{g}\left(Z_{0},Z_{N+1}\right)\right)+\mathcal{O}\left(\alpha_{0}^{3}\right)\,,\nonumber \\
z'_{I^{(N+1)}}-Z_{N+1} & = & \frac{\alpha_{0}}{\partial\rho(Z_{N+1})}-\alpha_{0}^{2}\left(\frac{\partial^{2}\rho(Z_{N+1})}{\left(\partial\rho(Z_{N+1})\right)^{3}}\right.\nonumber \\
 &  & \qquad\qquad\qquad\quad\left.{}+\frac{1}{\left(\partial\rho\left(Z_{N+1}\right)\right)^{2}}\partial_{N+1}\tilde{g}\left(Z_{N+1},Z_{0}\right)\right)+\mathcal{O}\left(\alpha_{0}^{3}\right)\,,\label{eq:zI0prime-Z0}\\
\bar{N}_{\;00}^{\prime rr} & = & \bar{N}_{00}^{rr}+\frac{\alpha_{0}}{\alpha_{r}}\left(g\left(z_{I^{\left(r\right)}},Z_{0}\right)-g\left(z_{I^{\left(r\right)}},Z_{N+1}\right)-g\left(Z_{r},Z_{0}\right)+g\left(Z_{r},Z_{N+1}\right)\right)\nonumber \\
 &  & \quad{}+\mathcal{O}(\alpha_{0}^{2})\,,\nonumber \\
\bar{N}_{\;00}^{\prime00} & = & \ln\left(-\frac{\alpha_{0}}{\partial\rho(Z_{0})}\right)-1+\frac{\alpha_{0}}{2}\frac{\partial^{2}\rho(Z_{0})}{\left(\partial\rho(Z_{0})\right)^{2}}+\frac{\alpha_{0}}{\partial\rho(Z_{0})}\partial_{Z_{0}}\tilde{g}\left(Z_{0},Z_{N+1}\right)+\mathcal{O}(\alpha_{0}^{2})\,,\nonumber \\
\bar{N}_{\;0\;0}^{\prime N+1N+1} & = & \ln\left(\frac{\alpha_{0}}{\partial\rho(Z_{N+1})}\right)-1\nonumber \\
 &  & \quad-\frac{\alpha_{0}}{2}\frac{\partial^{2}\rho(Z_{N+1})}{\left(\partial\rho(Z_{N+1})\right)^{2}}-\frac{\alpha_{0}}{\partial\rho(Z_{N+1})}\partial_{Z_{N+1}}\tilde{g}\left(Z_{N+1},Z_{0}\right)+\mathcal{O}(\alpha_{0}^{2})\,,\\
\partial^{2}\rho'\left(z'_{I}\right) & = & \partial^{2}\rho(z_{I})+\alpha_{0}\frac{\partial^{2}}{\partial z_{I}^{2}}\left(g\left(z_{I},Z_{0}\right)-g\left(z_{I},Z_{N+1}\right)\right)\nonumber \\
 &  & \hphantom{\partial^{2}\rho(z_{I})}{}-\alpha_{0}\frac{\partial^{3}\rho}{\partial^{2}\rho}\left(z_{I}\right)\frac{\partial}{\partial z_{I}}\left(g\left(z_{I},Z_{0}\right)-g\left(z_{I},Z_{N+1}\right)\right)+\mathcal{O}(\alpha_{0}^{2})\,,\nonumber \\
\partial^{2}\rho'\left(z'_{I^{(0)}}\right) & = & -\frac{\left(\partial\rho(Z_{0})\right)^{2}}{\alpha_{0}}+3\partial^{2}\rho(Z_{0})+2\partial\rho\left(Z_{0}\right)\partial_{Z_{0}}\tilde{g}\left(Z_{0},Z_{N+1}\right)+\mathcal{O}\left(\alpha_{0}\right)\,,\nonumber \\
\partial^{2}\rho'\left(z'_{I^{(N+1)}}\right) & = & \frac{\left(\partial\rho(Z_{N+1})\right)^{2}}{\alpha_{0}}+3\partial^{2}\rho(Z_{N+1})+2\partial\rho\left(Z_{N+1}\right)\partial_{Z_{N+1}}\tilde{g}\left(Z_{N+1},Z_{0}\right)+\mathcal{O}\left(\alpha_{0}\right)\,,\nonumber \\
\partial^{3}\rho'\left(z'_{I}\right) & = & \partial^{3}\rho(z_{I})+\mathcal{O}(\alpha_{0})\,,\nonumber \\
\partial^{3}\rho'\left(z'_{I^{(0)}}\right) & = & -\frac{2}{\alpha_{0}^{2}}\left(\partial\rho(Z_{0})\right)^{3}\nonumber \\
 &  & {}+\frac{6}{\alpha_{0}}\partial\rho(Z_{0})\partial^{2}\rho(Z_{0})+\frac{6}{\alpha_{0}}\left(\partial\rho(Z_{0})\right)^{2}\partial_{Z_{0}}\tilde{g}\left(Z_{0},Z_{N+1}\right)+\mathcal{O}\left(\alpha_{0}^{0}\right)\,,\nonumber \\
\partial^{3}\rho'\left(z'_{I^{(N+1)}}\right) & = & -\frac{2}{\alpha_{0}^{2}}\left(\partial\rho(Z_{N+1})\right)^{3}-\frac{6}{\alpha_{0}}\partial\rho(Z_{N+1})\partial^{2}\rho(Z_{N+1})\nonumber \\
 &  & \hphantom{-\frac{2}{\alpha_{0}^{2}}\left(\partial\rho(Z_{N+1})\right)^{3}}{}-\frac{6}{\alpha_{0}}\left(\partial\rho(Z_{N+1})\right)^{2}\partial_{Z_{N+1}}\tilde{g}\left(Z_{N+1},Z_{0}\right)+\mathcal{O}\left(\alpha_{0}^{0}\right)\,.
\end{eqnarray}
Here $z_{I}^{\prime}$ $(I=1,\ldots,2g-2+N)$, $z_{I^{(0)}}^{\prime}$,
$z_{I^{(N+1)}}^{\prime}$ are the interaction points for $\rho^{\prime}$,
which behave as $z_{I}^{\prime}\to z_{I}$, $z_{I^{(0)}}^{\prime}\to Z_{0}$,
$z_{I^{(N+1)}}^{\prime}\to Z_{N+1}$ in the limit $\alpha_{0}\to0$
respectively.

We also need the formula for the variation of $-W^{\prime}$, which
is the $-W$ in (\ref{eq:-W}) defined for $\rho^{\prime}$. Comparing
the behavior of the Schwarzian derivative for $z\sim z_{I}$ 
\begin{equation}
-2\{\rho,z\}\sim{\displaystyle \frac{3}{(z-z_{I})^{2}}+\frac{1}{z-z_{I}}\frac{\partial^{3}\rho(z_{I})}{\partial^{2}\rho(z_{I})}}\,,\label{eq:SchwarzzIloop}
\end{equation}
derived from $\partial\rho\left(z\right)\sim\partial^{2}\rho\left(z_{I}\right)\left(z-z_{I}\right)+\frac{1}{2}\partial^{3}\rho\left(z_{I}\right)\left(z-z_{I}\right)^{2}$
with 
\begin{equation}
-2\{\rho,z\}\sim\frac{3}{(z-z_{I})^{2}}+\frac{1}{z-z_{I}}\frac{\partial}{\partial z_{I}}(-W)\,,
\end{equation}
which is given in \cite{Ishibashi:2013nma} (eq.(B.15)), we obtain
the formula 
\begin{equation}
\frac{\partial(-W)}{\partial z_{I}}=\frac{\partial^{3}\rho(z_{I})}{\partial^{2}\rho(z_{I})}\,.
\end{equation}
{}From this, we get the expansion of $-W^{\prime}$ as 
\begin{eqnarray*}
\frac{\partial\left(-W^{\prime}\right)}{\partial z'_{I}} & = & \frac{\partial\left(-W\right)}{\partial z_{I}}+\mathcal{O}(\alpha_{0})\,,\\
\frac{\partial\left(-W^{\prime}\right)}{\partial z'_{I^{(0)}}} & = & \frac{2}{\alpha_{0}}\partial\rho(Z_{0})-2\partial_{Z_{0}}\tilde{g}\left(Z_{0},Z_{N+1}\right)+\mathcal{O}(\alpha_{0})\,,\\
\frac{\partial\left(-W'\right)}{\partial z'_{I^{(N+1)}}} & = & -\frac{2}{\alpha_{0}}\partial\rho(Z_{N+1})-2\partial_{Z_{N+1}}\tilde{g}\left(Z_{N+1},Z_{0}\right)+\mathcal{O}(\alpha_{0})\,.
\end{eqnarray*}

Using all these, it is straightforward to derive 
\begin{eqnarray}
\lefteqn{\left\langle \partial X^{-}\left(z\right)\right\rangle _{\rho}=\left.2i\partial_{Z_{0}}\partial_{\alpha_{0}}\left(-\frac{d-26}{24}\Gamma\left[g_{z\bar{z}}^{\mathrm{A}},\,-\frac{i}{2}\left(\rho^{\prime}+\bar{\rho}^{\prime}\right)\right]\right)\right|_{\alpha_{0}=0,\, Z_{0}=z}}\nonumber \\
 & = & \left.2i\partial_{Z_{0}}\partial_{\alpha_{0}}\left[-\frac{d-26}{24}\left(-2\sum_{r=0}^{N+1}\mathop{\mathrm{Re}}\bar{N}_{\ 00}^{\prime rr}-\frac{3}{2}\sum_{I}\ln|\partial^{2}\rho^{\prime}(z_{I}^{\prime})|^{2}-W^{\prime}\right)\right]\right|_{\alpha_{0}=0,\, Z_{0}=z}\nonumber \\
 & = & \frac{d-26}{24}2i\partial_{z}\left[\sum_{r=1}^{N}B_{r}\left(z,\bar{z}\right)+\sum_{I=1}^{2g-2+N}B_{I}\left(z,\bar{z}\right)\right]\,,\label{eq:dX-loop}
\end{eqnarray}
where 
\begin{eqnarray}
B_{r}\left(z,\bar{z}\right) & \equiv & \frac{1}{\alpha_{r}}\left(g\left(Z_{r},z\right)-g\left(z_{I^{\left(r\right)}},z\right)\right)+\mathrm{c.c.}\,,
\nonumber \\
B_{I}\left(z,\bar{z}\right) & \equiv & \frac{\partial^{3}\rho}{2\left(\partial^{2}\rho\right)^{2}}\left(z_{I}\right)\frac{\partial}{\partial z_{I}}g\left(z_{I},z\right)-\frac{3}{2\partial^{2}\rho\left(z_{I}\right)}\frac{\partial^{2}}{\partial z_{I}^{2}}g\left(z_{I},z\right)+\mathrm{c.c.}\,.\label{eq:BIz}
\end{eqnarray}
Notice that $\left\langle \partial X^{-}\left(z\right)\right\rangle _{\rho}$
is meromorphic with respect to $z$.

\subsection{Energy-momentum tensor}

Now let us check if the energy-momentum tensor $T^{X^{\pm}}$ of the
$X^{\pm}$ CFT defined in (\ref{eq:bosonicT-Xpm}) satisfies the desired
properties. 

We first consider the correlation function of the form 
\begin{equation}
\left\langle T^{X^{\pm}}\left(z\right)\prod_{r=1}^{N}e^{-ip_{r}^{+}X^{-}}(Z_{r},\bar{Z}_{r})\prod_{s=1}^{M}e^{-ip_{s}^{-}X^{+}}(w_{s},\bar{w}_{s})\right\rangle _{\hat{g}_{z\bar{z}}}^{X^{\pm}}\,,\label{eq:XpmT1}
\end{equation}
which can be evaluated as 
\begin{eqnarray}
\lefteqn{\left\langle T^{X^{\pm}}\left(z\right)\prod_{r=1}^{N}e^{-ip_{r}^{+}X^{-}}(Z_{r},\bar{Z}_{r})\prod_{s=1}^{M}e^{-ip_{s}^{-}X^{+}}(w_{s},\bar{w}_{s})\right\rangle _{\hat{g}_{z\bar{z}}}^{X^{\pm}}}\nonumber \\
 &  & =\left[-\frac{i}{2}\partial\rho(z)\sum_{s=1}^{M}\left(-ip_{s}^{-}\right)\begC1{\partial X^{-}}\conC{\left(z\right)}\endC1{X^{+}}\left(w_{s}\right)+\left\langle T^{X^{\pm}}(z)\right\rangle _{\rho}\right]\nonumber \\
 &  & \hphantom{=\quad}\times\left\langle \prod_{r=1}^{N}e^{-ip_{r}^{+}X^{-}}(Z_{r},\bar{Z}_{r})\prod_{s=1}^{M}e^{-ip_{s}^{-}X^{+}}(w_{s},\bar{w}_{s})\right\rangle _{\hat{g}_{z\bar{z}}}^{X^{\pm}}\,,\label{eq:XpmT12}
\end{eqnarray}
where 
\begin{eqnarray}
\left\langle T^{X^{\pm}}(z)\right\rangle _{\rho} & = & -\frac{i}{2}\partial\rho(z)\left\langle \partial X^{-}(z)\right\rangle _{\rho}-\frac{d-26}{12}\{\rho,z\}\nonumber \\
 &  & {}-\frac{1}{3}\left.\frac{\partial^{3}}{\partial z^{\prime3}}E(z',z)\right|_{z'=z}-\pi\omega(z)\frac{1}{\mathop{\mathrm{Im}}\Omega}\omega(z)~.\label{eq:vevTXpm}
\end{eqnarray}

Since (\ref{eq:XpmT1}) can be considered as the generating function
of correlation functions with one $T^{X^{\pm}}\left(z\right)$ insertion,
we can deduce various properties of $T^{X^{\pm}}$ from it. The first
thing we check is whether $T^{X^{\pm}}\left(z\right)$ is singular
at $z=z_{I}$. Even if there are no operator insertions at $z=z_{I}$,
(\ref{eq:dX-loop}) implies that $\partial X^{-}\left(z\right)$ is
singular there. It will be disastrous for BRST quantization of the
worldsheet theory, if $T^{X^{\pm}}\left(z\right)$ is singular at
such points. For $z\sim z_{I}$, one finds%
\footnote{On the right hand side of the first equation in (\ref{eq:dX-zI}),
the sum $\sum_{r}$ is over the $r$'s such that $z_{I}=z_{I^{\left(r\right)}}$.%
} 
\begin{eqnarray}
\left\langle \partial X^{-}(z)\right\rangle _{\rho} & \sim & -\frac{d-26}{24}2i\left[\frac{1}{(z-z_{I})^{3}}\frac{3}{\partial^{2}\rho(z_{I})}-\frac{1}{(z-z_{I})^{2}}\frac{1}{2}\frac{\partial^{3}\rho(z_{I})}{(\partial^{2}\rho(z_{I}))^{2}}+\sum_{r}\frac{1}{\alpha_{r}}\frac{1}{z-z_{I^{\left(r\right)}}}\right]\,,\nonumber \\
\partial\rho\left(z\right) & \sim & (z-z_{I})\partial^{2}\rho(z_{I})+\frac{1}{2}(z-z_{I})^{2}\partial^{3}\rho(z_{I})+\mathcal{O}\left((z-z_{I})^{3}\right)\,,\label{eq:dX-zI}
\end{eqnarray}
and 
\begin{equation}
-\frac{i}{2}\partial\rho(z)\left\langle \partial X^{-}(z)\right\rangle _{\rho}\sim-\frac{d-26}{24}\left[\frac{3}{(z-z_{I})^{3}}+\frac{1}{z-z_{I}}\frac{\partial^{3}\rho(z_{I})}{\partial^{2}\rho(z_{I})}+\mathcal{O}(1)\right]\,.
\end{equation}
With (\ref{eq:SchwarzzIloop}), (\ref{eq:XpmT12}) and (\ref{eq:vevTXpm}),
this implies that $T^{X^{\pm}}(z)$ is not singular at $z=z_{I}$.

The singular behaviors at $z=Z_{r}$ are obtained as 
\begin{eqnarray}
\left\langle \partial X^{-}(z)\right\rangle _{\rho} & \sim & \frac{d-26}{24}2i\nonumber \\
 &  & \times\left[\frac{\frac{1}{\alpha_{r}}}{z-Z_{r}}+\lim_{z\to Z_{r}}\left(\partial B_{r}\left(z\right)-\frac{\frac{1}{\alpha_{r}}}{z-Z_{r}}\right)+\sum_{s\ne r}\partial B_{s}\left(Z_{r}\right)+\sum_{I}\partial B_{I}\left(Z_{r}\right)\right]\,,\nonumber \\
\partial\rho\left(z\right) & \sim & \frac{\alpha_{r}}{z-Z_{r}}+\lim_{z\to Z_{r}}\left(\partial\rho\left(z\right)-\frac{\alpha_{r}}{z-Z_{r}}\right)\,,
\end{eqnarray}
and 
\begin{eqnarray}
\lefteqn{-\frac{i}{2}\partial\rho(z)\left\langle \partial X^{-}(z)\right\rangle _{\rho}}\nonumber \\
 & \sim & \frac{d-26}{24}\left[\frac{1}{\left(z-Z_{r}\right)^{2}}\right.\nonumber \\
 &  & \qquad{}+\frac{1}{z-Z_{r}}\left\{ \frac{1}{\alpha_{r}}\lim_{z\to Z_{r}}\left(\partial\rho\left(z\right)-\frac{\alpha_{r}}{z-Z_{r}}\right)\right.\nonumber \\
 &  & \hphantom{\quad+\frac{1}{z-Z_{r}}\quad}+\left.\left.\alpha_{r}\left(\lim_{z\to Z_{r}}\left(\partial B_{r}\left(z\right)-\frac{\frac{1}{\alpha_{r}}}{z-Z_{r}}\right)+\sum_{s\ne r}\partial B_{s}\left(z_{r}\right)+\sum_{I}\partial B_{I}\left(Z_{r}\right)\right)\right\} \right]\nonumber \\
 & \sim & \frac{d-26}{24}\left[\frac{1}{\left(z-Z_{r}\right)^{2}}+\frac{1}{z-Z_{r}}\frac{\partial}{\partial Z_{r}}\left(-2\sum_{s}\mathop{\mathrm{Re}}\bar{N}_{00}^{rr}-\frac{3}{2}\sum_{I}\ln\left|\partial\rho\left(z_{I}\right)\right|^{2}\right)\right.\nonumber \\
 &  & \hphantom{\frac{d-26}{24}\qquad}\left.+\frac{1}{z-Z_{r}}\sum_{I}\left(\frac{\partial z_{I}}{\partial Z_{r}}\frac{\partial\left(-W\right)}{\partial z_{I}}+\frac{\partial\bar{z}_{I}}{\partial Z_{r}}\frac{\partial\left(-W\right)}{\partial\bar{z}_{I}}\right)\right]\,.\label{eq:dX+dX-zZrloop}
\end{eqnarray}
In order to get (\ref{eq:dX+dX-zZrloop}), we have used 
\begin{eqnarray}
\lefteqn{\lim_{z\to Z_{r}}\left(\partial B_{r}\left(z\right)-\frac{\frac{1}{\alpha_{r}}}{z-Z_{r}}\right)+\sum_{s\ne r}\partial B_{s}\left(z_{r}\right)+\sum_{I}\partial B_{I}\left(Z_{r}\right)}\nonumber \\
 &  & =-\frac{\pi}{\alpha_{r}}\int_{P_{0}}^{Z_{r}}\omega\frac{1}{\mathop{\mathrm{Im}}\Omega}\omega\left(Z_{r}\right)-\frac{1}{\alpha_{r}}\frac{\partial}{\partial Z_{r}}g\left(z_{I^{\left(r\right)}},Z_{r}\right)+\sum_{s\ne r}\frac{1}{\alpha_{s}}\frac{\partial}{\partial Z_{r}}\left(g\left(Z_{s},Z_{r}\right)-g\left(z_{I^{\left(s\right)}},Z_{r}\right)\right)\nonumber \\
 &  & \hphantom{-}+\sum_{s}\frac{1}{\alpha_{s}}\frac{\partial}{\partial Z_{r}}\left(\bar{g}\left(Z_{s},Z_{r}\right)-\bar{g}\left(z_{I^{\left(s\right)}},Z_{r}\right)\right)+\sum_{I}\frac{\partial}{\partial Z_{r}}B_{I}\left(Z_{r}\right)\,,\nonumber \\
\lefteqn{\lim_{z\to Z_{r}}\left(\partial\rho\left(z\right)-\frac{\alpha_{r}}{z-Z_{r}}\right)=\sum_{s\ne r}\alpha_{s}\frac{\partial}{\partial Z_{r}}g\left(Z_{r},Z_{s}\right)-2\pi i\alpha_{r}\omega\left(Z_{r}\right)\frac{1}{\mathop{\mathrm{Im}}\Omega}\mathop{\mathrm{Im}}\int_{P_{0}}^{Z_{r}}\omega\,,}
\end{eqnarray}
and 
\begin{eqnarray}
\frac{\partial z_{I}}{\partial Z_{r}} & = & -\frac{\alpha_{r}}{\partial^{2}\rho\left(z_{I}\right)}\partial_{z_{I}}\partial_{Z_{r}}g\left(z_{I},Z_{r}\right)\,,\nonumber \\
\frac{\partial\bar{z}_{I}}{\partial Z_{r}} & = & -\frac{\alpha_{r}}{\partial^{2}\bar{\rho}\left(\bar{z}_{I}\right)}\partial_{\bar{z}_{I}}\partial_{Z_{r}}\bar{g}\left(\bar{z}_{I},Z_{r}\right)\,,\nonumber \\
\frac{\partial\bar{N}_{00}^{rr}}{\partial Z_{r}} & = & \frac{\partial}{\partial Z_{r}}g\left(z_{I^{\left(r\right)}},Z_{r}\right)-\sum_{s\ne r}\frac{\alpha_{s}}{\alpha_{r}}\frac{\partial}{\partial Z_{r}}g\left(Z_{r},Z_{s}\right)\nonumber \\
 &  & \quad+2\pi i\omega\left(Z_{r}\right)\frac{1}{\mathop{\mathrm{Im}}\Omega}\mathop{\mathrm{Im}}\int_{P_{0}}^{Z_{r}}\omega+\pi\int_{P_{0}}^{Z_{r}}\omega\frac{1}{\mathop{\mathrm{Im}}\Omega}\omega\left(Z_{r}\right)\,,\nonumber \\
\frac{\partial\bar{N}_{00}^{ss}}{\partial Z_{r}} & = & \frac{\alpha_{r}}{\alpha_{s}}\frac{\partial}{\partial Z_{r}}\left(g\left(z_{I^{\left(s\right)}},Z_{r}\right)-g\left(Z_{s},Z_{r}\right)\right)\,\left(s\ne r\right)\,,\nonumber \\
\frac{\partial\bar{N}_{00}^{ss*}}{\partial Z_{r}} & = & \frac{\alpha_{r}}{\alpha_{s}}\frac{\partial}{\partial Z_{r}}\left(\bar{g}\left(\bar{z}_{I^{\left(s\right)}},Z_{r}\right)-\bar{g}\left(\bar{Z_{s}},Z_{r}\right)\right)\,.
\end{eqnarray}
Now with 
\begin{equation}
-2\{\rho,z\}\sim-\frac{1}{(z-Z_{r})^{2}}+\frac{1}{z-Z_{r}}\frac{\partial(-W)}{\partial Z_{r}}\,,
\end{equation}
it is easy to derive 
\begin{equation}
\left\langle T^{X^{\pm}}(z)\right\rangle _{\rho}\sim\frac{1}{z-Z_{r}}\frac{d}{dZ_{r}}\left(-\frac{d-26}{24}\Gamma\right)\,.\label{eq:TXpmZr}
\end{equation}
(\ref{eq:TXpmZr}) implies 
\begin{equation}
T^{X^{\pm}}(z)e^{-ip_{r}^{+}X^{-}}(Z_{r},\bar{Z}_{r})\sim\frac{1}{z-Z_{r}}\frac{d}{dZ_{r}}e^{-ip_{r}^{+}X^{-}}(Z_{r},\bar{Z}_{r})\,.
\end{equation}
In the same way, the following OPE can be derived: 
\begin{equation}
T^{X^{\pm}}(z)e^{-ip_{r}^{+}X^{-}-ip_{r}^{-}X^{+}}(Z_{r},\bar{Z}_{r})\sim\left[\frac{-p_{r}^{+}p_{r}^{-}}{\left(z-Z_{r}\right)^{2}}+\frac{1}{z-Z_{r}}\frac{d}{dZ_{r}}\right]e^{-ip_{r}^{+}X^{-}-ip_{r}^{-}X^{+}}(Z_{r}.\bar{Z}_{r})\,.
\end{equation}

Finally we examine if the energy-momentum tensor $T^{X^{\pm}}\left(z\right)$
satisfies the Virasoro algebra. In order to do so, we first derive
the OPE of $\partial X^{-}\left(z\right)\partial X^{-}\left(z^{\prime}\right)$.
From the correlation function 
\begin{eqnarray}
\lefteqn{\left\langle \partial X^{-}\left(z\right)\partial X^{-}\left(Z_{0}\right)\right\rangle _{\rho}^{\mathrm{c}}=\frac{d-26}{24}\left(2i\right)^{2}\partial_{z}\partial_{Z_{0}}\partial_{\alpha_{0}}\left[\sum_{r=1}^{N}B_{r}^{\prime}\left(z\right)+\sum_{I=1}^{2h-2+N}B_{I}^{\prime}\left(z\right)\right]_{\alpha_{0}=0}}\nonumber \\
 &  & \sim\frac{d-26}{24}\left(2i\right)^{2}\partial_{z}\partial_{Z_{0}}\left[-\frac{2}{\left(\partial\rho\left(Z_{0}\right)\right)^{2}}\frac{1}{\left(z-Z_{0}\right)^{2}}-\frac{2\partial^{3}\rho}{\left(\partial\rho\right)^{3}}\left(Z_{0}\right)\frac{1}{z-Z_{0}}\right],
\end{eqnarray}
we get the OPE 
\begin{eqnarray}
\partial X^{-}(z)\partial X^{-}(z') & \sim & -\frac{d-26}{12}\partial_{z}\partial_{z'}\left[\frac{1}{(z-z')^{2}}\frac{1}{\partial X^{+}(z)\partial X^{+}(z')}\right]\nonumber \\
 & \sim & -\frac{d-26}{12}\left[-\frac{1}{(z-z')^{4}}\frac{6}{(\partial X^{+}(z'))^{2}}-\frac{1}{(z-z')^{3}}3\partial_{z'}\frac{1}{(\partial X^{+}(z'))^{2}}\right.\nonumber \\
 &  & \hphantom{-\frac{d-26}{12}\left[\right]\ }\left.{}-\frac{1}{(z-z')^{2}}\frac{1}{2}\partial_{z'}^{2}\frac{1}{(\partial X^{+}(z'))^{2}}\right]~.
\end{eqnarray}
With 
\begin{eqnarray}
\partial X^{+}\left(z\right)\partial X^{-}\left(z^{\prime}\right) & \sim & \frac{1}{\left(z-z^{\prime}\right)^{2}}\,,\nonumber \\
\partial X^{+}\left(z\right)\partial X^{+}\left(z^{\prime}\right) & \sim & \mbox{regular}\,,
\end{eqnarray}
one can deduce the OPE 
\begin{equation}
T^{X^{\pm}}\left(z\right)T^{X^{\pm}}\left(z^{\prime}\right)\sim\frac{\frac{28-d}{2}}{\left(z-z^{\prime}\right)^{4}}+\frac{1}{\left(z-z^{\prime}\right)^{2}}2T^{\pm}\left(z^{\prime}\right)+\frac{1}{z-z^{\prime}}\partial T^{X^{\pm}}\left(z^{\prime}\right)\,,
\end{equation}
which coincides with the Virasoro algebra with central charge $28-d$
as desired.

\section{Interaction points and the odd moduli\label{sec:Interaction-points-and}}

In this appendix, we present the basic facts about the interaction
points and the odd moduli in the light-cone supersheet formalism \cite{Berkovits:1985ji,Berkovits:1987gp,Aoki:1990yn,Mandelstam:1991tw}.

For the supersheet coordinate 
\begin{equation}
\rho\left(\mathbf{z}\right)=\rho_{b}\left(z\right)+\theta f\left(z\right)\ ,
\end{equation}
we define $\tilde{\mathbf{z}}_{I}$ such that $\partial\rho\left(\tilde{\mathbf{z}}_{I}\right)=\partial D\rho\left(\tilde{\mathbf{z}}_{I}\right)=0$.
Since 
\begin{eqnarray}
\partial\rho & = & \partial\rho_{b}+\theta\partial f\ ,\nonumber \\
\partial D\rho & = & \partial f+\theta\partial^{2}\rho_{b}\ ,
\end{eqnarray}
we find $\tilde{\mathbf{z}}_{I}=\left(\tilde{z}_{I},\tilde{\theta}_{I}\right)$
to be 
\begin{equation}
\tilde{z}_{I}=z_{I}^{\left(b\right)}\ ,\qquad\tilde{\theta}_{I}=-\frac{\partial f}{\partial^{2}\rho_{b}}\left(z_{I}^{\left(b\right)}\right)\ ,
\end{equation}
where $z_{I}^{\left(b\right)}$ is one of the interaction points for
$\rho_{b}\left(z\right)$, i.e. $\partial\rho_{b}\left(z_{I}^{\left(b\right)}\right)=0$.
The $\rho$ coordinate corresponding to $\tilde{\mathbf{z}}_{I}$
is given as 
\begin{equation}
\rho\left(\tilde{\mathbf{z}}_{I}\right)=\rho_{b}\left(z_{I}^{\left(b\right)}\right)-\frac{\partial ff}{\partial^{2}\rho_{b}}\left(z_{I}^{\left(b\right)}\right)\ .
\end{equation}

The interaction point $\mathbf{z}_{I}=\left(z_{I},\theta_{I}\right)$
which is superconformal covariant is the one such that for some Grassmann
odd $\tilde{\xi}_{I}$ 
\begin{equation}
\hat{\rho}\left(\mathbf{z}\right)=\rho\left(\mathbf{z}\right)-\rho\left(\tilde{\mathbf{z}}_{I}\right)-\left(\rho\left(\mathbf{z}\right)-\rho\left(\tilde{\mathbf{z}}_{I}\right)\right)^{-\frac{1}{4}}\xi\tilde{\xi}_{I}\ ,
\end{equation}
can be expanded as 
\begin{equation}
\hat{\rho}\left(\mathbf{z}\right)=\frac{1}{2}\partial^{2}\hat{\rho}\left(\mathbf{z}_{I}\right)\left(\mathbf{z}-\mathbf{z}_{I}\right)^{2}+\cdots\ ,
\end{equation}
around $\mathbf{z}\sim\mathbf{z}_{I}$. Here $\xi=\left(\partial\rho\right)^{-\frac{1}{2}}D\rho$.
Expanding $\rho\left(\mathbf{z}\right)$ as 
\begin{equation}
\rho\left(\mathbf{z}\right)=\rho\left(\tilde{\mathbf{z}}_{I}\right)+\tilde{\theta}\alpha+\frac{1}{2}c^{2}\tilde{z}^{2}+\tilde{z}^{2}\tilde{\theta}\beta+\tilde{z}^{2}a+\cdots\ ,
\end{equation}
where 
\begin{equation}
\tilde{z}=\mathbf{z}-\tilde{\mathbf{z}}_{I}\ ,\qquad\tilde{\theta}=\theta-\tilde{\theta}_{I}\ ,
\end{equation}
we find 
\begin{eqnarray}
\mathbf{z}_{I}-\tilde{\mathbf{z}}_{I} & = & -\frac{1}{c^{4}}\alpha\beta\ ,\nonumber \\
\theta_{I}-\tilde{\theta}_{I} & = & \frac{a}{c^{4}}\alpha\ ,\label{eq:boldzI}
\end{eqnarray}
and 
\begin{eqnarray}
\tilde{\xi}_{I} & = & 2^{-\frac{1}{4}}c^{-\frac{1}{2}}\alpha\nonumber \\
 & = & 2^{-\frac{1}{4}}\left(\partial^{2}\rho_{b}\right)^{-\frac{1}{4}}\left(f+\frac{f\partial f\partial^{2}f}{4\left(\partial^{2}\rho_{b}\right)^{2}}\right)\left(z_{I}^{\left(b\right)}\right)\ .
\end{eqnarray}
$\tilde{\xi}_{I}$'s are proportional to the odd supermoduli parameters
in the light-cone gauge parametrization. For our purpose, it is convenient
to define 
\begin{equation}
\xi_{I}\equiv2\left(\partial^{2}\rho_{b}\right)^{-\frac{1}{4}}\left(f+\frac{f\partial f\partial^{2}f}{4\left(\partial^{2}\rho_{b}\right)^{2}}\right)\left(z_{I}^{\left(b\right)}\right)\ .
\end{equation}
From (\ref{eq:boldzI}), we get 
\begin{equation}
\rho\left(\mathbf{z}_{I}\right)=\rho\left(\tilde{\mathbf{z}}_{I}\right)+\left(\theta_{I}-\tilde{\theta}_{I}\right)D\rho\left(\tilde{\mathbf{z}}_{I}\right)\ ,
\end{equation}
but since $\alpha=D\rho\left(\tilde{\mathbf{z}}_{I}\right)$, we find
\begin{equation}
\rho\left(\mathbf{z}_{I}\right)=\rho\left(\tilde{\mathbf{z}}_{I}\right)\ .
\end{equation}

\section{Calculations of correlation functions of supersymmetric $X^{\pm}$
CFT\label{sec:Calculations-of-the}}

In this appendix, we present some details of the calculation of the
correlation functions of supersymmetric $X^{\pm}$ CFT.

$\left\langle D\mathcal{X}^{-}\left(\mathbf{z}\right)\right\rangle _{\mathbf{\rho}}$
in (\ref{eq:DX-rho}) can be expressed as 
\begin{eqnarray}
\left\langle D\mathcal{X}^{-}\left(\mathbf{z}\right)\right\rangle _{\mathbf{\rho}} & = & \frac{d-10}{8}2iD\left[\frac{1}{2}\left(\sum_{r}B_{r}\left(z,\bar{z}\right)+\sum_{I}B_{I}\left(z,\bar{z}\right)\right)\right.\nonumber \\
 &  & \hphantom{\frac{\hat{c}}{8}2iD\left[\right.}\left.{}+N\left(\mathbf{z},\bar{\mathbf{z}}\right)+\sum_{r}N_{r}\left(\mathbf{z},\bar{\mathbf{z}}\right)+\sum_{I}N_{I}\left(\mathbf{z},\bar{\mathbf{z}}\right)\right]\,,\label{eq:DX-}
\end{eqnarray}
where 
\begin{eqnarray}
 &  & \sum_{r}B_{r}\left(Z_{0},\bar{Z}_{0}\right)+\sum_{I}B_{I}\left(Z_{0},\bar{Z}_{0}\right)-\sum_{r}B_{r}\left(Z_{N+1},\bar{Z}_{N+1}\right)-\sum_{I}B_{I}\left(Z_{N+1},\bar{Z}_{N+1}\right)\nonumber \\
 &  & \qquad=\left.\partial_{\alpha_{0}}\left(-\Gamma^{\prime}\right)\right|_{\alpha_{0}=0}\,,\nonumber \\
 &  & N\left(\mathbf{Z}_{0},\bar{\mathbf{Z}}_{0}\right)=\left.\partial_{\alpha_{0}}\left(-\delta\Gamma_{0}^{\prime}-\delta\Gamma_{I^{\left(0\right)}}^{\prime}\right)\right|_{\alpha_{0}=0}\,,\nonumber \\
 &  & N_{r}\left(\mathbf{Z}_{0},\bar{\mathbf{Z}}_{0}\right)-N_{r}\left(\mathbf{Z}_{N+1},\bar{\mathbf{Z}}_{N+1}\right)=\left.\partial_{\alpha_{0}}\left(-\delta\Gamma_{r}^{\prime}\right)\right|_{\alpha_{0}=0}\,,
\nonumber \\
 &  & N_{I}\left(\mathbf{Z}_{0},\bar{\mathbf{Z}}_{0}\right)-N_{I}\left(\mathbf{Z}_{N+1},\bar{\mathbf{Z}}_{N+1}\right)=\left.\partial_{\alpha_{0}}\left(-\delta\Gamma_{I}^{\prime}\right)\right|_{\alpha_{0}=0}\,.\label{eq:NI}
\end{eqnarray}

It is straightforward to calculate these terms. $B_{r}\left(z,\bar{z}\right)$
and $B_{I}\left(z,\bar{z}\right)$ essentially coincide with the bosonic
ones 
\begin{eqnarray}
B_{r}\left(z,\bar{z}\right) & = & \frac{1}{\alpha_{r}}\left(g\left(Z_{r},z\right)-g\left(z_{I^{\left(r\right)}},z\right)\right)+\mathrm{c.c.}\,,\nonumber \\
B_{I}\left(z,\bar{z}\right) & = & \frac{\partial^{3}\rho_{b}}{2\left(\partial^{2}\rho_{b}\right)^{2}}\left(z_{I}\right)\frac{\partial}{\partial z_{I}}g\left(z_{I},z\right)-\frac{3}{2\partial^{2}\rho_{b}\left(z_{I}\right)}\frac{\partial^{2}}{\partial z_{I}^{2}}g\left(z_{I},z\right)+\mathrm{c.c.}\,.\nonumber \\
g\left(z,w\right) & = & \ln E(z,w)-2\pi i\int_{P_{0}}^{z}\omega\frac{1}{\mathop{\mathrm{Im}}\Omega}\mathop{\mathrm{Im}}\int_{P_{0}}^{w}\omega\,.
\end{eqnarray}
In order to evaluate $N\left(\mathbf{z},\bar{\mathbf{z}}\right)$,
we need $z_{I^{\left(0\right)}}^{\prime\left(b\right)}-Z_{0}$ up
to $\mathcal{O}\left(\alpha_{0}^{2}\right)$, which is given by the
second equation in (\ref{eq:zI0prime-Z0}). Using this, it is straightforward
to get 
\begin{eqnarray}
N\left(\mathbf{z},\bar{\mathbf{z}}\right) & = & -\frac{2\partial^{2}ff}{\left(\partial\rho_{b}\right)^{3}}\left(z\right)+\frac{4\partial^{2}\rho_{b}\partial ff}{\left(\partial\rho_{b}\right)^{4}}\left(z\right)\nonumber \\
 &  & \quad+\theta\left(-\frac{\partial^{2}f\partial ff}{2\left(\partial\rho_{b}\right)^{4}}+\frac{2\partial^{2}f}{\left(\partial\rho_{b}\right)^{2}}-\frac{4\partial^{2}\rho_{b}\partial f}{\left(\partial\rho_{b}\right)^{3}}+\frac{5\left(\partial^{2}\rho_{b}\right)^{2}f}{2\left(\partial\rho_{b}\right)^{4}}-\frac{\partial^{3}\rho_{b}f}{\left(\partial\rho_{b}\right)^{3}}\right)\left(z\right)\nonumber \\
 &  & +\mathrm{c.c.}\,.\label{eq:Nboldz}
\end{eqnarray}
This expression does not involve $Z_{N+1}$ or $\Theta_{N+1}$. In
the same way, one can show that $\left.\partial_{\alpha_{0}}\left(-\delta\Gamma_{N+1}^{\prime}-\delta\Gamma_{I^{\left(N+1\right)}}^{\prime}\right)\right|_{\alpha_{0}=0}$
does not involve $Z_{0}$ or $\Theta_{0}$ and we do not have to include
the contributions from this term in (\ref{eq:DX-}). $N_{r}\left(\mathbf{z},\bar{\mathbf{z}}\right)$
and $N_{I}\left(\mathbf{z},\bar{\mathbf{z}}\right)$ can be evaluated
by simply calculating the right hand sides of 
in (\ref{eq:NI}). The results are in the form on the left hand sides
of these equations and we obtain 
\begin{eqnarray}
\lefteqn{N_{r}\left(\mathbf{z},\bar{\mathbf{z}}\right)}\nonumber \\
 & = & \frac{1}{2\alpha_{r}}\left[\left\{ -\frac{\partial^{2}ff}{\left(\partial^{2}\rho_{b}\right)^{2}}+\frac{\partial^{3}\rho_{b}\partial ff}{\left(\partial^{2}\rho_{b}\right)^{3}}\right\} \left(z_{I^{\left(r\right)}}^{\left(b\right)}\right)\partial_{z_{I^{\left(r\right)}}^{\left(b\right)}}g\left(z_{I^{\left(r\right)}}^{\left(b\right)},z\right)-\frac{\partial ff}{\left(\partial^{2}\rho_{b}\right)^{2}}\partial_{z_{I^{\left(r\right)}}^{\left(b\right)}}^{2}g\left(z_{I^{\left(r\right)}}^{\left(b\right)},z\right)\right.\nonumber \\
 &  & \hphantom{\frac{1}{2\alpha_{r}}\quad}\left.+\theta\left\{ \frac{\partial f}{\partial^{2}\rho_{b}}\left(z_{I^{\left(r\right)}}^{\left(b\right)}\right)S_{\delta}\left(z_{I^{\left(r\right)}}^{\left(b\right)},z\right)-\frac{f}{\partial^{2}\rho_{b}}\left(z_{I^{\left(r\right)}}^{\left(b\right)}\right)\partial_{z_{I^{\left(r\right)}}^{\left(b\right)}}S_{\delta}\left(z_{I^{\left(r\right)}}^{\left(b\right)},z\right)\right\} \right]\nonumber \\
 &  & {}+\mathrm{c.c.}\,,
\nonumber \\
\lefteqn{N_{I}\left(\mathbf{z},\bar{\mathbf{z}}\right)}\nonumber \\
 & = & \left\{ \frac{\partial^{4}f\partial^{2}f\partial ff}{12\left(\partial^{2}\rho_{b}\right)^{5}}-\frac{2\left(\partial^{4}ff+\partial^{3}f\partial f\right)}{3\left(\partial^{2}\rho_{b}\right)^{3}}-\frac{\partial^{3}\rho_{b}\partial^{3}f\partial^{2}f\partial ff}{3\left(\partial^{2}\rho_{b}\right)^{6}}\right.\nonumber \\
 &  & \ {}+\frac{\partial^{3}\rho_{b}\left(\frac{7}{3}\partial^{3}ff+\partial^{2}f\partial f\right)}{\left(\partial^{2}\rho_{b}\right)^{4}}-\frac{15\left(\partial^{3}\rho_{b}\right)^{2}\partial^{2}ff}{4\left(\partial^{2}\rho_{b}\right)^{5}}\nonumber \\
 &  & \ \left.{}+\frac{17\partial^{4}\rho_{b}\partial^{2}ff}{12\left(\partial^{2}\rho_{b}\right)^{4}}+\left(\frac{3\left(\partial^{3}\rho_{b}\right)^{3}}{\left(\partial^{2}\rho_{b}\right)^{6}}-\frac{11\partial^{4}\rho_{b}\partial^{3}\rho_{b}}{4\left(\partial^{2}\rho_{b}\right)^{5}}+\frac{5\partial^{5}\rho_{b}}{12\left(\partial^{2}\rho_{b}\right)^{4}}\right)\partial ff\right\} \left(z_{I}^{\left(b\right)}\right)\partial_{z_{I}^{\left(b\right)}}g\left(z_{I}^{\left(b\right)},z\right)\nonumber \\
 &  & {}+\left\{ \frac{\partial^{3}f\partial^{2}f\partial ff}{3\left(\partial^{2}\rho_{b}\right)^{5}}-\frac{4\partial^{3}ff}{3\left(\partial^{2}\rho_{b}\right)^{3}}\right.\nonumber \\
 &  & \hphantom{+\ }\left.{}+\frac{3\partial^{3}\rho_{b}\partial^{2}ff}{\left(\partial^{2}\rho_{b}\right)^{4}}-\frac{3\left(\partial^{3}\rho_{b}\right)^{2}\partial ff}{\left(\partial^{2}\rho_{b}\right)^{5}}+\frac{5\partial^{4}\rho_{b}\partial ff}{4\left(\partial^{2}\rho_{b}\right)^{4}}\right\} \left(z_{I}^{\left(b\right)}\right)\partial_{z_{I}^{\left(b\right)}}^{2}g\left(z_{I}^{\left(b\right)},z\right)\nonumber \\
 &  & {}+\left\{ -\frac{\partial^{2}ff}{\left(\partial^{2}\rho_{b}\right)^{3}}+\frac{3\partial^{3}\rho_{b}\partial ff}{2\left(\partial^{2}\rho_{b}\right)^{4}}\right\} \left(z_{I}^{\left(b\right)}\right)\partial_{z_{I}^{\left(b\right)}}^{3}g\left(z_{I}^{\left(b\right)},z\right)-\frac{5\partial ff}{12\left(\partial^{2}\rho_{b}\right)^{3}}\left(z_{I}^{\left(b\right)}\right)\partial_{z_{I}^{\left(b\right)}}^{4}g\left(z_{I}^{\left(b\right)},z\right)\nonumber \\
 &  & {}+\theta\left[\left\{ -\frac{\partial^{3}f\partial^{2}f\partial f}{12\left(\partial^{2}\rho_{b}\right)^{4}}+\frac{2\partial^{3}f}{3\left(\partial^{2}\rho_{b}\right)^{2}}-\frac{\partial^{3}\rho_{b}\partial^{2}f}{\left(\partial^{2}\rho_{b}\right)^{3}}+\frac{3\left(\partial^{3}\rho_{b}\right)^{2}\partial f}{4\left(\partial^{2}\rho_{b}\right)^{4}}-\frac{5\partial^{4}\rho_{b}\partial f}{12\left(\partial^{2}\rho_{b}\right)^{3}}\right\} \left(z_{I}^{\left(b\right)}\right)S_{\delta}\left(z_{I}^{\left(b\right)},z\right)\right.\nonumber \\
 &  & \hphantom{\quad+\theta\quad}+\left\{ \frac{\partial^{3}f\partial^{2}ff}{12\left(\partial^{2}\rho_{b}\right)^{4}}-\frac{3\left(\partial^{3}\rho_{b}\right)^{2}f}{4\left(\partial^{2}\rho_{b}\right)^{4}}+\frac{5\partial^{4}\rho_{b}f}{12\left(\partial^{2}\rho_{b}\right)^{3}}\right\} \left(z_{I}^{\left(b\right)}\right)\partial_{z_{I}^{\left(b\right)}}S_{\delta}\left(z_{I}^{\left(b\right)},z\right)\nonumber \\
 &  & \hphantom{\quad+\theta\quad}+\left\{ -\frac{\partial^{3}f\partial ff}{12\left(\partial^{2}\rho_{b}\right)^{4}}+\frac{\partial^{3}\rho_{b}f}{\left(\partial^{2}\rho_{b}\right)^{3}}\right\} \left(z_{I}^{\left(b\right)}\right)\partial_{z_{I}^{\left(b\right)}}^{2}S_{\delta}\left(z_{I}^{\left(b\right)},z\right)\nonumber \\
 &  & \hphantom{\quad+\theta\quad}+\left.\left\{ \frac{\partial^{2}f\partial ff}{12\left(\partial^{2}\rho_{b}\right)^{4}}-\frac{2f}{3\left(\partial^{2}\rho_{b}\right)^{2}}\right\} \left(z_{I}^{\left(b\right)}\right)\partial_{z_{I}^{\left(b\right)}}^{3}S_{\delta}\left(z_{I}^{\left(b\right)},z\right)\right]\nonumber \\
 &  & {}+\mathrm{c.c.}\,.\label{eq:NIboldz}
\end{eqnarray}

A few comments are in order: 
\begin{itemize}
\item $\left\langle D\mathcal{X}^{-}\left(\mathbf{z}\right)\right\rangle _{\mathbf{\rho}}$
given in (\ref{eq:DX-}) is meromorphic with respect to $\mathbf{z}$. 
\item $\mathcal{X}^{\pm}$ involves auxiliary field $F^{\pm}$ as its component.
With the action (\ref{eq:superSXpm}), the equation of motion implies
$F^{+}=0$. $F^{+}$ can be nonzero at $z=Z_{r}$ where $\mathcal{X}^{-}$
is inserted. (\ref{eq:DX-}) implies that the expectation value of
$F^{-}$ vanishes even with the sources. 
\end{itemize}

\subsection{An expression in terms of the superfield $\rho\left(\mathbf{z}\right)$}

It is possible to rewrite $\left\langle D\mathcal{X}^{-}\left(\mathbf{z}\right)\right\rangle _{\mathbf{\rho}}$
given in (\ref{eq:DX-}) in terms of the superfield $\rho\left(\mathbf{z}\right)$,
the covariant derivative and the fermionic coordinate $\theta$ as
\begin{equation}
\left\langle D\mathcal{X}^{-}\left(\mathbf{z}\right)\right\rangle _{\mathbf{\rho}}=\frac{d-10}{8}2iD\mathop{\mathrm{Re}}\left[\sum_{r}\left(\frac{1}{\alpha_{r}}g(Z_{r},z)+\mathcal{K}_{1}(\tilde{\mathbf{z}}_{I^{(r)}};\mathbf{z})\right)+\mathcal{K}_{2}(\mathbf{z})+\sum_{I}\mathcal{K}_{3}(\tilde{\mathbf{z}}_{I};\mathbf{z})\right]\,,\label{eq:DX-2}
\end{equation}
where 
\begin{eqnarray}
 &  & \mathcal{K}_{1}\left(\tilde{\mathbf{z}}_{I^{(r)}};\mathbf{z}\right)\nonumber \\
 &  & \quad=\frac{1}{\alpha_{r}}\left[-g\left(\tilde{\mathbf{z}}_{I^{(r)}},\mathbf{z}\right)-\partial g\left(\tilde{\mathbf{z}}_{I^{(r)}},\mathbf{z}\right)\frac{\partial^{2}D\rho D\rho}{(\partial^{2}\rho)}\left(\tilde{\mathbf{z}}_{I^{(r)}}\right)+\partial D\left(g\left(\tilde{\mathbf{z}}_{I^{(r)}},\mathbf{z}\right)\right)\frac{D\rho}{\partial^{2}\rho}\left(\tilde{\mathbf{z}}_{I^{(r)}}\right)\right]\,,\nonumber \\
 &  & \mathcal{K}_{2}(\mathbf{z})=\left(8\frac{\partial^{2}\rho\partial D\rho D\rho}{(\partial\rho)^{4}}-4\frac{\partial^{2}D\rho D\rho}{(\partial\rho)^{3}}\right)(\mathbf{z})\nonumber \\
 &  & \hphantom{\mathcal{K}_{2}(\mathbf{z})=}{}+\theta\left(-3\frac{(\partial^{2}\rho)^{2}}{(\partial\rho)^{4}}+2\frac{\partial^{3}\rho}{(\partial\rho)^{3}}+3\frac{\partial^{2}D\rho\partial D\rho}{(\partial\rho)^{4}}\right)D\rho(\mathbf{z})~,\nonumber \\
 &  & \mathcal{K}_{3}\left(\mathbf{z}_{I};\mathbf{z}\right)\nonumber \\
 &  & \ =D\left(g\left(\tilde{\mathbf{z}}_{I},\mathbf{z}\right)\right)\left(2\frac{\partial^{3}\rho\partial^{2}D\rho}{(\partial^{2}\rho)^{3}}-\frac{4}{3}\frac{\partial^{3}D\rho}{(\partial^{2}\rho)^{2}}\right)\left(\tilde{\mathbf{z}}_{I}\right)\nonumber \\
 &  & \quad{}+\partial\left(g\left(\tilde{\mathbf{z}}_{I},\mathbf{z}\right)\right)\left\{ \frac{14}{3}\frac{\partial^{3}D\rho\partial^{3}\rho D\rho}{(\partial^{2}\rho)^{4}}-\frac{4}{3}\frac{\partial^{4}D\rho D\rho}{(\partial^{2}\rho)^{3}}\right.\nonumber \\
 &  & \hphantom{+\partial\left(g\left(\tilde{\mathbf{z}}_{I},\mathbf{z}\right)\right)\left(\right)\ }\left.{}+\left(-\frac{15}{2}\frac{(\partial^{3}\rho)^{2}}{(\partial^{2}\rho)^{5}}+\frac{17}{6}\frac{\partial^{4}\rho}{(\partial^{2}\rho)^{4}}\right)\partial^{2}D\rho D\rho+\frac{1}{2}\frac{\partial^{3}\rho}{(\partial^{2}\rho)^{2}}\right\} \left(\tilde{\mathbf{z}}_{I}\right)\nonumber \\
 &  & \quad{}+\partial D\left(g\left(\tilde{\mathbf{z}}_{I},\mathbf{z}\right)\right)\left(-\frac{1}{6}\frac{\partial^{3}D\rho\partial^{2}D\rho D\rho}{(\partial^{2}\rho)^{4}}+\frac{3}{2}\frac{(\partial^{3}\rho)^{2}D\rho}{(\partial^{2}\rho)^{4}}-\frac{5}{6}\frac{\partial^{4}\rho D\rho}{(\partial^{2}\rho)^{3}}\right)\left(\tilde{\mathbf{z}}_{I}\right)\nonumber \\
 &  & \quad{}+\partial^{2}\left(g\left(\tilde{\mathbf{z}}_{I},\mathbf{z}\right)\right)\left(6\frac{\partial^{3}\rho\partial^{2}D\rho D\rho}{(\partial^{2}\rho)^{4}}-\frac{8}{3}\frac{\partial^{3}D\rho D\rho}{(\partial^{2}\rho)^{3}}-\frac{3}{2}\frac{1}{\partial^{2}\rho}\right)\left(\tilde{\mathbf{z}}_{I}\right)\nonumber \\
 &  & \quad{}-2\partial^{2}D\left(g\left(\tilde{\mathbf{z}}_{I},\mathbf{z}\right)\right)\frac{\partial^{3}\rho D\rho}{(\partial^{2}\rho)^{3}}\left(\tilde{\mathbf{z}}_{I}\right)-2\partial^{3}\left(g\left(\tilde{\mathbf{z}}_{I},\mathbf{z}\right)\right)\frac{\partial^{2}D\rho D\rho}{(\partial^{2}\rho)^{3}}\left(\tilde{\mathbf{z}}_{I}\right)\nonumber \\
 &  & \quad{}+\frac{4}{3}\partial^{3}D\left(g\left(\tilde{\mathbf{z}}_{I},\mathbf{z}\right)\right)\frac{D\rho}{(\partial^{2}\rho)^{2}}\left(\tilde{\mathbf{z}}_{I}\right)\nonumber \\
 &  & \quad{}+\tilde{\theta}_{I}\left[D\left(g\left(\tilde{\mathbf{z}}_{I},\mathbf{z}\right)\right)\left(\frac{1}{2}\frac{(\partial^{3}\rho)^{2}}{(\partial^{2}\rho)^{3}}-\frac{1}{2}\frac{\partial^{4}\rho}{(\partial^{2}\rho)^{2}}-\frac{1}{2}\frac{\partial^{3}D\rho\partial^{2}D\rho}{(\partial^{2}\rho)^{3}}\right)\left(\tilde{\mathbf{z}}_{I}\right)\right.\nonumber \\
 &  & \quad\hphantom{+\left[\tilde{\theta}_{I}\right]}{}+\partial\left(g\left(\tilde{\mathbf{z}}_{I},\mathbf{z}\right)\right)\left\{ -3\frac{\partial^{3}\rho\partial^{3}D\rho\partial^{2}D\rho D\rho}{(\partial^{2}\rho)^{5}}+\frac{\partial^{4}D\rho\partial^{2}D\rho D\rho}{(\partial^{2}\rho)^{4}}\right.\nonumber \\
 &  & \qquad\hphantom{\left[+\tilde{\theta}_{I}\right]+\partial\left(g\left(\tilde{\mathbf{z}}_{I},\mathbf{z}\right)\right)\ }{}+\left(\frac{3}{2}\frac{(\partial^{3}\rho)^{3}}{(\partial^{2}\rho)^{5}}-2\frac{\partial^{4}\rho\partial^{3}\rho}{(\partial^{2}\rho)^{4}}+\frac{1}{2}\frac{\partial^{5}\rho}{(\partial^{2}\rho)^{3}}\right)D\rho\nonumber \\
 &  & \qquad\hphantom{\left[+\tilde{\theta}_{I}\right]+\partial\left(g\left(\tilde{\mathbf{z}}_{I},\mathbf{z}\right)\right)\ }\left.{}+\frac{\partial^{3}\rho\partial^{2}D\rho}{(\partial^{2}\rho)^{3}}-\frac{1}{2}\frac{\partial^{3}D\rho}{(\partial^{2}\rho)^{2}}\right\} \left(\tilde{\mathbf{z}}_{I}\right)\nonumber \\
 &  & \qquad\hphantom{\left[+\tilde{\theta}_{I}\right]}{}+\partial D\left(g\left(\tilde{\mathbf{z}}_{I},\mathbf{z}\right)\right)\left\{ -\frac{3}{2}\frac{\partial^{3}\rho\partial^{3}D\rho D\rho}{(\partial^{2}\rho)^{4}}{}+\frac{1}{2}\frac{\partial^{4}D\rho D\rho}{(\partial^{2}\rho)^{3}}\right.\nonumber \\
 &  & \qquad\hphantom{\left[+\tilde{\theta}_{I}\right]+\partial D\left(g\left(\tilde{\mathbf{z}}_{I},\mathbf{z}\right)\right)\ }\left.{}+\left(\frac{3}{2}\frac{(\partial^{3}\rho)^{2}}{(\partial^{2}\rho)^{5}}-\frac{1}{2}\frac{\partial^{4}\rho}{(\partial^{2}\rho)^{4}}\right)\partial^{2}D\rho D\rho-\frac{1}{2}\frac{\partial^{3}\rho}{(\partial^{2}\rho)^{2}}\right\} \left(\tilde{\mathbf{z}}_{I}\right)\nonumber \\
 &  & \qquad\hphantom{\left[+\tilde{\theta}_{I}\right]}{}+\partial^{2}\left(g\left(\tilde{\mathbf{z}}_{I},\mathbf{z}\right)\right)\left\{ \left(-\frac{3}{2}\frac{(\partial^{3}\rho)^{2}}{(\partial^{2}\rho)^{4}}+\frac{\partial^{4}\rho}{(\partial^{2}\rho)^{3}}\right)D\rho+\frac{3}{2}\frac{\partial^{3}D\rho\partial^{2}D\rho D\rho}{(\partial^{2}\rho)^{4}}-\frac{3}{2}\frac{\partial^{2}D\rho}{(\partial^{2}\rho)^{2}}\right\} \left(\tilde{\mathbf{z}}_{I}\right)\nonumber \\
 &  & \qquad\hphantom{\left[+\tilde{\theta}_{I}\right]}{}+\partial^{2}D\left(g\left(\tilde{\mathbf{z}}_{I},\mathbf{z}\right)\right)\left(\frac{1}{2}\frac{\partial^{3}D\rho D\rho}{(\partial^{2}\rho)^{3}}+\frac{3}{2}\frac{1}{\partial^{2}\rho}\right)\left(\tilde{\mathbf{z}}_{I}\right)\nonumber \\
 &  & \qquad\hphantom{\left[+\tilde{\theta}_{I}\right]}{}+\partial^{3}\left(g\left(\tilde{\mathbf{z}}_{I},\mathbf{z}\right)\right)\frac{\partial^{3}\rho D\rho}{(\partial^{2}\rho)^{3}}\left(\tilde{\mathbf{z}}_{I}\right)-\frac{1}{2}\partial^{3}D\left(g\left(\tilde{\mathbf{z}}_{I},\mathbf{z}\right)\right)\frac{\partial^{2}D\rho D\rho}{(\partial^{2}\rho)^{3}}\left(\tilde{\mathbf{z}}_{I}\right)\nonumber \\
 &  & \qquad\hphantom{\left[+\tilde{\theta}_{I}\right]}\left.{}-\frac{1}{2}\partial^{4}(g\left(\tilde{\mathbf{z}}_{I},\mathbf{z}\right))\frac{D\rho}{(\partial^{2}\rho)^{2}}\left(\tilde{\mathbf{z}}_{I}\right)\right]\,,
\end{eqnarray}
where the derivatives act on the first arguments of $g\left(\tilde{\mathbf{z}}_{I^{(r)}},\mathbf{z}\right)$
and $g\left(\tilde{\mathbf{z}}_{I},\mathbf{z}\right)$. This expression
shall be useful in doing calculations.

\subsection{Energy-momentum tensor}

Now we can discuss the properties of the energy-momentum tensor superfield
$T^{\mathcal{X}^{\pm}}(\mathbf{z})$ given in (\ref{eq:superT-Xpm}).
For example, the correlation functions with one $T^{\mathcal{X}^{\pm}}\left(\mathbf{z}\right)$
insertion can be given by 
\begin{eqnarray}
\lefteqn{\left\langle T^{\mathcal{X}^{\pm}}\left(\mathbf{z}\right)\prod_{r=1}^{N}e^{-ip_{r}^{+}\mathcal{X}^{-}}(\mathbf{Z}_{r},\bar{\mathbf{Z}}_{r})\prod_{s=1}^{M}e^{-ip_{s}^{-}\mathcal{X}^{+}}(\mathbf{w}_{s},\bar{\mathbf{w}}_{s})\right\rangle _{\hat{g}_{z\bar{z}}}^{\mathcal{X}^{\pm}}}\nonumber \\
 &  & \quad=\left[\frac{1}{2}\left(-\frac{i}{2}\partial\rho\left(\mathbf{z}\right)\right)\sum_{s=1}^{M}\left(-ip_{s}^{-}\right)\begC1{D\mathcal{X}^{-}}\conC{\left(\mathbf{z}\right)}\endC1{\mathcal{X}^{+}}\left(\mathbf{w}_{s}\right)\right.\nonumber \\
 &  & \hphantom{\quad=}\left.\qquad{}+\frac{1}{2}\left(-\frac{i}{2}D\rho\left(\mathbf{z}\right)\right)\sum_{s=1}^{M}\left(-ip_{s}^{-}\right)\begC1{\partial\mathcal{X}^{-}}\conC{\left(\mathbf{z}\right)}\endC1{\mathcal{X}^{+}}\left(\mathbf{w}_{s}\right)
+\left\langle T^{\mathcal{X}^{\pm}}\left(\mathbf{z}\right)\right\rangle _{\rho}\vphantom{\frac{1}{2}\left(-\frac{i}{2}\partial\rho\left(\mathbf{z}\right)\right)\sum_{s=1}^{M}\left(-ip_{s}^{-}\right)\begC1{DX^{-}}\conC{\left(\mathbf{z}\right)}\endC1{X^{+}}\left(\mathbf{w}_{s}\right)}\right]\nonumber \\
 &  & \hphantom{\quad=}\times\left\langle \prod_{r=1}^{N}e^{-ip_{r}^{+}\mathcal{X}^{-}}(\mathbf{Z}_{r},\bar{\mathbf{Z}}_{r})\prod_{s=1}^{M}e^{-ip_{s}^{-}\mathcal{X}^{+}}(\mathbf{w}_{s},\bar{\mathbf{w}}_{s})\right\rangle _{\hat{g}_{z\bar{z}}}^{X^{\pm}}\,,\label{eq:superXpmT1}
\end{eqnarray}
where 
\begin{eqnarray}
\left\langle T^{\mathcal{X}^{\pm}}\left(\mathbf{z}\right)\right\rangle _{\rho} & = & \frac{1}{2}\left(-\frac{i}{2}\partial\rho\left(\mathbf{z}\right)\left\langle D\mathcal{X}^{-}\left(\mathbf{z}\right)\right\rangle _{\rho}-\frac{i}{2}D\rho\left(\mathbf{z}\right)\left\langle \partial\mathcal{X}^{-}\left(\mathbf{z}\right)\right\rangle _{\rho}\right)-\frac{d-10}{4}S\left(\mathbf{z},\mbox{\mathversion{bold}\ensuremath{\rho}}\right)\nonumber \\
 &  & \quad{}+\frac{1}{2}\lim_{\mathbf{w}\to\mathbf{z}}\left(\begC1{\partial\mathcal{X}^{+}}\conC{\left(\mathbf{w}\right)}\endC1{D\mathcal{X}^{-}}\left(\mathbf{z}\right)-\partial_{\mathbf{w}}D_{\mathbf{z}}\ln\left(\mathbf{w}-\mathbf{z}\right)\right)\nonumber \\
 &  & \quad{}+\frac{1}{2}\lim_{\mathbf{w}\to\mathbf{z}}\left(\begC1{D\mathcal{X}^{+}}\conC{\left(\mathbf{w}\right)}\endC1{\partial\mathcal{X}^{-}}\left(\mathbf{z}\right)-D_{\mathbf{w}}\partial_{\mathbf{z}}\ln\left(\mathbf{w}-\mathbf{z}\right)\right)\,.
\end{eqnarray}

It is straightforward to calculate correlation functions of $T^{\mathcal{X}^{\pm}}\left(\mathbf{z}\right)$
and show the properties of the supersymmetric $X^{\pm}$ CFT mentioned
at the end of subsection \ref{sub:Correlation-functions-of}, using
the expression (\ref{eq:DX-}) or (\ref{eq:DX-2}). Since these calculations
are not so illuminating, we do not reproduce them here.

 \bibliographystyle{utphys}
\bibliography{SFTOct05_14}

\providecommand{\href}[2]{#2}\begingroup\raggedright\begin{thebibliography}{10}

\bibitem{Baba:2009ns}
Y.~Baba, N.~Ishibashi, and K.~Murakami, ``{Light-Cone Gauge String Field Theory
  in Noncritical Dimensions},''
  \href{http://dx.doi.org/10.1088/1126-6708/2009/12/010}{{\em JHEP} {\bf 12}
  (2009)  010},
\href{http://arxiv.org/abs/0909.4675}{{\tt arXiv:0909.4675 [hep-th]}}.

\bibitem{Baba:2009fi}
Y.~Baba, N.~Ishibashi, and K.~Murakami, ``{Light-cone Gauge NSR Strings in
  Noncritical Dimensions},''
  \href{http://dx.doi.org/10.1007/JHEP01(2010)119}{{\em JHEP} {\bf 01} (2010)
  119},
\href{http://arxiv.org/abs/0911.3704}{{\tt arXiv:0911.3704 [hep-th]}}.

\bibitem{Baba:2009kr}
Y.~Baba, N.~Ishibashi, and K.~Murakami, ``{Light-Cone Gauge Superstring Field
  Theory and Dimensional Regularization},''
  \href{http://dx.doi.org/10.1088/1126-6708/2009/10/035}{{\em JHEP} {\bf 10}
  (2009)  035},
\href{http://arxiv.org/abs/0906.3577}{{\tt arXiv:0906.3577 [hep-th]}}.

\bibitem{Baba:2009zm}
Y.~Baba, N.~Ishibashi, and K.~Murakami, ``{Light-cone Gauge Superstring Field
  Theory and Dimensional Regularization II},''
  \href{http://dx.doi.org/10.1007/JHEP08(2010)102}{{\em JHEP} {\bf 08} (2010)
  102},
\href{http://arxiv.org/abs/0912.4811}{{\tt arXiv:0912.4811 [hep-th]}}.

\bibitem{Ishibashi:2010nq}
N.~Ishibashi and K.~Murakami, ``{Light-cone Gauge NSR Strings in Noncritical
  Dimensions II -- Ramond Sector},''
  \href{http://dx.doi.org/10.1007/JHEP01(2011)008}{{\em JHEP} {\bf 01} (2011)
  008},
\href{http://arxiv.org/abs/1011.0112}{{\tt arXiv:1011.0112 [hep-th]}}.

\bibitem{Ishibashi:2011fy}
N.~Ishibashi and K.~Murakami, ``{Spacetime Fermions in Light-cone Gauge
  Superstring Field Theory and Dimensional Regularization},''
  \href{http://dx.doi.org/10.1007/JHEP07(2011)090}{{\em JHEP} {\bf 07} (2011)
  090},
\href{http://arxiv.org/abs/1103.2220}{{\tt arXiv:1103.2220 [hep-th]}}.

\bibitem{Greensite:1986gv}
J.~Greensite and F.~R. Klinkhamer, ``{NEW INTERACTIONS FOR SUPERSTRINGS},''
\href{http://dx.doi.org/10.1016/0550-3213(87)90256-2}{{\em Nucl. Phys.} {\bf
  B281} (1987)  269}.

\bibitem{Greensite:1987hm}
J.~Greensite and F.~R. Klinkhamer, ``{SUPERSTRING AMPLITUDES AND CONTACT
  INTERACTIONS},''
\href{http://dx.doi.org/10.1016/0550-3213(88)90622-0}{{\em Nucl. Phys.} {\bf
  B304} (1988)  108}.

\bibitem{Greensite:1987sm}
J.~Greensite and F.~R. Klinkhamer, ``{CONTACT INTERACTIONS IN CLOSED
  SUPERSTRING FIELD THEORY},''
\href{http://dx.doi.org/10.1016/0550-3213(87)90485-8}{{\em Nucl. Phys.} {\bf
  B291} (1987)  557}.

\bibitem{Green:1987qu}
M.~B. Green and N.~Seiberg, ``{CONTACT INTERACTIONS IN SUPERSTRING THEORY},''
\href{http://dx.doi.org/10.1016/0550-3213(88)90549-4}{{\em Nucl. Phys.} {\bf
  B299} (1988)  559}.

\bibitem{Wendt:1987zh}
C.~Wendt, ``{SCATTERING AMPLITUDES AND CONTACT INTERACTIONS IN WITTEN'S
  SUPERSTRING FIELD THEORY},''
\href{http://dx.doi.org/10.1016/0550-3213(89)90118-1}{{\em Nucl. Phys.} {\bf
  B314} (1989)  209}.

\bibitem{Ishibashi:2013nma}
N.~Ishibashi and K.~Murakami, ``{Multiloop Amplitudes of Light-cone Gauge
  Bosonic String Field Theory in Noncritical Dimensions},''
  \href{http://dx.doi.org/10.1007/JHEP09(2013)053}{{\em JHEP} {\bf 09} (2013)
  053},
\href{http://arxiv.org/abs/1307.6001}{{\tt arXiv:1307.6001 [hep-th]}}.

\bibitem{D'Hoker:1988ta}
E.~D'Hoker and D.~H. Phong, ``{The Geometry of String Perturbation Theory},''
\href{http://dx.doi.org/10.1103/RevModPhys.60.917}{{\em Rev. Mod. Phys.} {\bf
  60} (1988)  917}.

\bibitem{Mandelstam:1985ww}
S.~Mandelstam, ``{THE INTERACTING STRING PICTURE AND FUNCTIONAL
  INTEGRATION},''. Lectures given at Workshop on Unified String Theories, Santa
  Barbara, CA, Jul 29 - Aug 16, 1985.

\bibitem{AlvarezGaume:1987vm}
L.~Alvarez-Gaume, J.~B. Bost, G.~W. Moore, P.~C. Nelson, and C.~Vafa,
  ``{Bosonization on higher genus Riemann surfaces},''
\href{http://dx.doi.org/10.1007/BF01218489}{{\em Commun. Math. Phys.} {\bf 112}
  (1987)  503}.

\bibitem{Verlinde:1986kw}
E.~P. Verlinde and H.~L. Verlinde, ``{Chiral bosonization, determinants and the
  string partition function},''
\href{http://dx.doi.org/10.1016/0550-3213(87)90219-7}{{\em Nucl. Phys.} {\bf
  B288} (1987)  357}.

\bibitem{Dugan:1987qe}
M.~J. Dugan and H.~Sonoda, ``{FUNCTIONAL DETERMINANTS ON RIEMANN SURFACES},''
\href{http://dx.doi.org/10.1016/0550-3213(87)90378-6}{{\em Nucl. Phys.} {\bf
  B289} (1987)  227}.

\bibitem{Sonoda:1987ra}
H.~Sonoda, ``{FUNCTIONAL DETERMINANTS ON PUNCTURED RIEMANN SURFACES AND THEIR
  APPLICATION TO STRING THEORY},''
\href{http://dx.doi.org/10.1016/0550-3213(87)90578-5}{{\em Nucl. Phys.} {\bf
  B294} (1987)  157}.

\bibitem{Wentworth:1991}
R.~Wentworth, ``The asymptotics of the arakelov-green's function and faltings'
  delta invariant,'' {\em Commun. Math. Phys.} {\bf 137} (1991)  427.

\bibitem{Wentworth:2008}
R.~A. Wentworth, ``Precise constants in bosonization formulas on riemann
  surfaces. i,'' {\em Commun. Math. Phys.} {\bf 282} (2008)  339.

\bibitem{Faltings:1984}
G.~Faltings, ``Calculus on arithmetic surfaces,'' {\em Ann. of Math.} {\bf 119}
  (1984)  387.

\bibitem{Berkovits:1985ji}
N.~Berkovits, ``{CALCULATION OF SCATTERING AMPLITUDES FOR THE NEVEU-SCHWARZ
  MODEL USING SUPERSHEET FUNCTIONAL INTEGRATION},''
\href{http://dx.doi.org/10.1016/0550-3213(86)90070-2}{{\em Nucl. Phys.} {\bf
  B276} (1986)  650}.

\bibitem{Berkovits:1987gp}
N.~Berkovits, ``{SUPERSHEET FUNCTIONAL INTEGRATION AND THE INTERACTING
  NEVEU-SCHWARZ STRING},''
\href{http://dx.doi.org/10.1016/0550-3213(88)90642-6}{{\em Nucl. Phys.} {\bf
  B304} (1988)  537}.

\bibitem{Aoki:1990yn}
K.~Aoki, E.~D'Hoker, and D.~H. Phong, ``{UNITARITY OF CLOSED SUPERSTRING
  PERTURBATION THEORY},''
\href{http://dx.doi.org/10.1016/0550-3213(90)90575-X}{{\em Nucl. Phys.} {\bf
  B342} (1990)  149--230}.

\bibitem{Berkovits1990}
N.~Berkovits, ``{Supersheet Functional Integration and the Calculation of Nsr
  Scattering Amplitudes Involving Arbitrarily Many External Ramond Strings},''
\href{http://dx.doi.org/10.1016/0550-3213(90)90088-U}{{\em Nucl. Phys.} {\bf
  B331} (1990)  659}.

\bibitem{Berkovits1992a}
N.~Berkovits, ``{The Heterotic Green-Schwarz superstring on an N=(2,0)
  superworldsheet},''
  \href{http://dx.doi.org/10.1016/0550-3213(92)90591-X}{{\em Nucl. Phys.} {\bf
  B379} (1992)  96--120},
\href{http://arxiv.org/abs/hep-th/9201004}{{\tt arXiv:hep-th/9201004
  [hep-th]}}.

\bibitem{Berkovits1993}
N.~Berkovits, ``{Calculation of Green-Schwarz superstring amplitudes using the
  N=2 twistor string formalism},''
  \href{http://dx.doi.org/10.1016/0550-3213(93)90209-8}{{\em Nucl. Phys.} {\bf
  B395} (1993)  77--118},
\href{http://arxiv.org/abs/hep-th/9208035}{{\tt arXiv:hep-th/9208035
  [hep-th]}}.

\bibitem{Berkovits2002}
N.~J. Berkovits and J.~M. Maldacena, ``{N=2 superconformal description of
  superstring in Ramond-Ramond plane wave backgrounds},''
  \href{http://dx.doi.org/10.1088/1126-6708/2002/10/059}{{\em JHEP} {\bf 10}
  (2002)  059},
\href{http://arxiv.org/abs/hep-th/0208092}{{\tt arXiv:hep-th/0208092
  [hep-th]}}.

\bibitem{Green1983b}
M.~B. Green and J.~H. Schwarz, ``{Superstring Interactions},''
\href{http://dx.doi.org/10.1016/0550-3213(83)90475-3}{{\em Nucl. Phys.} {\bf
  B218} (1983)  43--88}.

\bibitem{Green1983}
M.~B. Green, J.~H. Schwarz, and L.~Brink, ``{Superfield Theory of Type II
  Superstrings},''
\href{http://dx.doi.org/10.1016/0550-3213(83)90651-X}{{\em Nucl. Phys.} {\bf
  B219} (1983)  437--478}.

\bibitem{D'Hoker:1989ae}
E.~D'Hoker and D.~H. Phong, ``{FUNCTIONAL DETERMINANTS ON MANDELSTAM
  DIAGRAMS},''
\href{http://dx.doi.org/10.1007/BF01218453}{{\em Commun. Math. Phys.} {\bf 124}
  (1989)  629--645}.

\bibitem{Mandelstam:1991tw}
S.~Mandelstam, ``{The n loop string amplitude: Explicit formulas, finiteness
  and absence of ambiguities},''
\href{http://dx.doi.org/10.1016/0370-2693(92)90961-3}{{\em Phys. Lett.} {\bf
  B277} (1992)  82--88}.

\end{thebibliography}\endgroup

\end{document}